\documentclass[defaultstyle,11pt]{thesis}

\usepackage{hyperref}		

\usepackage{graphicx,amsfonts,amssymb,amsmath,epsfig,amsthm,multicol,fancyhdr, slashed,eucal}
\usepackage{mathrsfs}
\usepackage{mathtools}
\usepackage{xcolor}
\usepackage{simplewick} 
\usepackage{comment}

\usepackage{cite}
\usepackage[utf8]{inputenc}

\usepackage[title]{appendix}

\usepackage{csquotes}

\graphicspath{{home/}}

\numberwithin{equation}{section}

\DeclareTextFontCommand{\emph}{\em}

\newcommand{\bo}{\boldsymbol}
\newcommand{\smallfrac}[2]{{\textstyle\frac{#1}{#2}}}
\newcommand{\BE}{\begin{equation}}
\newcommand{\EE}{\end{equation}}
\newcommand{\mrm}{\mathrm}

\newcommand{\dd}{\mathrm{d}}

\newcommand{\me}{\mathrm{e}}
\newcommand{\mcal}{\mathcal}
\newcommand{\mn}{\mathnormal}
\newcommand{\del}{\partial}
\newcommand{\eps}{\epsilon}
\newcommand{\veps}{\varepsilon}

\newcommand{\tot}{\mrm{tot}}
\newcommand{\mLam}{\mn\Lambda}

\newcommand{\mPhi}{\mn\Phi}
\newcommand{\mDelt}{\mn\Delta}
\newcommand{\mSig}{\mn\Sigma}
\newcommand{\fdelphi}{ \frac{\delta}{\delta \phi} }

\newcommand{\fdelAB}[2]{ \frac{\delta #1}{\delta #2} }

\newcommand{\ac}[1]{{\small{\bf\color{cyan} {#1}}}}
\newcommand{\ren}{\mrm{r}}
\newcommand{\con}{\mrm{c}}
\newcommand{\MSbar}{\overline{\mrm{MS}}}
\newcommand{\barpsi}{\overline{\psi}}

\newcommand{\Bnab}{
	\mbox{$
		\raisebox{+.21ex}{\scalebox{0.8}{\rotatebox[origin=c]{7}{$\lambda$}}}\hspace{-0.6em}
		\nabla
		\hspace{-0.871em}\raisebox{.1ex}{\scalebox{1.79}{$\color{white}.$}}
		\hspace{-0.185em}\raisebox{.64ex}{\scalebox{0.65}{$\color{white}.$}}
		\hspace{-.309em}\raisebox{.47ex}{\scalebox{0.5}{\rotatebox[origin=c]{+26.6}{$\color{white}\bo{|}$}}}
		\hspace{-.342em}\raisebox{.58ex}{\scalebox{0.5}{\rotatebox[origin=c]{+26.6}{$\color{white}\bo{|}$}}}
		\hspace{-0.251em}\raisebox{.18ex}{\scalebox{0.97}{\rotatebox[origin=c]{-26.6}{$\color{white}\bo{|}$}}}
		\hspace{-1.14em}\raisebox{.18ex}{\scalebox{0.95}{\rotatebox[origin=c]{+26.6}{$\color{white}\bo{|}$}}}
		\hspace{+.3em}
	$}{}
}
\newcommand{\sBnab}{\mbox{$\scalebox{0.8}{$\Bnab$}$}}

\newcommand{\lsim}{\mathrel{\raisebox{-.6ex}{$\stackrel{\textstyle<}{\sim}$}}}
\newcommand{\gsim}{\mathrel{\raisebox{-.6ex}{$\stackrel{\textstyle>}{\sim}$}}}

\newcommand{\nn}{\nonumber}

\newcommand{\fig}[1]{figure~\ref{#1}}

\long\def\/*#1*/{}
\definecolor{red}{rgb}{1.0, 0, 0}

\newcommand{\op}[0]{\ensuremath{\mathcal{O}}}
\newcommand{\Rop}[0]{\ensuremath{\mathcal{R}}}
\newcommand{\symop}[0]{\ensuremath{\mathcal{A}}}

\title{Novel Approaches to Renormalization Group Transformations in the Continuum and on the Lattice}

\author{Andrea A.}{Carosso}

\otherdegrees{B.S., B.A., University of Delaware, 2014 \\
	      M.S., University of Colorado Boulder, 2017}

\degree{Doctor of Philosophy}		
	{Ph.D., Physics}		

\dept{Department of}			
	{Physics}		

\advisor{Prof.}				
	{Anna Hasenfratz}			

\reader{Prof. Ethan Neil}		

\abstract{  

This thesis is, broadly speaking, on the subject of the Renormalization Group (RG), that is, the systematic means by which we understand how the physics of different energy or length scales interact with one another, and the dependence of physical quantities on the scale of their description. RG is of fundamental importance in the physical sciences; applications of RG range from the problem of modeling turbulence all the way to particle physics and quantum theories of gravity. RG grew out of quantum field theory, where it provided the conceptual tools necessary for a deeper understanding of renormalization, the apparent sensitivity of low-energy processes to high-energy physics. In statistical physics, RG played a central role in explaining the nontrivial phenomena associated with systems living at their critical points. By introducing the notion of \textit{fixed points} in phase diagrams, RG was able to describe the origin of critical behavior.

The thesis that follows is, in particular, about \textit{new} methods of achieving RG transformations, in both a continuum spacetime background and on a lattice discretization thereof. The subject is explored from the point of view of euclidean quantum field theory, or perhaps more accurately, statistical field theory. As a thesis grounded on the computational method of lattice simulation, I emphasize the role of lattice formulations throughout the work, especially in the first two chapters. In the first, I describe the essential aspects of lattice theory and its symbiosis with RG. In the second, I present a new, \textit{continuous} approach to RG on the lattice, based on a numerical tool called Gradient Flow (GF). Simulation results from quartic scalar field theory in 2 and 3 dimensions ($\phi^4_d$) and 4-dimensional 12-flavor SU(3) gauge theory, will be presented. In the third and fourth chapters, the focus becomes more analytic. Chapter 3 is an introductory review of Functional Renormalization Group (FRG). In chapter 4, I introduce the concept of Stochastic RG (SRG) by working out the relationship between FRG and stochastic processes.
}

\dedication[Dedication]{
\emph{In memoria di mia nonna, Teresita}.
}

\acknowledgements{	\OnePageChapter	
	
I am greatly indebted to several people for their help and support in carrying out this work; it would not have happened the way it did without them:

First, I thank my advisor, Anna Hasenfratz; if I have learned anything about the renormalization group it has been due, in large part, to many lessons and discussions with Anna (and a long time spent squinting and ruminating over the numerous RG flow diagrams she drew up on the fly). Next, I thank Ethan Neil, not only for instructing me in the fine art of data analysis, but also for being willing to both listen to, and to provide valuable feedback on, many of my dubious ideas. I would also like to thank Tom DeGrand for his advice and support, and for the many discussions we have had over the years; I have particularly enjoyed our mutual interest in the history of physics.

Next, I would like to thank some of my peers. I thank Dan Hackett for innumerable conversations in which wild speculations on physics were forwarded, and for always being willing to help me with difficult debugging problems. I also  thank Will Jay and Oscar Henriksson, with whom a wealth of interesting physics and mathematics has been pondered.

And from the heart, lastly, I thank my family and friends for their steadfast love and support, and for having been there for me throughout my various waverings.

}

\ToCisShort	

\LoFisShort	

\LoTisShort	

\begin{document}

\chapter{Lattice theory}

\begin{displayquote}
\textit{``I listened to K. Hepp (1963 - 64) and others describe their results in axiomatic field theory; I didn't understand what they said in detail but I got the message that I should think in position space rather than momentum space.'' -- K. Wilson, Nobel Lecture, 1982 \cite{Wilson:1993dy}}
\end{displayquote}

\section{Introduction}

What do we mean by a lattice theory? Suppose we wish to model a physical system of fields, which may be strongly interacting, in order to render the problem of solving it tractable by means of, say, a computer simulation.\footnote{In many cases the theory being studied is not directly realized in nature (as far as we know), but for the sake of exposition, here we imagine it is --- lattice QCD is an example of a realistic theory. It is a virtue of lattice theory, however, to be able to study the physics of (most) any model one wishes in a nonperturbative way.} The system possesses various physical properties, which we characterize by mathematical quantities. One class of properties of particular importance is the following. We imagine that there is a principle of locality, such that the interactions of separate chunks of the field weaken with increasing separation. By ``interaction,'' we choose to mean \textit{correlation}, that the values which characterize local properties of the system become less correlated as we look at ever more distant pairs of chunks. We characterize this locality by what is called the \textit{correlation length} $\xi$, a number with units of distance.

Now, to construct our model, we consider a natural idea. Perhaps the problem will become tractable by \textit{approximating} the continuous spacetime background by a discrete lattice of points, restricting the physical entities (fields) to take values only on those points (or on the links connecting them, in the case of gauge theory), and discretizing the interactions in some way. We call this the ``lattice model'' of the physical system. The separation between the points we denote by $a$, the lattice spacing. The lattice model, if it's a good model, should be able to predict approximations to the properties of the real system. That is, given some set of input parameters $g$, the model should ultimately produce numerical quantities in rough agreement with those of the real system, including a correlation length $\hat \xi$, which we choose to be dimensionless and such that $a \hat \xi$ should approximate the physical value $\xi$.

One might initially believe that all there is to do is choose a value for $a$ and values for $g$, plug it into our simulation and -- voil\'a! -- obtain a description of the system in rough agreement with the real system: $a \hat \xi \approx \xi$. But $\hat \xi = \hat \xi(g)$ is a function of the input parameters, so picking random values for $a$ and $g$ will not generally yield the correct $\xi$; they only match for certain combinations of $a$ and $g$. Furthermore, we know we'll probably need to pick small values of $a$, that is, small with respect to $\xi$, to approach the true values, since we expect that modeling a field theory by only a few lattice sites will generally produce terrible approximations (and of course, the number of sites we simulate with must be finite). Suppose, then, that we choose a value for $a$, and then \textit{scan} the space of $g$ until $a \hat \xi \approx \xi$ is achieved with some desired degree of accuracy. That's perfectly fine. But notice that this statement is equivalent to the following one. For any choice of $g$, and given the \textit{empirical} value of $\xi$, a value of $a$ follows: $a = \xi / \hat \xi(g)$. We say that the pair $(g,\xi)$ \textit{sets the scale} of the simulation. This means we can construct our model in terms of entirely dimensionless quantities, measure $\hat \xi$, and determine $a$ by comparison with $\xi$. This latter approach is far more useful in practice. One reason is that, ultimately, we expect the model to better approximate the physics as $a$ becomes smaller and smaller, but simulations with small parameter values can be less efficient than ones with $O(1)$ parameters, generally, and simulation with $a=0$ in all functions is certainly a non-starter.  Thus in our simulations, we define the fields and any other quantities as dimensionless by scaling out the (to be determined) spacing $a$. Once we have determined $a$, we can measure any observable we wish and multiply it by appropriate powers of $a$ to obtain dimensionful predictions which approximate the physical system's properties, to a precision determined by $a$.



In many cases the real system is continuous, so we are often interested in obtaining a limit $a \to 0$, or at least $a$ so small that there is no discernible difference between what we observe in experiment and what we simulate. But this must therefore correspond to a particular limit $g \to g_*$ of the model parameters where $\hat \xi(g_*) = \infty$.\footnote{Because there is usually more than one parameter, there are usually many points in the space of parameters that constitute a continuum limit, and one therefore speaks of the \textit{critical surface} in parameter space, as they often form a submanifold.} If such a limit exists, we call it the \textit{continuum limit} of the lattice theory. If there does not exist any such limit (i.e. point in the space of $g$ parameters), then the lattice theory has no continuum limit, and therefore cannot describe any physics that is known to be continuous. In many cases, however, the system being simulated is actually discrete, for example in condensed matter systems like ferromagnets. In such cases, the existence of a continuum limit is nevertheless an essential aspect in the explanation of its critical properties, as we will come to understand throughout this chapter. It is even possible that the quantum description of gravity will have a fundamental discreteness about it, but whatever it is, it must possess a nontrivial long-distance limit in which it reproduces General Relativity; the existence of such a limit is related to the existence of a continuum limit. But we shall not in this thesis concern ourselves with theories of gravity.

We have determined that the spacing $a$, the physical value $\xi$, and the lattice parameters $g$, are intimately related. The manner in which they are related is therefore of paramount importance in lattice theory. Their relation is, for historical reasons, called the \textit{renormalization group}. Often the relation is characterized by an inversion of sorts, giving parameters as functions of the spacing: $g = g(a)$. If the model has a continuum limit, and if this limit occurs for small values $g_*$, then we may use perturbation theory to study the approach to the continuum. In the event that the limit occurs for $g_* = 0$, the theory is called \textit{asymptotically free}. The enterprise of lattice QCD is based on the assumption of asymptotic freedom: $a \to 0$ as $g \to 0$. But continuum limits need not always occur for small $g$. When $g_*$ is large, a nonperturbative means of determining the continuum limit is necessary. Assuming the physical $\xi$ is always finite, then the continuum limit is characterized by the phenomenon of $\hat \xi \to \infty$. Thus, by simulating a lattice model at many $g$ values until $\hat \xi$ is observed to get ever larger in some region of parameter space, one may nonperturbatively approaching continuum limits, in principle. The problem in practice with this approach is that such an array of simulations can become extremely costly computationally, so other methods must be devised, not to mention the inherent limitation of working in a finite volume, that $\hat \xi \leq N$ where $N$ is the number of lattice sites along each direction. One such method is based on the notion of \textit{finite-size scaling}, and we will explicitly see it in action when we use the Binder cumulant to locate the continuum limit of a particular scalar model.

\section{Discretization and lattice models}

We now give a description of how one typically constructs a lattice theory. We mostly follow the presentation of Montvay and M\"unster \cite{Montvay:1994cy}, except specifying to $d=3$ rather than 4 in the discussion of renormalized quantities. As such, the material in this subsection is largely just review, and many demonstrations are omitted.

\subsection{Lattice actions} In this work we will often focus on the infamous scalar field theory in dimension $d$ with quartic interaction, denoted $\phi^4_d$, and determined by the continuum action
\BE \label{phi4d_action}
S(\varphi) = \int_{\mathbb{R}^d} \dd^d x \Bigg( \frac{1}{2} (\del \varphi)^2(x) + \frac{m_0^2}{2} \varphi^2(x) + \frac{g_0}{4!} \varphi^4(x) \Bigg).
\EE
To discretize the theory, we define a square lattice $\mathfrak{L}_d := (a \mathbb{Z}_N)^d$ where $a$ is the lattice spacing, $N$ denotes the number of sites in each direction, and $\mathbb{Z}_N$ is the integers mod $N$. Sites of the lattice with dimension of distance are written as $x = an$ for $n\in\mathbb{Z}_N$. The dimensionful lattice field will be denoted $\varphi(x)$. We also typically will assume periodic boundary conditions for the fields, $\varphi(x+N a \mrm{e}_\mu) = \varphi(x)$, for $\me_\mu$ the unit vector in the $\mu$ direction, $\mu = 1,\dots,d$. The corresponding lattice action is obtained by choosing a discretization of the spatial derivatives. The simplest choice is the forward-difference operator $a \hat \del_\mu f(x) = f(x+a\me_\mu)-f(x)$, which yields the action
\BE
S(\varphi) = a^d \sum_{x \in \mathfrak{L}_d} \Bigg( \frac{1}{2} \sum_{\mu = 1}^d (\hat \del_\mu \varphi (x) )^2 + \frac{m_0^2}{2} \varphi^2(x) + \frac{g_0}{4!} \varphi^4(x) \Bigg).
\EE
This is not the only choice of discretization, however. Intuitively, \textit{any} terms differing from the continuum action by $O(a)$ are valid lattice actions, as they would then imply the same continuum limit.\footnote{A more precise statement is that any action in the same \textit{universality class} constitutes a valid discretization. What is needed is that the long-distance properties of both the continuum action and the lattice discretization are identical.} It will even become apparent that adding higher-order terms like $\varphi^6$ are valid as well!

The lattice action above is not yet dimensionless, as it depends explicitly on the spacing $a$, and as such is not convenient for simulations. One option is to scale $a$ out of all quantities in the action and simulate with that. Another option that is most popular is to instead define the simulation action by\footnote{Most of the time we will not bother with dimless site index labels $n$ for $\hat \varphi$, where $x= an$, using $x$ as the argument for both dimless and dimful fields. We also write $x + \mu$ to mean $x + a \me_\mu$, where $\me_\mu$ is a unit vector along the $\mu$ direction.}
\BE \label{phi4_lat_action}
S(\hat \varphi) = \sum_{x \in \mathfrak{L}_d} \Big( - \beta \sum_{\mu = 1}^d \hat \varphi(x) \hat \varphi(x+\mu)  + \hat \varphi^2(x) + \lambda (\hat \varphi^2(x) - 1)^2 \Big),
\EE
which is equivalent to the dimensionful lattice action by the relations
\BE
\hat \varphi = a^{d_\phi} \sqrt{\beta} \; \varphi, \quad \frac{1}{2} (am_0)^2 = \frac{1-2\lambda}{\beta} - d, \quad g_0 = \frac{4! \lambda}{\beta^2} a^{d-4},
\EE
where $d_\phi = d/2-1$ is the canonical mass dimension of the (position space) field $\varphi$. The first term in the simulation action is of the form of a nearest-neighbor interaction, the same kind which defines the Ising model. The action furthermore possesses a $\mathbb{Z}_2$ symmetry $\varphi \mapsto - \varphi$. In the limit $\lambda \to \infty$, the partition function becomes that of the Ising model, since for larger $\lambda$ values, the contribution of configurations with $\hat \varphi^2 \neq 1$ becomes vanishingly small, thereby constraining the field to have unit size. Alternatively, one may go the opposite direction, and starting from an Ising model derive a scalar field theory by appropriately changing variables in the partition function, see \cite{Kopietz:2010zz}. It turns out that they are related in an even more general way, that of \textit{universality}, which we will describe when we get to RG.

For perturbative calculations, it is convenient to work in momentum space. The continuum action in momentum space is obtained by Fourier transformation and is given by
\begin{align}
S(\varphi) = \frac{1}{2} & \int_{\mathbb{R}^d} \frac{\dd^d p}{(2\pi)^d} \varphi(p) \big( p^2 + m_0^2 \big) \varphi(-p) \nonumber \\
& + \frac{g_0}{4!} \Big( \prod_{i=1}^4 \int_{\mathbb{R}^d}  \frac{\dd^d p_i}{(2\pi)^d} \; \varphi(p_i) \Big) (2 \pi)^d \delta(p_1 + p_2 + p_3 + p_4).
\end{align}
A continuum action needs to be regularized, for example, with a momentum cutoff $\mLam$ or by continuing $d$ onto the complex plane. This becomes apparent if one does perturbation theory without a regulator, where one encounters singular loop integrals (see eq. (\ref{mass_divergence})), but it should be noted that such singularities occur even for free theories in observables at zero distance. Although we will focus on lattice regularization and sharp cutoffs in the rest of this chapter, we will work with smooth cutoffs in the continuum in chapters 3 and 4. The dimful\footnote{``Dimensionful'' and ``dimensionless'' are sometimes shortened to ``dimful'' and ``dimless'' in this work.} lattice action in momentum space is given by
\begin{align}
S(\varphi) = \frac{1}{2 V} & \sum_{p \in \mathfrak{B}_d} \varphi(p) \big( \hat p^2 + m_0^2 \big) \varphi(-p) \nonumber \\
& + \frac{g_0}{4! V^4} \Big( \prod_{i=1}^4 \sum_{p_i \in \mathfrak{B}_d} \varphi(p_i) \Big) V \hat \delta(p_1 + p_2 + p_3 + p_4),
\end{align}
where $V = L^d = (aN)^d$ is the lattice volume, $\mathfrak{B}_d \cong (2 \pi / a)^d \mathbb{Z}^d_N$ is the Brillouin zone, $\hat \delta(p)$ is the (dimless) Kronecker delta, and
\BE
\hat p_\mu = \frac{2}{a} \sin \frac{p_\mu a}{2}
\EE
is the often-encountered lattice momentum function.

\subsection{Observables} All the observable quantities predicted by a quantum field theory have expressions as expectation values of functions of the fundamental field. Furthermore, these are the quantities directly measured in lattice simulations. 

The observables of the theory are the expectation values
\BE
\langle \mcal{O}(\varphi) \rangle_S = \frac{1}{Z} \int \mathscr{D} \varphi \; \mcal{O}(\varphi) \me^{-S(\varphi)}, \quad \mrm{where} \quad Z = \int \mathscr{D} \varphi \; \me^{-S(\varphi)}
\EE
is the partition function. From the statistical physics perspective, the factor $\me^{-S}$ is a Boltzmann weight over the configuration space of $\varphi$, and $Z$ is the factor which normalizes the probability distribution. The free energy of the system with respect to that of the corresponding free theory is given by
\BE
F - F_0 = - \ln Z / Z_0,
\EE
where $Z_0$ is the free theory partition function. The expression on the r.h.s. is equal to a sum over all connected vacuum diagrams, a result known as the \textit{linked cluster theorem} \cite{Kopietz:2010zz, ZinnJustin:2002ru}. By separating the quadratic (gaussian) part of the action from the interaction $V(\varphi)$, the observables admit a perturbative representation
\BE \label{pert_series}
\langle \mcal{O}(\varphi) \rangle_S = \frac{Z_0}{Z} \sum_{n=0}^\infty \frac{(-1)^n}{n!} \langle \mcal{O}(\varphi) [V(\varphi)]^n \rangle_{0},
\EE
where $\langle \cdot \rangle_0$ denotes expectations of the free theory. The role of the leading factor of $Z_0/Z$ is to  divide out all the vacuum bubbles. The observables can be considered as generated by a certain functional, the sourced partition function
\BE
Z(J) = \int \mathscr{D} \varphi \; \me^{-S(\varphi) + J \circ \varphi},
\EE
where\footnote{This notation is sometimes used in the literature to reduce clutter. It will be generalized to a functional tensor notation in chapter 4.}
\BE
J \circ \varphi := \int \dd^d x J(x) \varphi(x).
\EE
On a lattice, the integral would be replaced by a summation. The $n$-point functions are given in terms of $Z(J)$ by\footnote{$G^{(n)}$ is used rather than $Z^{(n)}$ because the latter would lead to great confusion once $Z$ factors are introduced.}
\BE
\langle \varphi(x_1) \cdots \varphi(x_n) \rangle = \frac{\delta}{\delta J(x_1)} \cdots \frac{\delta}{\delta J(x_n)} \; Z(J) \Big|_{J=0} = G^{(n)}(x_1, \dots , x_n).
\EE
Such observables do not typically decay as the separations $|x_i - x_j| \to \infty$, but the \textit{connected} observables do decay (under the assumptions of the cluster decomposition principle \cite{ZinnJustin:2002ru}):
\BE
\langle \varphi(x_1) \cdots \varphi(x_n) \rangle^\mrm{c} = \frac{\delta}{\delta J(x_1)} \cdots \frac{\delta}{\delta J(x_n)} \; W(J) \Big|_{J=0} = W^{(n)}(x_1, \dots , x_n),
\EE
where the generator of connected $n$-point functions is defined by
\BE
\me^{W(J)} := \frac{1}{Z_0} \int \mathscr{D} \varphi \; \me^{-S(\varphi) + J \circ \varphi} = Z(J) / Z_0.
\EE
The normalization by the free partition function guarantees that $W^{(0)} = -F + F_0$. 

The last generating functional we consider, for now, is the 1PI generator $\Gamma(v)$ defined as the Legendre transform of $W(J)$:
\BE
\Gamma(v) := v \circ J(v) - W(J(v)),
\EE
where $v(x) = \langle \varphi(x) \rangle_J$ is the vacuum expectation value (``vev'') of the field from the sourced action. Notice that we may write $v$ as
\BE
v(x) = \frac{\delta W(J)}{\delta J(x)},
\EE
to compute
\BE
\fdelAB{\Gamma(v)}{v(x)} = J(x) + v \circ \fdelAB{J}{v(x)} - \fdelAB{W(J)}{J} \circ \fdelAB{J}{v(x)} = J(x).
\EE
We can interpret this equation as providing a quantum equation of evolution of the vev $v(x)$, determined by the ``quantum effective action'' $\Gamma(v)$. The derivatives of $\Gamma(v)$ are the vertex functions $\Gamma^{(n)}$, which play a central role in renormalization theory. By differentiating the previous equation by $J(y)$ and using the chain rule, we find
\BE \label{Gamma2W2}
(\Gamma^{(2)} W^{(2)})(x,y) = \delta(x-y),
\EE
where the product $\Gamma^{(2)} W^{(2)}$ makes sense as matrix multiplication. $\delta(x-y)$ are the components of the identity matrix in the position basis. It follows that
\BE
\Gamma^{(2)} = [W^{(2)}]^{-1}.
\EE
The relation between the $\Gamma^{(n)}$ and $W^{(n)}$ follow from repeated differentiation of this formula with respect to $\varphi$. The relation of higher $n$-point functions can be deduced by repeated differentiation of eq. (\ref{Gamma2W2}).

Denoting by $\Delta$ the free propagator of the theory, $W^{(2)}$ has a series of the form
\BE
W^{(2)} = \frac{1}{\Delta^{-1} + \Sigma} = \mDelt - \mDelt \Sigma \mDelt + \mDelt \Sigma \mDelt \Sigma \mDelt + O(\Sigma^3),
\EE
where $\Sigma$ is the self-energy matrix. Hence,
\BE
\Gamma^{(2)} = \Delta^{-1} + \Sigma,
\EE
which is an essential quantity in any field theory. In momentum space, the propagator is diagonal by translation invariance, meaning that
\BE
\Gamma^{(2)}(p_1,p_2) = (2\pi)^d \delta(p_1 + p_2) \Gamma^{(2)}(p_2),
\EE
where the notation $\Gamma^{(2)}(p)$ is convenient for the non-delta part.\footnote{Similarly, in position space the components of the 2-point function are often written as $G^{(2)}(x,y) = G^{(2)}(x-y)$.}
An important observable in a $\phi^4$ theory is the 4-point vertex function
\BE
\Gamma^{(4)}(p_1,p_2,p_3,p_4) =- \frac{W^{(4)}(p_1,p_2,p_3,p_4)}{W^{(2)}(p_1) W^{(2)}(p_2) W^{(2)}(p_3) W^{(2)}(p_4)},
\EE
which determines the $2 \to 2$ scattering amplitudes of particles, and therefore characterizes the strength of the interaction of the particles in the theory. It is proportional to a momentum-conserving delta function.

\subsection{Renormalized couplings} When one carries out the computation of observables determined by eq. (\ref{pert_series}), one often encounters \textit{cutoff sensitivity}: subleading orders in the perturbative series generally contain terms proportional to powers of $1/a$ or $\ln a \mu$ for some mass scale $\mu$, which diverge as $a \to 0$. Historically, this led to the development of the theory of \textit{renormalization}, which had roots in the first work on field theory by the founders of quantum mechanics in the 1930's, and which was given a firm foundation by F. Dyson around 1950  \cite{Dyson:1949bp, Dyson:1997gy}. Under renormalization, one systematically ``eliminates'' the sensitivity to the cutoff by defining renormalized parameters and reexpressing the perturbation series in terms of these parameters. If, by defining a finite number of such renormalized parameters, the resulting series has \textit{no} cutoff sensitivity, meaning that all $a$-dependence is order $O(a^2 \ln^\ell a \mu)$, then the theory is called \textit{perturbatively renormalizable}. It was thought for many years, up until the mid 1970's, that quantum field theories needed to be renormalizable in order to be serious candidates for the description of real physics. The advent of Wilsonian RG and the notion of effective field theory were to eventually undermine such philosophies \cite{Cao:1993gpm, Cao:1997my}. Nevertheless, renormalization is still important in field theory: it is the procedure by which the parameters used to define a theory are related to experimentally determined parameters, which is a necessary step in any physical science.

Continuum conventions typically take the coefficient of $p^2$ in $\Gamma^{(2)}(p)$ to be 1, a procedure called \textit{wave function renormalization}. Since the measured propagator does not typically satisfy this condition, one defines the $Z$ \textit{factor} of $\varphi$ by
\BE
Z_\phi = \frac{\dd}{\dd p^2}  \Gamma^{(2)}(p)\Big|_{p^2=0},
\EE
and then defines the \textit{renormalized field} $\varphi_\ren := \varphi / \sqrt{Z_\phi}$, so that
\BE
1 = \frac{\dd \Gamma^{(2)}_\ren(p)}{\dd p^2}\Big|_{p^2=0}.
\EE
The renormalized connected functions and 1PI functions then satisfy
\BE
W_\ren^{(n)} = Z_\phi^{-\frac{n}{2}} W^{(n)}_0, \quad \Gamma_\ren^{(n)} = Z_\phi^{\frac{n}{2}} \Gamma^{(n)}_0,
\EE
where a 0-subscript has been put on the bare $n$-point functions to further distinguish them from renormalized $n$-point functions. Thus, the only difference in the values of these functions is a proportionality factor by some power of $Z_\phi$.

In a euclidean theory, the correlation length $\xi$ is determined by the inverse of the renormalized mass (the smallest eigenenergy above zero in the spectrum of the theory), which is defined by
\BE
m^2_\ren = \Gamma^{(2)}_\ren(p)|_{p=0}.
\EE
It is a physical quantity, as it sets the rate of exponential decay of correlations among distant parts of the system. Another observable of interest is the dimful renormalized coupling
\BE
\lambda_\ren = \Gamma^{(4)}_\ren(p_1, p_2, p_3, p_4)|_{p_i^2 = 0},
\EE
which characterizes the strength of the interactions between particles, as mentioned in the last subsection. Both of the renormalized couplings are \textit{long-distance} quantities, since they are defined at zero momentum. 
These couplings can be thought of as the observable counterparts to the bare couplings $m_0^2, \; g_0$, which are the input parameters of the lattice model, since they agree at leading order in perturbation theory, as we will soon observe.

The renormalized couplings of the theory are totally determined by the bare couplings and the cutoff. As such, we can write them as functions thereof:
\begin{align}
Z_\phi & = \smallfrac{\dd}{\dd p^2} \Gamma^{(2)}_0(p; m_0, g_0, a) |_{p=0} , \nonumber \\
m^2_\ren & = \Gamma^{(2)}_\ren(p; m_0,g_0, a)|_{p=0}, \nonumber \\
\lambda_\ren & = \Gamma^{(4)}_\ren(\bo p; m_0, g_0, a)|_{\bo p = 0},
\end{align}
where $\bo p$ is the 4-tuple of momenta. In the next section, we will describe RG in somewhat general terms, but we shall follow along with the example of $\phi^4_3$. To that end, let us find expressions for the renormalized couplings above.

In perturbation theory, observables are determined from eq. (\ref{pert_series}) with the lattice action $S(\varphi)$ and applying Wick's theorem for gaussian integrals. The bare connected 2-point function is found to be
\BE
W^{(2)}_0(p) = \mDelt(p) - \frac{\lambda_0}{2} \mDelt(p) \mDelt(-p) \sum_\ell \mDelt(\ell) + O(\lambda_0^2),
\EE
where the free propagator is
\BE
\mDelt(p) = \frac{1}{\hat p^2 + m_0^2}.
\EE
The inverse of $W^{(2)}_0(p)$ gives us $\Gamma^{(2)}_0(p)$:
\BE
\Gamma^{(2)}_0(p) = \mDelt^{-1}(p) + \frac{\lambda_0}{2 V} \sum_\ell \mDelt(\ell) + O(\lambda_0^2) = \hat p^2 + m_0^2 + \frac{\lambda_0}{2 V} \sum_\ell \mDelt(\ell) + O(\lambda_0^2).
\EE
By expanding the lattice momenta $\hat p$ in $a$,
\BE
\hat p^2 = p^2 - \frac{a^2}{12} \sum_\mu p_\mu^4 + O(a^4 p^6),
\EE
we see that there is no change to the $p^2$ coefficient at 1-loop order, which means that $Z_\phi = 1 + O(\lambda_0^2)$. The renormalized mass, however, has a first order contribution like
\BE
m_\ren^2 = m_0^2 + \frac{\lambda_0}{2 V} \sum_\ell \mDelt(\ell) + O(\lambda_0^2).
\EE
The connected 4-point function at 1-loop order is
\begin{align}
W^{(4)}_0 & (\bo p) = V \hat \delta (p_\tot) \mDelt(p_1) \cdots \mDelt(p_4) \nonumber \\
& \times \Big[ - \lambda_0 + \frac{\lambda_0^2}{2 V} \sum_\ell \mDelt(\ell) \sum_{i=1}^4 \mDelt(p_i) + \frac{\lambda_0^2}{2 V} \sum_{i=1}^3 \sum_\ell \mDelt(\ell) \mDelt(\ell + p_{\sigma_i}) + O(\lambda_0^3) \Big].
\end{align}
Dividing by four factors of $W^{(2)}$ and expanding the denominator in $\lambda_0$ cancels the second term above, which is not 1PI. One then obtains the 1PI function, and evaluation at zero external momenta then yields (minus) the renormalized coupling, or
\BE
\lambda_\ren = \lambda_0 - \frac{3 \lambda_0^2}{2 V} \sum_\ell \mDelt(\ell) \mDelt(\ell) + O(\lambda_0^3).
\EE
We remark that corresponding to the dimful equations above are the dimless relations
\begin{align}
\hat m_\ren^2 & = \hat m_0^2 + \frac{\hat \lambda_0}{2 N^d} \sum_\ell a^2 \mDelt(\ell) + O(\hat \lambda_0^2), \nonumber \\
\hat \lambda_\ren & = \hat \lambda_0 - \frac{3 \hat \lambda_0^2}{2 N^d} \sum_\ell a^4 \mDelt(\ell) \mDelt(\ell) + O(\hat \lambda_0^3),
\end{align}
obtained by letting $a$ give dimension to all quantities, e.g. $\hat m_\ren = m_\ren a, \; \hat \lambda_\ren = a^{4-d} \lambda_\ren $. Up to  factors of $\beta$ coming from $\hat \varphi = a^{d_\phi} \sqrt{\beta} \varphi$, these couplings are directly measured in lattice simulations. Notice that the dimless renormalized couplings are therefore determined solely by the choice of dimless simulation parameters (and the lattice size $N$).

The evaluation of lattice loop integrals is generally more difficult than those of the continuum, and one resorts to expansion in $\hat m_0$ and numerical integrations for exact results, under the assumption that small $\hat m_0$ indeed is the interesting limit. The $\hat m_0$ expansions are typically asymptotic series, since the coefficients of the would-be Taylor expansion are often singular at some order.

To make our lives easier, we evaluate these integrals in the naive continuum limit, where deviations from the continuum result due to the lattice arise from the expansion of $\hat p^2$ in $p^2 a^2$. The renormalized mass with a sharp cutoff $\mLam = a^{-1}$ evaluates to
\BE \label{mass_divergence}
m_\ren^2 = m_0^2 + \frac{\lambda_0}{2} \; \Omega_3 \Big[ \frac{1}{a} - m_0 \arctan \frac{1}{a m_0}	\Big] + O(\lambda_0^2),
\EE
where $\Omega_d = S_{d-1} / (2 \pi)^d$ is a common factor arising in loop integrals; $S_{n}$ is the $n$-sphere surface area. For $d=3$, $\Omega_3 = 1/2\pi^2$. In perturbation theory, one is ultimately interested in replacing $m_0$ by $m_\ren$ in the series of other observables, so we expand the expression in powers of $m_0$:
\BE
m_\ren^2 = m_0^2 + \frac{\lambda_0}{4 \pi^2 a} \Big[ 1 - \frac{\pi}{2} a m_0 + O((a m_0)^2) \Big] + O(\lambda_0^2).
\EE
Multiplying by $a^2$ leads to
\BE \label{mren}
\hat m_\ren^2 = \hat m_0^2 + \frac{\hat \lambda_0}{4 \pi^2} \Big[ 1 - \frac{\pi}{2} \hat m_0 + O(\hat m_0^2) \Big] + O(\hat \lambda_0^2).
\EE
The continuum limit $a \to 0$ of the lattice model occurs for $\hat m_\ren = m_\ren a \to 0$ with $m_\ren \neq 0$. We see that the limit is equivalent to
\BE
\hat m_0^2 \to - \frac{1}{4 \pi^2} \hat \lambda_0 + O(\hat \lambda_0^2, \hat \lambda_0 \hat m_0).
\EE
In other words, we can approach the continuum limit of the model by fixing $\hat \lambda_0$ and \textit{tuning} $\hat m_0^2$ to some particular value, which to first order in perturbation theory is given as above.\footnote{Since a perturbative estimate may not always be reliable, this way of choosing simulation parameters is not taken, in practice.}

The dimful renormalized coupling is similarly given by
\BE
\lambda_\ren = \lambda_0 - \frac{3 \lambda_0^2}{8 \pi^2} \Big[ \frac{1}{m_0} \arctan \frac{1}{am_0} - \frac{a}{1 + a^2 m_0^2} \Big] + O(\lambda_0^3),
\EE
which (asymptotically) expands to
\BE
\lambda_\ren = \lambda_0 - \frac{3}{8 \pi^2} \frac{\lambda_0^2}{m_0} \Big[ \frac{\pi}{2} - 2 \hat m_0 + O(\hat m_0^3) \Big] + O(\lambda_0^3),
\EE
and multiplying by $a$ we find
\BE \label{lamren}
\hat \lambda_\ren = \hat \lambda_0 - \frac{3}{8 \pi^2} \frac{\hat \lambda_0^2}{\hat m_0} \Big[ \frac{\pi}{2}  - 2 \hat m_0 + O(\hat m_0^3) \Big] + O(\hat  \lambda_0^3).
\EE
I remark that a more useful dimensionless coupling to define is $g_\ren := \lambda_\ren / m_\ren$, as we will see at the end of the following section.

Once the series representation of the renormalized couplings has been obtained, one can invert them to obtain the bare couplings as series in renormalized ones. This allows all observables to be reexpressed in terms of renormalized parameters. When such a reexpression leads to a total elimination of cutoff-sensitivity, a theory is called \textit{renormalizable}, as mentioned before. In $\phi^4_3$ theory, there are in fact only two primitive diagrams that are cutoff-sensitive, which are renormalized by the mass $\hat m_\ren^2$ and the wave function $Z$-factor. This scenario is an instance of \textit{super-renormalizability}.  It will turn out, however, that in order to talk about the infrared properties of $\phi^4_3$, it is nevertheless important to define the renormalized coupling $\lambda_\ren$ and to express the perturbation series in terms of $\lambda_\ren$, or $g_\ren$.

\section{Perturbative renormalization group}

The term ``renormalization group'' was first used in 1953 by Stueckelberg and Petermann \cite{Petermann:1953wpa} to describe the transformations which relate renormalized couplings defined at various scales in QED4. The next year, Gell-Mann and Low introduced their analysis of the scale-dependent coupling of QED \cite{GellMann:1954fq}, which introduced the concept of the \textit{beta function}. The method of Gell-Mann and Low may be termed \textit{perturbative renormalization group}, as it concerns itself with equations derivable only in a perturbative context. Perturbative RG was brought to its final form by Callan \cite{Callan:1970yg} and Symanzik \cite{Symanzik:1970rt} in 1970, right as Wilson was starting to put his theory of RG together. Wilson's philosophy was inherently nonperturbative, even though many of its instances involved perturbation theory. On the lattice, nonetheless, it is useful to begin with an understanding of perturbative RG, as it applies well in many theories, including QCD. 

By comparing the measured value of $\hat m_\ren$ with the empirical correlation length, we can determine the lattice spacing by $a = \hat m_\ren / m_\ren = \xi / \hat \xi$, which again is an example of setting the scale. At fixed empirical scale $m_\ren$, a change in the cutoff $a$ therefore amounts to a change in $\hat m_\ren$, which is itself a function of $\hat m_0, \; \hat g_0$. Hence, a change in the cutoff is tantamount to a change in the bare parameters; this relationship is called the \textit{bare renormalization group}, which we describe below. Alternatively, we can consider the bare parameters to be fixed, and look at the change in renormalized observables as the renormalized mass $m_\ren$ is changed. This second perspective implies the \textit{Callan-Symanzik} equations. By studying these two faces of RG, we may form a picture of the behavior of a theory in the space of bare or renormalized parameters.

\subsection{Bare RG equations} In any lattice observable, we can in principle replace bare parameter dependence by renormalized parameter dependence, by using the equations which define them:
\BE
H^{(n)}_\ren(\hat g_\ren, \hat m_\ren) := \Gamma_\ren^{(n)}(\hat g_0(\hat g_\ren, \hat m_\ren), \hat m_0(\hat g_\ren, \hat m_\ren)).
\EE
Comparison with the correlation length determines the spacing $a$, and we can then define
\BE
\tilde H^{(n)}_\ren(g_\ren, m_\ren, a) := H^{(n)}_\ren(g_\ren a^{d_g}, m_\ren a),
\EE
where $d_g$ is minus the mass dimension of $g_\ren$. If the theory is \textit{perturbatively renormalizable}, then these functions have the nontrivial property of having a limit as $a \to 0$,
\BE
\tilde H^{(n)}_\ren(g_\ren, m_\ren, a) = \tilde H^{(n)}_\ren(g_\ren, m_\ren, 0) + O(a^2 \ln^\ell a),
\EE
where $\ell$ is some positive integer determined perturbatively. Renormalizability then implies (at fixed $g_\ren, m_\ren$)
\BE
a \frac{\del}{\del a} \tilde H^{(n)}_\ren(g_\ren, m_\ren, a) = O(a^2 \ln^\ell a).
\EE
The terms on the r.h.s. are called \textit{scaling violations}. Since the various $n$-point functions above are numerically equal, $\tilde H^{(n)} = H^{(n)} = \Gamma^{(n)}$, we can write the differential renormalizability statement in terms of the $\Gamma^{(n)}$,
\BE
a \frac{\dd}{\dd a}  \Gamma_\ren^{(n)}(\hat g_0(\hat g_\ren, \hat m_\ren), \hat m_0(\hat g_\ren, \hat m_\ren)) = O(a^2 \ln^\ell a),
\EE
where the $a$-dependence is implicit in $\hat g_\ren, \hat m_\ren$; we could therefore write the arguments of $\Gamma^{(n)}_\ren$ as $\hat g_0(a), \hat m_0(a)$. Such functions describe the family of bare parameters which all yield the same physics. It will be convenient to replace $\hat m_0$ by $\hat m_\ren$, which can be done in principle by solving the equation defining $\hat m_\ren$ for $\hat m_0$, to obtain functions $\tilde \Gamma^{(n)}_\ren(\hat g_0, \hat m_\ren)$, yielding
\BE
a \frac{\dd}{\dd a} \tilde \Gamma_\ren^{(n)}(\hat g_0(a), \hat m_\ren(a)) = O(a^2 \ln^\ell a).
\EE
Writing $\tilde \Gamma^{(n)}_\ren = Z_\phi^{n/2} \tilde \Gamma^{(n)}_0$, and then using the chain rule, while recalling that $\hat m_\ren = a m_\ren$, we find the \textit{bare RG equations}
\BE
\Big( \hat m_\ren \frac{\del}{\del \hat m_\ren} - \beta_\mrm{latt} \frac{\del}{\del \hat g_0} + n \gamma_\mrm{latt} \Big) \tilde \Gamma^{(n)}_0( \hat g_0, \hat m_\ren) \Big|_{g_\ren, m_\ren} = O(a^2 \ln^\ell a),
\EE
where the lattice beta function and anomalous field dimension are defined by\footnote{The sign is chosen so that decreasing $a$ is equivalent to increasing $m_\ren$ in the renormalized RG equations.}
\BE
\beta_\mrm{latt} := - a \frac{\dd \hat g_0}{\dd a} \Big|_{g_\ren, m_\ren} , \quad  \gamma_\mrm{latt} := \frac{a}{2} \frac{\dd \ln Z_\phi}{\dd a}\Big|_{g_\ren, m_\ren}.
\EE
The total derivatives here become partials when the bare parameters are expressed in terms of $(g_\ren, m_\ren, a)$ via $(\hat g_\ren, \hat m_\ren)$. A further consequence of perturbative renormalizability is that $\beta_\mrm{latt} = \beta_\mrm{latt}(\hat g_0)$ is a pure function of $\hat g_0$, up to scaling violations. If this function is known, the equation may be integrated to obtain $\hat g_0(a)$. Knowledge of the beta function is essential to understanding the approach to the continuum limit of a lattice theory, as we will soon see.

\subsection{Callan-Symanzik equations} A complimentary scenario is to consider the bare coupling $\hat g_0$ as a fixed parameter and to vary $\hat m_\ren$ via $m_\ren$. Varying $\hat m_\ren$ is equivalent to varying $\hat m_0$ at fixed $\hat g_0$ in a lattice simulation. From the relation
\BE
H^{(n)}_\ren(\hat g_\ren, \hat m_\ren) = \tilde \Gamma_\ren^{(n)}(\hat g_0, \hat m_\ren),
\EE
the total derivative of the l.h.s. with respect to $\hat m_\ren$ is
\BE
\Big( m_\ren \frac{\del}{\del m_\ren} + \beta_\ren \frac{\del}{\del \hat g_\ren} \Big) H^{(n)}_\ren(\hat g_\ren, \hat m_\ren), \quad \beta_\ren := m_\ren \frac{\dd \hat g_\ren}{\dd m_\ren} \Big|_{\hat g_0, a},
\EE
while that of the r.h.s. is
\BE
n \gamma_\ren \tilde \Gamma_\ren^{(n)}(\hat g_0, \hat m_\ren) + \Delta \tilde \Gamma_0^{(n)}(\hat g_0, \hat m_\ren),
\EE
where
\BE
\gamma_\ren := \frac{ m_\ren}{2} \frac{\dd \ln Z_\phi}{\dd m_\ren}\Big|_{\hat g_0}, \quad  \Delta \tilde \Gamma_0^{(n)}(\hat g_0, \hat m_\ren) := Z_\phi^{\frac{n}{2}} m_\ren \frac{\del}{\del m_\ren} \tilde \Gamma_0^{(n)}(\hat g_0, \hat m_\ren).
\EE
Writing everything in terms of $H^{(n)}$, we find the \textit{Callan-Symanzik equations} of $\phi^4_d$,
\BE
\Big( \hat m_\ren \frac{\del}{\del \hat m_\ren} + \beta_\ren \frac{\del}{\del \hat g_\ren} - n \gamma_\ren \Big) H^{(n)}_\ren(\hat g_\ren, \hat m_\ren) = \Delta \tilde \Gamma_0^{(n)}(\hat g_0, \hat m_\ren).
\EE
The r.h.s. is an observable which involves an insertion of the renormalized $\phi^2$ operator. A more thorough analysis of renormalizability must also include such insertions, but here we just report that the correlations of observables with arbitrary numbers of insertions of $\phi^2$ are also perturbatively renormalizable in $\phi^4_d$ theories \cite{ZinnJustin:2002ru}.\footnote{The CS equation in this form may look different from the forms we've grown used to due to the presence of the $\Delta \Gamma$ term. But this is a result of having used $\bo p = 0$ as the subtraction scale in the renormalization conditions, rather than some scale $\mu > 0$. See \cite{ZinnJustin:2002ru} for details.}

To sum up the previous two subsections, we have seen that the existence of a perturbatively renormalizable theory implies certain RG equations which describe the variation of observables, whether they're bare or renormalized ones, as the dimless correlation length is varied via $\hat m_\ren$. Being first order PDE's, they may be solved by the method of characteristics in the limit that we ignore scaling violations. These solutions constitute the \textit{scaling forms} of the observables in the continuum limit, an observation of far-reaching explanatory power in both field theory and critical phenomena.

\subsection{Continuum limits} 

For lattice simulations, the primary utility of beta functions is that they tell us how to simulate closer to the continuum limit, as we now describe. A general renormalized beta function will have the perturbative form
\BE
\beta_\ren(\hat g_\ren) = \hat m_\ren \frac{\dd \hat g_\ren}{\dd \hat m_\ren} \Big|_{\hat g_0,a} = \beta_1 \hat g_\ren + \beta_2 \hat g_\ren^2 + \beta_3 \hat g_\ren^3 + O(\hat g_\ren^4).
\EE
The sign of the beta function determines whether $\hat g_\ren$ decreases or increases as the cutoff is varied at fixed $\hat g_0$. As the continuum limit $\hat m_\ren \to 0$ is approached, we see that the behavior of $\hat g_\ren$ is determined by the zeros $\hat g_*$ of $\beta_\ren$. Such values are called \textit{fixed points} of the theory. Notice from the perturbative expression above that $\hat g_* = 0$ is always a fixed point, at least when the expansion above is valid. This is called the \textit{gaussian} fixed point (GFP). If $\beta_\ren$ is positive near the GFP, then as $\hat m_\ren \to 0$, the renormalized coupling approaches zero, and we say the theory is \textit{trivial}. In general, if the slope near a fixed point $\hat g_*$ is positive, then it attracts the renormalized coupling in the continuum limit, and we call such a fixed point an \textit{infrared} fixed point (IRFP). If the slope is negative, on the other hand, then $\hat g_\ren$ repels away from $\hat g_*$ in the continuum limit. These are called \textit{ultraviolet} fixed points (UVFP).

If we consider these cases from the bare RG perspective (where $\hat g_\ren$ is held fixed), then the bare coupling $\hat g_0$ behaves in the ``opposite'' way.
This matches our intuition that $\hat g_0$, as a UV quantity, should behave in an ``opposite'' way as $\hat g_\ren$, an IR quantity. Qualitatively, an IRFP repels $\hat g_0$, whereas a UVFP attracts it, in the continuum limit. If a renormalized beta function is monotonic and vanishes at $\hat g_\ren = 0$, then the behavior of the theory is relatively simple. If positive, one would approach a trivial theory $\hat g_\ren = 0$ in the continuum, and if negative, $\hat g_\ren$ grows in the continuum limit.

As a concrete example, we consider $\phi^4_3$, which has a nontrivial RG diagram, exhibiting both kinds of fixed points. For the parallel discussion of $\phi^4_4$, see Montvay and M\"unster sections 1.7 and 2.4. To compute the renormalized beta function, begin with the perturbative expression for the renormalized coupling,  eq. (\ref{lamren}):
\BE \label{lambda_ren}
\lambda_\ren = \lambda_0 - \frac{3}{8 \pi^2} \frac{\lambda_0^2}{m_0} \Big[ \frac{\pi}{2}  - 2 \hat m_0 + O(\hat m_0^2) \Big] + O(\hat  \lambda_0^3).
\EE
To study the variation as the continuum limit is approached, we replace $\hat m_0$ with $\hat m_\ren$ in eq. (\ref{lambda_ren}), valid at this order in perturbation theory. Since the renormalized coupling is a long-distance quantity, it is natural to give $\lambda_\ren$ dimension with the renormalized mass, defining the dimensionless coupling by \cite{Binney:1992vn, ZinnJustin:2002ru}
\BE \label{dimless_gren}
g_\ren := \frac{\lambda_\ren}{m_\ren} = \frac{\lambda_0}{m_\ren} - \frac{3}{8 \pi^2} \frac{\lambda_0^2}{m_\ren^2} \Big[ \frac{\pi}{2}  - 2 \hat m_\ren + O(\hat m_\ren^2, \hat \lambda_0 \hat m_\ren^2) \Big] + O(\lambda_0^3).
\EE
The reason for this definition is also suggested in perturbation theory, where this turns out to be the natural renormalized expansion parameter. In three dimensions, some power-counting and graph theory imply that the mass dimension of a Feynman diagram contributing to an $E$-point vertex function at order $V$ in $\lambda_0$ will be
\BE
\delta(E,V) = 3 - \smallfrac{1}{2} E - V.
\EE
This tells us two important facts. First, the asymptotic dependence on the UV cutoff $\Lambda = 1/a$ decreases with increasing external points ($E$) and with increasing order in perturbation theory ($V$). In fact, there are only 2 primitive diagrams in the theory which diverge as $\Lambda \to \infty$, the snail and the sunset diagrams that appear in $\Gamma^{(2)}$; this fact makes $\phi^4_3$ an example of a \textit{superrenormalizable} theory. The second fact we learn from $\delta (E,V)$ is that, if we factor out $m_0$  from every loop integral and change momentum variables $p = m_0 \bar p$, then upper limits of integrals become $\Lambda / m_0 = 1/\hat m_0$, and the dimensionless integral gets multiplied by a factor of $m_0^{\delta(E,V)}$. Since the first two terms, $3-E/2$, are independent of $V$, they factor out of the entire perturbation series. Meanwhile, the remaining expansion is in powers of $\lambda_0 / m_0$. Thus the generic observables will have a series that looks schematically like
\BE
\Gamma^{(E)} = \Gamma^{(E)}_\mrm{tree} + m^{3-E/2}_0 \sum_{V = 1}^\infty A_{E,V}(1/\hat m_0) \big( \lambda_0 m_0^{-1} \big)^{V}
\EE
and all the coefficients $A_{E,V}(1/\hat m_0)$ are finite as $\hat m_0 \to 0$ except the snail and sunset diagrams. Replacing the bare parameters by their renormalized counterparts yields series in $g_\ren$, apart from the over-all multiplication by $m_\ren^{3-E/2}$.

To compute the renormalized beta function $\beta_\ren(g_\ren)$, we compute from eq. (\ref{dimless_gren})
\BE
\hat m_\ren \frac{\dd g_\ren}{\dd \hat m_\ren} \Big|_{\hat \lambda_0} = - \frac{\lambda_0}{m_\ren} + \frac{3}{4 \pi^2} \frac{\lambda_0^2}{m_\ren^2} \Big[ \frac{\pi}{2}  + O(\hat m_\ren) \Big] + O(\lambda_0^3).
\EE
Solving for $\lambda_0$ in terms of $\lambda_\ren$ then yields
\BE
\beta_\ren(g_\ren) = - g_\ren + \frac{3}{16 \pi} g_\ren^2 + O(g_\ren^3, g_\ren^2 \hat m_\ren).
\EE
The terms proportional to $\hat m_\ren$ vanish in the continuum limit (they are an example of scaling violations). To compute the bare beta function, we need the derivative of $\hat \lambda_0$ at fixed $g_\ren$. Using eq. (\ref{dimless_gren}) again, but being mindful of the $O(\hat m_\ren)$ part of the 1-loop term, and using the chain rule, we compute
\BE
\beta_0(\hat \lambda_0) = \hat \lambda_0 - \frac{3}{4 \pi} \hat \lambda_0^2 + O(\hat \lambda_0^3, \hat \lambda_0^2 \hat m_\ren).
\EE
From $\beta_\ren(g_\ren)$, we learn that an IRFP exists around $g_* = 8 \pi / 3$,\footnote{This parameter does not seem very small. However, its every occurrence in the perturbation series above comes with a factor of $1/(2 \pi)^3$, so the effective expansion parameter is in fact $1/3 \pi^2$, which is small \cite{Binney:1992vn}.} while the gaussian fixed point is a UVFP. Thus, at fixed bare coupling, $g_\ren$ tends to $g_*$ in the continuum limit, whereas at fixed $g_\ren$,  $\hat \lambda_0$ tends to zero in the continuum. The fact that $g_\ren \to g_*$ as one approaches the continuum, no matter what $\hat \lambda_0$ one begins with, is an example of \textit{universality} at the IRFP. Moreover, all critical quantities, like exponents and amplitude ratios, are expressible as functions of $g_*$, and therefore are also universal \cite{ZinnJustin:2002ru}. In 4 dimensions, the parallel analysis leads one to the conclusion that $g_\ren \to g_* = 0$ in the continuum limit, a result that has found further evidence from much more systematic analytic calculations \cite{Baker:1981zz,Aizenman:1982ze,Frohlich:1982tw,Gawedzki:1985ic,Luscher:1987ay} as well as lattice simulations \cite{Freedman:1981wr,Hasenfratz:1987eh,Kim:1992rw}. This is an example of \textit{triviality} in a quantum field theory.

This has all been perturbative, and confined to a few couplings. One may rightly wonder whether this picture holds nonperturbatively, or when there are \textit{many} couplings. Furthermore, the lingering question about how this plays out for \textit{nonrenormalizable} theories suggests itself: how should we understand situations where operators are present in the action for which perturbative renormalizability fails? In a sense, the key to a deeper understanding of RG rests in finding an answer to these questions. The insight of Wilson which led to an answer was to formulate RG in an entirely nonperturbative way with the help of the concept of block spins and theory space. It was also through his formulation that the application of the Callan-Symanzik equations to critical phenomena became apparent.

\section{Block-spin RG}

In the 1950's and 60's it became clear that the traditional approach to critical phenomena, namely, Landau mean field theory \cite{Landau:1937obd}, was inadequate to describe the long-established experimental fact of nongaussian scaling of thermodynamic properties in statistical systems near their critical points \cite{Wilson:1993dy,Cao:1999pw}. Progress was made with the pursuit of high-temperature series expansions by Domb, Fisher, and others. In 1965, Widom \cite{doi:10.1063/1.1696618} proposed a \textit{scaling hypothesis} for the thermodynamic free energy which was able to reproduce some of the observed scaling laws. But these hypotheses lacked any deep theoretical basis. The concept of ``block-spins'' emerged in the late 60's as a promising avenue to theoretically understand such scaling, beginning with a suggestion by Buckingham \cite{Cao:1999pw}, and separately (though more fully) by Kadanoff in 1966 \cite{Kadanoff:1966wm}. Kadanoff's work then formed the basis of Wilson's theory of RG, which he introduced in 1971 \cite{Wilson:1971bg,Wilson:1971dh,Wilson:1971dc}, and which finally provided a compelling theoretical explanation for the aforementioned critical properties.\footnote{The line of progress hitherto described is, of course, a narrow view of a much broader field of contributions and research in the late 60's. As Wilson notes in his Nobel lecture \cite{Wilson:1993dy}, independent work on the relationship between field theory and critical phenomena was carried out during the same time period by Gribov, Migdal, Symanzik, Polyakov, Dyson, and others. It should be noted that some of these parallel developments have recently been exploited in the conformal bootstrap program \cite{Poland:2018epd} with striking success.} The numerical implementation of block-spin RG was later carried out in the 1980's by Swendsen, Wilson, and others, in a framework known as Monte Carlo Renormalization Group (MCRG) \cite{Swendsen:1979gn,Pawley:1984et}. MCRG has since become a commonplace tool in the study of RG properties of lattice systems.

\subsection{Block-spin transformations}

Starting with a lattice of spins $\varphi(x)$, a new set of \textit{blocked} spins $\varphi_b(x)$ is defined by local averages of the old ones,
\BE
\varphi_b(x_b) := (B_b \varphi)(x) = \frac{b^\Delta}{b^d} \sum_{\veps} \varphi(x + \veps),
\EE
where $\veps$ is a vector pointing to each neighbor of $x$ within a distance $b$, which is called the ``scale factor,'' and $\Delta$ is called the \textit{scaling dimension} of $\varphi$, which we will discuss soon. The index $x_b$ refers to the site of a blocked lattice superimposed on the original one, located at some chosen site within the block of original sites. Unless the initial system had an infinite volume, the blocked spins must live on a smaller lattice. The blocking operator $B_b$ defined by
\BE
B_b(x,y) = \frac{b^\Delta}{b^d} \sum_\veps \delta(x+\veps, y)
\EE
will be useful to keep in mind later in this work.

The blocking transformation on the fields induces a transformation of the action on the level of the partition function by introducing a delta function which sets new spins equal to blocked spins,
\BE
Z = \sum_\varphi \me^{-S(\varphi)} = \sum_\varphi \frac{1}{V}\sum_{\varphi_b} \delta(\varphi_b - B_b \varphi) \; \me^{-S(\varphi)} = \sum_{\varphi_b} \; \me^{-S_b(\varphi_b)}.
\EE
The last equality defines the blocked action. It generally does not equal the original action; it will contain many terms which were not present before, and the terms that were already present will have different values of their couplings. Kadanoff's approach was limited by not considering these extra terms. For example, if $S$ had only a nearest neighbor interaction
\BE
- J\sum_{x, \mu} \varphi(x) \varphi(x+\mu),
\EE
then the new action will have a different value $J'$ as well as new terms involving next-nearest neighbors, next-next-nearest neighbors, etc., and even higher-powered interactions like\footnote{Technically, if $\varphi$ takes values in all of $\mathbb{R}$, then the higher order terms in $\varphi$ are only generated when there are interacting terms, like $\varphi^4$ or $\varphi^6$, in the initial action. If the spins are constrained to have unit size, $|\varphi| = 1$, as in the Ising model, then the nearest neighbor term is sufficient to generate such higher order interactions. But the blocking transformation is different in the Ising model; one must project the blocked spin back to unit norm.}
\BE
\sum_{x} \sum_{\mu_1, \dots, \mu_j} \varphi(x) \varphi(x + \mu_1) \cdots \varphi(x + \mu_j),
\EE
for every $j = 2n, \; n \in \mathbb{Z}_+$. In fact, it will typically contain \textit{all} possible terms consistent with the symmetries of the system. We will explicitly compute a few such terms in chapter 3 when discussing functional RG. Part of Wilson's breakthrough was to recognize the relative importance of all these extra terms in the effective action.

If the correlation length of the system being described is $\xi$, then the original lattice spacing is $a = \xi / \hat \xi$, with $\hat \xi$ calculated in the original theory. By definition, the blocked lattice has a spacing $a_b = b a$, so the dimless correlation length of the blocked theory must be $\hat \xi_b = \hat \xi / b$. Thus the blocked theory will generally have a reduced dimless correlation length, which means that fewer degrees of freedom are strongly correlated across the lattice. The philosophy of both Kadanoff and Wilson was that the blocking transformation therefore reduces the complexity of many-body systems by systematically reducing the number of degrees of freedom being taken into account, without changing the physics \cite{Wilson:1973jj} (because the partition function is invariant), a philosophy which could be called the \textit{pragmatic} view of RG. Because critical phenomena are characterized by large correlation lengths, block-spin RG proves to be a useful tool.

The blocking transformation on the spins induces a transformation of the Boltzmann factor, or equivalently the action, as noted above. Thus, we can regard it as a map on the space of actions, parameterized by the number $n$ of iterations of the transformation determined by $b$, which produces a sequence, or \textit{flow},\footnote{``Flow'' may be misleading here, since the transformations are discrete. Continuous RG transformations will be described in the next three chapters.} on action space,
\BE
S_0 \to S_1 \to S_{2} \to \cdots \to S_{n}.
\EE
If $S_0$ had couplings $\bo g = (g_i)$, then the couplings in $S_{n}$ are denoted $\bo g_n$. Now, as $n\to \infty$, one eventually (for generic actions) approaches $\hat \xi_n \to 0$, namely, a trivially decoupled lattice system. In the vicinity of $\hat \xi_n = 0$, the action no longer changes much after each iteration. Actions which are exactly invariant under RG transformations are called \textit{fixed points}, denoted $S_*$. From the relation $\hat \xi_b = \hat \xi / b$, we observe that the only actions which can be fixed points must have either $\hat \xi = 0$ or $\infty$. The former type are called \textit{zero-correlation length} fixed points while the latter are called \textit{critical} fixed points, since they are the ones of use in the account of critical phenomena. Zero-correlation length fixed points act as sinks for RG trajectories, since any initial theory with $\hat \xi < \infty$ will eventually run into it, at least in the generic case where there are no limit cycles or other exotic behaviors. From $a = \xi / \hat \xi$, we also see that the critical fixed points correspond to zero lattice spacing systems (if $\xi \neq 0$), i.e. the continuum limit, consistent with the analysis of perturbative RG in the previous section.

\subsection{Correlator scaling laws}

One of the striking experimental discoveries of modern physics is that the correlation functions of statistical systems at criticality can exhibit nontrivial power law behavior, rather than a (typical) exponential decay, which is a manifestation of the long-distance correlations of critical systems. For spin systems, the critical spin-spin correlation function is observed to behave like\footnote{The nontrivial part of the correlator may be understood intuitively as an expression of scale-dependence of the interaction by writing $A / z^\eta = A(z)$, with $A(z) = A' a^\eta/ z^\eta = A'(1 - \eta \ln z / a + \dots)$, which modifies the free-field behavior \cite{Cao:1999pw}.}
\BE
G(z) := \langle \varphi(z) \varphi(0) \rangle = \frac{A}{z^{d-2+\eta}},
\EE
where the constant $A$ has mass dimension $-\eta$, since the dimension of the spins is $d_\phi = d/2-1$. The exponent $\eta$ is equal to zero in mean field theory \cite{Kopietz:2010zz}. The empirical fact that $\eta \neq 0$ for many systems constituted a major theoretical problem in the 60's. With the advent of RG, however, it finally found an explanation \cite{Kadanoff:1966wm,Wilson:1971bg,Cardy:1996xt}.

Let us compute $G(z)$ in the blocked theory with action $S_b$, without assuming any kind of $z$-dependence. Since $a_b = ba$, the dimensionless distance $\hat z_b = \hat z / b$ between blocked spins corresponds to a distance $\hat z$ between original spins,
\BE
G_b(\hat z_b) = \langle \hat \varphi_b(\hat z_b) \hat \varphi_b(0) \rangle_{S_b} = \langle B_b \hat \varphi (\hat z) B_b \hat \varphi(0) \rangle_{S_0} = \frac{b^{2\Delta}}{b^{2d}} \sum_{\veps \veps'} \langle \hat \varphi (\hat z + \veps) \hat \varphi(\veps') \rangle_{S_0}.
\EE
At large distances one expects the approximation $G(z) \approx [G(z+\veps) + G(z-\veps)]/2$ to get better and better, which leads to the asymptotic relation
\BE
G_b(\hat z/b) \sim b^{2\Delta} G(\hat z).
\EE
Now, if the action $S$ had couplings $\bo g = (g_i)$, then the blocked action typically has different ones $\bo g_n$, but the \textit{function of} these couplings $G(\hat z; \bo g)$ is the same in either case (assuming we include all possible couplings in the set $\bo g$), so 
\BE
G(\hat z/b; \bo g_b) \sim b^{2\Delta} G(\hat z; \bo g).
\EE
This relation holds for any pair of successive blocking steps. Let us now assume that we are in the vicinity of a fixed point of the RG transformation, meaning that $\bo g_b \approx \bo g \approx \bo g_*$, implying
\BE
G(\hat z/b; \bo g_*) \sim b^{2\Delta} G(\hat z; \bo g_*).
\EE
But this means $G$ is homogeneous of degree $2\Delta$. Thus, at large distances,
\BE
G(\hat z; \bo g_*) \sim \frac{A_*}{\hat z^{2\Delta}},
\EE
which produces the empirical result when $\Delta = d/2 - 1 + \eta/2 = d_\phi + \gamma_\phi$. $d_\phi$ is the canonical mass dimension of the field $\varphi$ in position space, so $\gamma_\phi = \eta/2$ is called the \textit{anomalous} dimension of $\varphi$. This anomalous dimension coincides with the one defined in the previous section in the context of field theory.\footnote{This may not be obvious. The bare RG equations of the $\Gamma^{(n)}_0$ imply a nontrivial scaling behavior in $a$ as $a \to 0$ that is power law-like with exponent $\gamma_\phi$, which for $\Gamma^{(2)}_0$ leads to the identification of $\eta = 2\gamma_\phi$.}

We remark that short-distance observables of the original theory are not quite invariant under a blocking transformation, in the following sense. The nearest-neighbor observable
\BE
\langle \varphi(x) \varphi(x+\veps) \rangle,
\EE
with $\veps < b$, has no direct counterpart in the blocked theory: those neighbors have been integrated out; the nearest-neighbor on the blocked lattice relates spins that are ``farther apart.'' By contrast, the correlator analysis above implied that $G(\hat z)$ at large distances \textit{is} invariant, up to a proportionality with the previous blocking step. This is why one says that RG transformations typically only preserve long-distance observables. We note that if the RG transformation could be made to be continuous, then one could meaningfully discuss infinitesimal variations of the short-distance observables, at least in the continuum. We will discuss this in chapter 4.

\subsection{Fixed points}

For ease of notation let $\bo g' = \bo g_b$. The blocked couplings may be expressed as functions of the previous ones:
\BE
\bo g' = \bo R_b(\bo g).
\EE
Near a fixed-point, assuming analyticity of $\bo R_b(\bo g_*)$, we may linearize the transformation,
\BE \label{linearized_couplings}
\bo g' = \bo g_* + \bo T_b(\bo g_*) (\bo g - \bo g_*) + O((\bo g - \bo g_*)^2), \quad \mrm{or} \quad \delta \bo g' = \bo T_b(\bo g_*) \delta \bo g + O(\delta \bo g^2),
\EE
where $\bo T_b(\bo g)$ is called the RG ``stability matrix,'' with components
\BE
\bo T_b(\bo g) = \frac{\del \bo g'}{\del \bo g}.
\EE
Let $\bo v_a$ be the left-eigenvectors of $\bo T$, i.e.
\BE
\bo v_a^\top \bo T_b(\bo g_*) = \lambda_a \bo v^\top_a,
\EE
and define the \textit{scaling variables} $u_\alpha$ by $u_a := \bo v_a^\top \delta \bo g$, so that the linearized transformation eq. (\ref{linearized_couplings}) implies
\BE
u_a' = \lambda_a u_a,
\EE
where $\lambda_a$ depends on $b$. Although the various couplings $\bo g$ will mix under the RG transformation, the scaling variables do not. A practical requirement of block-spin transformations is the composition property $\bo R_{b'} ( \bo R_b (\bo g)) = \bo R_{b'b}(\bo g)$, which then implies that the eigenvalues satisfy $\lambda_a(b') \lambda_a(b) = \lambda_a(b'b)$, which is solved for $\lambda_a(b) = b^{y_a}$, for some $b$-independent constants $y_a$ \cite{Kopietz:2010zz}. The $y_a$ are referred to as the \textit{RG eigenvalues} of the fixed point $\bo g_*$. 

We can write an arbitrary action as a scalar product of couplings $\bo g$ with action operators $\bo S = (S_i)$ as $S = \bo g^\top \bo S$. Denoting the fixed point action by $S_*$, an arbitrary deviation of an action from $S_*$ may then be written as
\BE
S - S_* = \bo g^\top \bo S - S_* = \delta \bo g^\top \bo S = \sum_a \delta \bo g^\top \bo v_a \; \bo v_a^\top \bo S = \sum_a u_a \mcal{R}_a,
\EE
where the \textit{scaling operators} have been defined, $\mcal{R}_a := \bo v_a^\top \bo S$, and we have used completeness of the left-eigenvectors. The fixed point values of the scaling variables are therefore zero, $u_{*a} = 0$. Performing an RG transformation beginning with $S$ close to $S_*$, we obtain
\BE
S ' = S_* + \sum_a b^{y_a} u_a \mcal{R}_a.
\EE
In particular, if $S = S_* + u_a \mcal{R}_a$ for some particular $a$, then the blocked action will again only involve $\mcal{R}_a$. We then can distinguish three scenarios for the behavior of a perturbation from the fixed point action:
\begin{itemize}
\item $y_a < 0$: the perturbation decays with blocking iterations, and is called \textit{irrelevant},
\item $y_a = 0$: the perturbation is independent of iterations, and is called \textit{exactly marginal},\footnote{I include ``exactly'' because one often talks about ``marginally'' irrelevant and relevant operators to mean ones which are marginal at a gaussian fixed point but become either irrelevant or relevant at a nearby fixed point.}
\item $y_a > 0$: the perturbation increases with iterations, and is called \textit{relevant}.
\end{itemize}
The relative sizes of the $u_a$ present in any given action determine how closely RG transformations will map it towards a fixed point. The negative RG eigenvalues diminish with iterations, so they do not prevent the approach to the fixed point. The exactly marginal operators, interestingly, are invariant, and therefore a perturbation by a marginal operator constitutes a \textit{new} fixed point. Generally, then, we see that the set of RG fixed points differing by marginal operators form a fixed point \textit{submanifold} in the space of actions. The positive eigenvalues, on the other hand, will steer the flow \textit{away} from the fixed point. Hence, the distance of closest approach to the fixed point depends strongly on what the values of the relevant scaling variables are; the smaller they are, the longer it takes for those terms to ``kick in.'' For initial actions that are ``tuned'' such that $u_\mrm{rel} = 0$, RG will map the action directly into the fixed point.

The region in parameter space that flows directly into the fixed point under RG transformations is called the \textit{basin of attraction} of the fixed point, and is therefore the surface $u_\mrm{rel} = 0$. On this surface, the irrelevant variables are unconstrained, and theories defined by actions which differ only by irrelevant variables have the same long-distance properties. This is the phenomenon of \textit{universality}. It explains the empirical fact that many different physical systems can have the same critical exponents (RG eigenvalues) near a second-order phase transition. For example, the Ising universality class in three dimensions describes  not only the critical behavior of certain ferromagnets, but also such diverse situations as the liquid-gas transition in xenon, critical points of binary fluids, the atomic arrangement transition in copper-zinc alloys, and superfluid helium transitions \cite{Peskin:1995ev}. Generally, one expects theories with exactly the same symmetries, in the same dimension, to belong to the same universality class. The defining symmetry of the Ising universality class is $\mathbb{Z}_2$ transformations of the order parameter.

In the correlator analysis above, it was assumed that the RG transformation had a fixed point to begin with. This will only be true if $\Delta$ is chosen carefully, and since we saw above that only $\Delta = d/2 - 1 + \eta/2$ led to the empirical value, it comes as no surprise. This may seem like an undesirable tuning of the blocking transformation, and from that perspective it is. However, once $\Delta$ is picked correctly, the scaling dimensions of any other local operators may be determined, in principle, by studying the scaling of correlation functions. If $\mcal{R}_a(\varphi; \hat x)$ is a local  scaling operator with corresponding RG eigenvalue $y_a$, then near the fixed point one has \cite{Cardy:1996xt}
\BE \label{corr_scaling}
\langle \mcal{R}_a(\varphi_b; \hat z / b) \mcal{R}_a(\varphi_b; 0)  \rangle_{S_b} \sim b^{2\Delta_a} \langle \mcal{R}_a(\varphi; \hat z) \mcal{R}_a(\varphi; 0) \rangle_{S},
\EE
where $\Delta_a = d - y_a$ is the \textit{scaling dimension} of $\mcal{R}_a$. A derivation of this formula in the context of functional RG is given in chapter 3, eq. (\ref{scalingops}). In practice, analytically, this formula is not in fact of much use; perturbative RG methods are more typically used, and recently the conformal bootstrap \cite{Poland:2018epd} has seen many successes. On the lattice, scaling dimensions may be systematically computed using MCRG techniques, as described below. In chapter 2, however, we will finally put eq. (\ref{corr_scaling}) to use in lattice simulations, but not with a blocking transformation, per se.

\subsection{Synthesis} Let us suppose we begin with an action which has been tuned in the manner described above. Since the partition function is invariant under the blocking, and since the blocking preserves the long-distance observables, it follows that the correlations of the system will exhibit the correlations of the fixed point theory, up to rescaling of the fields. Since the fixed point theory displays possibly nontrivial scaling behavior, as we saw with the correlator $G(z)$ above, we finally see how the block-spin RG formalism can explain critical phenomena.

For statistical systems that really do have a lattice spacing, due to a microscopic cutoff such as an inter-atomic spacing in a ferromagnet, the picture is the following. The relevant parameters correspond to temperature $T$ and external magnetic field $H$. For simplicity, we imagine that $H$ vanishes identically. The critical surface of the system then corresponds to $T = T_c$. Thus, buy tuning the ``temperature knob'' to $T \approx T_c$, one induces critical behavior in the system. In terms of the correlation length $\hat \xi = 1/m_\ren a$, the finite atomic spacing $a \neq 0$ means that one is tuning the renormalized mass to zero. The same procedure is accomplished in lattice simulations of spin systems to approach criticality.

In field theory, one typically speaks of criticality as being a ``continuum limit,'' because the relevant situation is presumably $a \to 0$ at $m_\ren \neq 0$, at least for a massive field theory. But the approach to this limit is achieved in the same way: tune the bare parameters so as to achieve $\hat \xi \to \infty$. If the fixed point theory exists, then so do theories all along the critical surface, since they are equivalent under RG transformations. Now, the renormalized theory (with $a = 0$) corresponds to a theory living on the critical surface, and therefore exists if there is a fixed point. This is the statement of nonperturbative renormalizability. Since the irrelevant variables near the fixed point play a subleading role, it is permissible to consider analyses involving only the variables with the largest RG eigenvalues, to a first approximation. This accounts for the success of the Callan-Symanzik-type of RG described in the previous section, so long as perturbation theory is valid. In particular, the case which holds $a$ fixed and sends $m_\ren \to 0$ allows one to describe critical statistical systems using perturbative RG.

\begin{figure}
\centering
\begin{minipage}{.48\textwidth}
\includegraphics[width=0.8\textwidth]{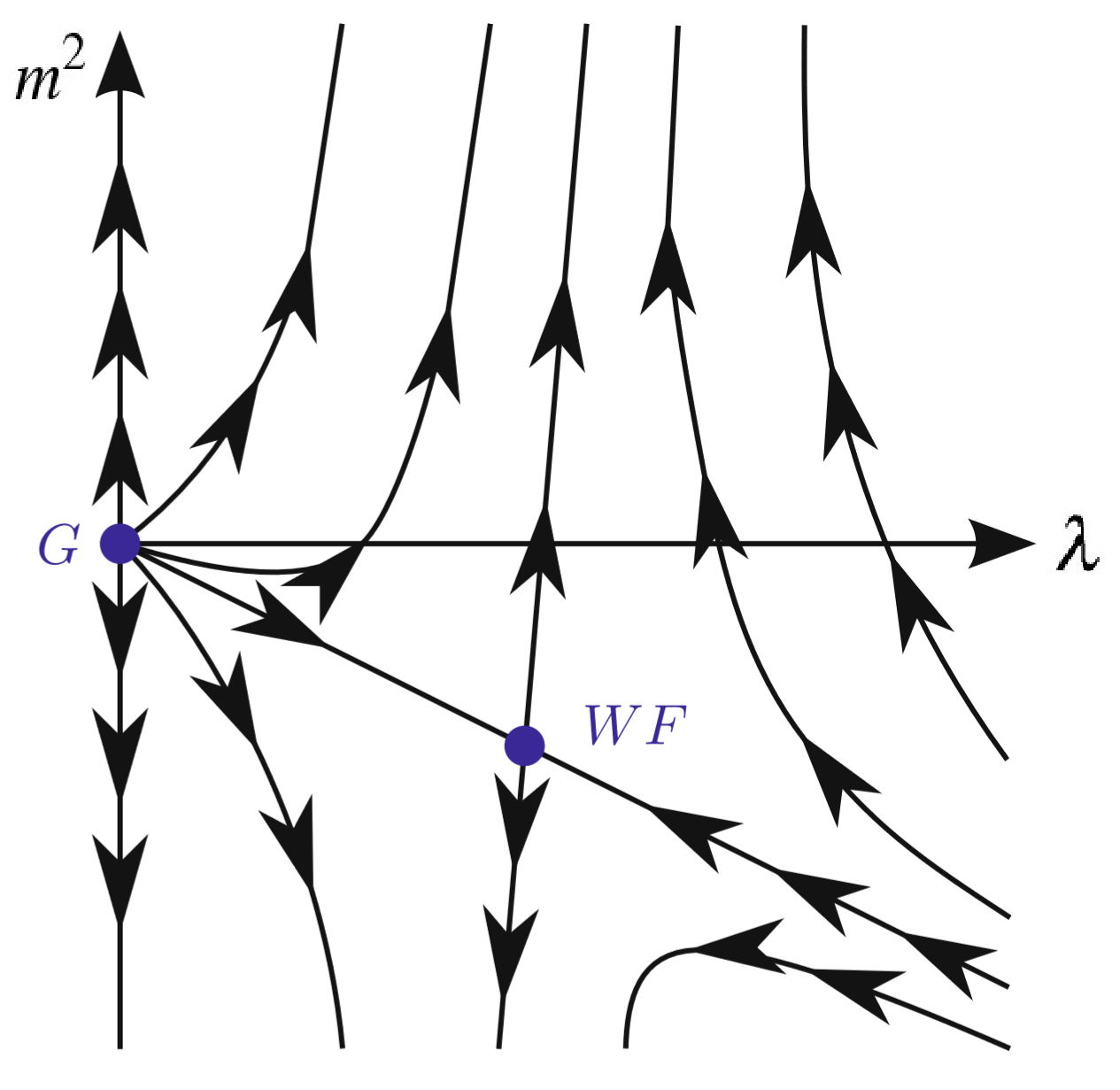}
\end{minipage}\hfill
\begin{minipage}{.48\textwidth}
\includegraphics[width=1.0\textwidth]{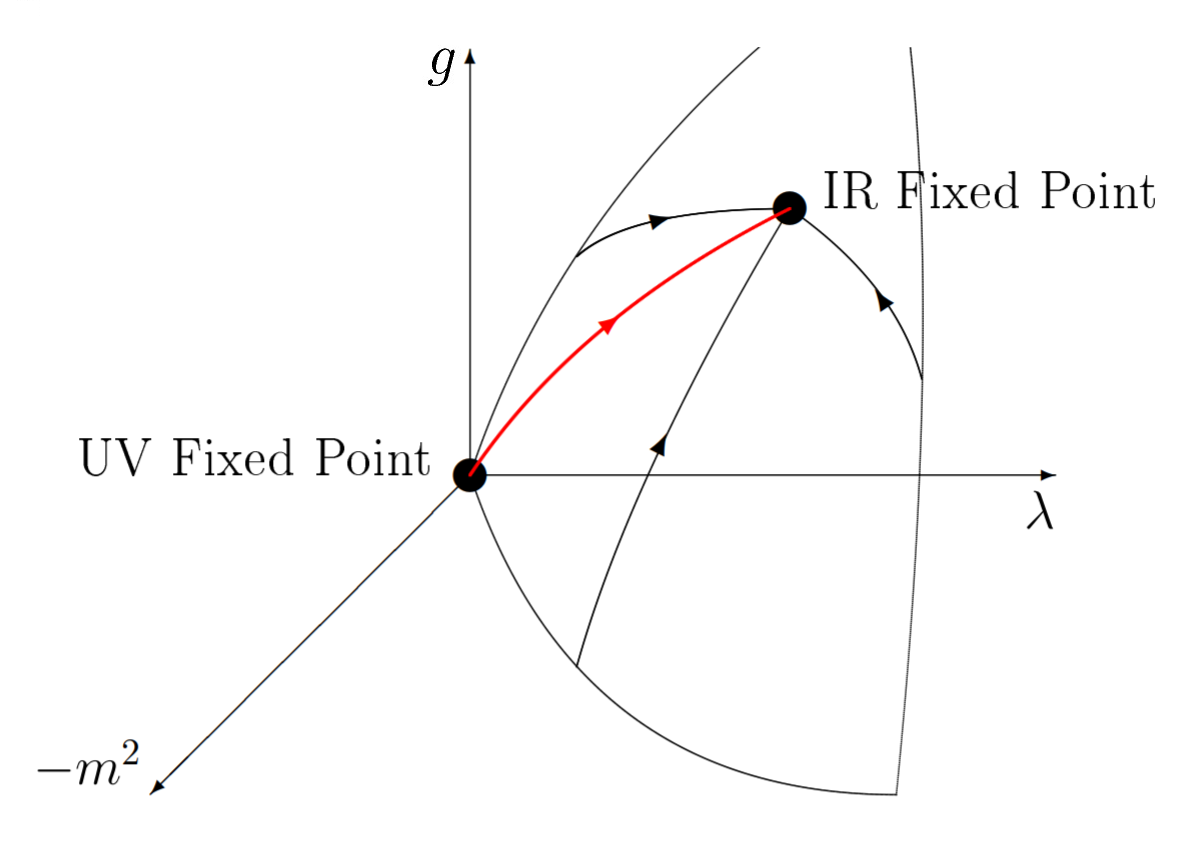}
\end{minipage}
\caption{\small{(Left) Projection of the RG flow in $\phi^4_3$ to the relevant hyperplane. Adapted from \cite{Kopietz:2010zz}. (Right) RG flow of $\phi^4_3$ with a third axis representing all irrelevant interactions (besides $\lambda$). Trajectories that do not begin on the critical surface initially approach the IRFP but eventually veer away in the directions orthogonal to the surface at the IRFP. Adapted from \cite{Capponi:2016yjz}.}}
\label{fig:WFFPs}
\end{figure}

In $\phi^4_3$ theory with vanishing external field $H$, the two most relevant parameters are the mass and the quartic coupling. We saw that the theory possesses an IRFP with nonzero coupling, the Wilson-Fisher fixed point (WFFP), and a UVFP with vanishing coupling, the gaussian fixed point (GFP). The critical surface of the GFP is the subspace defined by $m_0^2 = 0, \lambda_0 = 0$ (and all couplings $g_n$ of degree $n>2$ in $\phi$ also vanishing), since those variables are relevant at the GFP. If $\lambda_0 \neq 0$ (or $g_n \neq 0$), however, the IR behavior is dominated by the WFFP. The critical surface is determined by $\hat \xi = \infty$ ($m_0^2 =0$ is not sufficient with nonzero $\lambda_0, g_n$); all bare actions along this surface flow into the WFFP under RG iterations, as depicted in figure \ref{fig:WFFPs}. Since all higher-order interactions, such as $\phi^6, p^2 \phi^4, \phi^8,$ etc., are irrelevant at the WFFP,\footnote{This identification is somewhat loose; to each of these operators corresponds a \textit{scaling operator} that is irrelevant.} we observe that a large class of scalar field theories are governed by the same fixed point. Since those irrelevant operators coincide with the nonrenormalizable interactions in perturbation theory, we now know how to think of them: they are ultimately unproblematic because they do not significantly alter the long-distance properties of the theory.  Putting this knowledge to use is the program of \textit{effective field theory}, which we will briefly summarize in chapter 3. We close with a quote from Wilson:

\begin{displayquote}
\textit{``I go to graduate school in physics, and I take the first course in quantum field theory, and I’m totally disgusted with the way it’s related. They’re discussing something called renormalization group, and it’s a set of recipes, and I’m supposed to accept that these recipes work — no way. I made a resolution, I would learn to do the problems that they assigned, I would learn how to turn in answers that they would accept, holding my nose all the time, and someday I was going to understand what was really going on. And it took me ten years, but through the renormalization group work I finally convinced myself that there was a reasonable explanation for what was taught in that course.''} -- Reported in P. Ginsparg's \textit{Renormalized After-Dinner Anecdotes} at the ``Celebrating the Science of Kenneth Geddes Wilson" symposium in 2013 \cite{Ginsparg:2014fya}.
\end{displayquote}

\subsection{MCRG} The most systematic implementation of the block-spin RG transformation is via Swendsen's Monte Carlo Renormalization Group \cite{Swendsen:1979gn, Pawley:1984et}, which extracts estimates of critical exponents from a computation of the discrete RG stability matrix introduced above. Consider the expectation value of an action operator $S_k$ after a blocking step $S \to S'$,
\BE
\langle S_k' \rangle_{S'} = \frac{1}{Z(g')} \sum_{\varphi} S_k' \me^{-S'(g')} = - \frac{\del}{\del g_k'} \log Z(g').
\EE
From the invariance of the partition function, $Z(g) = Z(g')$, we can differentiate with respect to the couplings $g_j$ at the previous blocking step, to obtain
\BE
\frac{\del}{\del g_j} \langle S_k' \rangle_{S'} = - \langle S_k' S_j \rangle_{S}^\con,
\EE
where $S'_k$ in an expectation value with respect to $S$ is understood as evaluation of $S_k$ on the blocked field $\varphi_b$. Alternatively, we can use the chain rule to differentiate with respect to $g'$:
\BE
\sum_h  \frac{\del g'_h}{\del g_j} \frac{\del}{\del g'_h} \langle S_k' \rangle_{S'} = -\sum_h \frac{\del g'_h}{\del g_j} \langle S_k' S_h' \rangle_{S'}^\con.
\EE
Putting it all together we obtain
\BE \label{Swendsen_Eqns}
\langle S_k' S_j \rangle_{S}^\con = \sum_h \langle S_k' S_h' \rangle_{S}^\con \; T_{hj},
\EE
where the RG stability matrix $T_{hj} = \del g'_h / \del g_j$ enters. Since the observables on both sides may be explicitly computed in a simulation, one can compute the matrix $\bo T$ at blocking scale $b$ by numerically inverting the matrix equation above.

If the bare action is sufficiently close to the critical surface, then repeated blocking transformations carry one toward the RG fixed point. In its vicinity, the stability matrix will approach its fixed point value, and the diagonalization of $\bo T$ will have eigenvalues $b^{y_\alpha}$. Since $b$ is fixed by definition of the blocking, one can extract estimates for the RG eigenvalues using MCRG. The method is limited in practice by the number of iterations one can do given a finite simulation volume. Choosing $b=2$ leads to a halving of the linear size of the lattice with every iteration. Nonetheless, the MCRG method has been applied in numerous systems and has been quite successful \cite{Swendsen:1981rb,Pawley:1984et,Lang:1986pd,Hasenfratz:2009ea}.

\section{Lattice gauge theory and fermions}

In the next chapter we will see an example of a gauge theory in 4 dimensions, including fermions, so here we give a brief description of such theories on the lattice.

\subsection{Gauge theory} On the lattice, gauge fields are group-valued variables (or ``link variables'') $U_\mu(x) \in G$ living on links connecting adjacent sites, where the pair $(x,\mu)$ identifies the field on the link connecting site $x$ with $x+\mu$, and $\mu = 1, ... ,d$. Often $G$ is a Lie group, like U(1) or SU($N$), but discrete groups like $\mathbb{Z}_N$ or crystal groups are also sometimes considered. We denote the Lie algebra of $G$ by $\mathfrak{g}$. Gauge transformations arise from the change of variables
\BE
U_\mu(x) = \Omega(x) U'_\mu(x) \Omega^\dag(x+\mu), \quad \Omega(x) \in G,
\EE
so that traces of products of links which form a closed loop are gauge invariant. The simplest such product is the plaquette around every elementary square of the lattice,
\BE
U_{\mu\nu}(x) = U_{\mu}(x) U_{\nu}(x+\mu) U^\dag_\mu (x+\nu) U^\dag_\nu(x).
\EE
It is then sensible to construct an action as a positive definite sum over all plaquettes. This is called the \textit{Wilson action} after it was introduced in 1974 \cite{Wilson:1974sk},\footnote{Although the corresponding action for the discrete gauge group $\mathbb{Z}_2$ was written down 3 years earlier by Wegner \cite{Wegner:1984qt}.} and is given by
\BE \label{Waction}
S_W(U) = \frac{\beta}{N} \sum_x \sum_{\mu < \nu} \mrm{Re} \; \mrm{tr} \big[ \mathbb{I} - U_{\mu\nu}(x) \big],
\EE
where $N = \mrm{dim} \; G$. To formally obtain the continuum theory, one defines the Lie algebra-valued vector potentials $A_\mu(x) \in \mathfrak{g}$ by
\BE
U_\mu(x) = \exp a A_\mu(x),
\EE
and expands in $a$. Using the Baker-Campbell-Hausdorff formula, one finds the expansion
\BE
U_{\mu\nu}(x) = 2 \mathbb{I} + a^4 F_{\mu\nu}(x) F_{\mu\nu}(x) + O(a^5),
\EE
where
\BE
F_{\mu\nu}(x) = \del_\mu A_\nu(x) - \del_\nu A_\mu(x) + [A_\mu(x), A_\nu(x)]
\EE
is the continuum field strength tensor, which takes values in $\mathfrak{g}$.\footnote{We typically use the convention of anti-hermitian elements $X \in \mathfrak{g}$. To obtain the hermitian gauge field one lets $A_\mu = i \tilde A_\mu$. Furthermore, the presence of $g_0^2$ in the Wilson action is related to the perturbative convention of having it in the fermion coupling term $i g_0 \bar\psi \tilde{\slashed{A}}'  \psi$ by a rescaling $\tilde A_\mu' = \tilde A_\mu / g_0$. Such a rescaling makes the canonical dimension of the gauge field $d_A = 1$ in every dimension.} The Wilson action becomes
\BE
S_{W}(U(A)) = - \frac{ a^d}{2 g_0^2} \sum_x \Big[ \sum_{\mu\nu} \mrm{tr} [ F_{\mu\nu}(x) F_{\mu\nu}(x) ] + O(a) \Big].
\EE
for $\beta = 2 N / g_0^2$. The leading term is the Yang-Mills action $S_{\mrm{YM}}(A)$ in the naive continuum limit $a \to 0$, that is,
\BE
S_{\mrm{YM}}(A) = - \frac{1}{2 g_0^2} \int \dd^d x \;  \mrm{tr} [ F_{\mu\nu}(x) F_{\mu\nu}(x) ],
\EE
which defines the pure-gluonic sector of QCD.

The correlation length of a lattice gauge theory is determined by the lightest mass in its spectrum. For a pure gauge theory, this must refer to the lightest glueball state; a ``glueball'' is a bound state of gluons. To measure this one would have to compute the plaquette-plaquette correlator, which is a difficult task as the measurement is strongly affected by signal-to-noise problems \cite{DeGrand:2006zz}. Moreover, we do not even know experimentally what the mass of this state would be, since the real world includes fermions. The study of pure lattice gauge theory in 4 dimensions is therefore somewhat academic. However, several methods to set the scale in a semi-realistic way (i.e., using experimental measurements) have been put forth over the years.

A common method to set the scale is using the \textit{Sommer parameter} \cite{Sommer:1993ce}, which amounts to a measurement of the static quark potential $V(R)$. To measure this, one first defines the \textit{Wilson loop} operator by
\BE
W(C) := \mrm{tr} \prod_{\ell \in C} U(\ell),
\EE
where $C$ is a closed loop and $\ell \in C$ are the link labels along $C$. For a rectangular loop $C_{RT}$ of spatial size $R$ and temporal extent $T$, one can argue from a spectral decomposition that the static quark potential is given by
\BE
V(R) = - \lim_{T \to \infty} \frac{1}{T} \log \langle W(C_{RT}) \rangle.
\EE
The expected form of $V(R)$ is that of a linearly confining theory,
\BE
V(R) = A + \frac{B}{R} + \sigma R,
\EE
where $\sigma$ is the \textit{string tension}. The force corresponding to this potential is given by $F(R) = -V'(R)$. From separate studies of the nonrelativistic Schr\"odinger equation for heavy quarks, together with input from experimental data, it has been determined that $F(R) R^2|_{R_0} = 1.65$ occurs at a distance $R_0 \approx 0.5$ fm. Now, $R_0/a$ may be expressed in terms of the dimensionless coefficients $B, \; \sigma / a^2$ of $aV(R)$. Since $aV(R)$ can be measured on the lattice, one fits the measured potential to obtain an estimate of $R_0 / a$. By plugging in $R_0 = 0.5$ fm, one then has an estimate for $a$ in fm.

In practice, one is often simulating with dynamical fermions, the bound states of which are hadrons. Another common way to set the scale is then to input a well-known hadron mass in $\hat M_h = M_h a$, e.g. the mass of the $\rho$ meson or the $\Omega$ baryon \cite{Gattringer:2010zz, Sommer:2014mea}. A more recent procedure for setting the scale which utilizes Gradient Flow will be discussed in the next chapter.

Perturbative calculations in Yang-Mills theory suggest that the theory is asymptotically free \cite{Gross:1973id, Politzer:1973fx}, meaning that the coupling $g_0$ decreases as one probes higher energies. This result is confirmed in lattice perturbation theory, where one expands the Wilson action in powers of $a$, which allows observables to be expressed as series in $g_0$. One defines a renormalized coupling $g_\ren$,\footnote{There are various ways to define a renormalized coupling in this theory. In perturbation theory, one method is called \textit{momentum space subtraction} (MOM) \cite{Montvay:1994cy}, which uses the gluon propagator and the 1PI $\psi\psi A$ vertex $\Gamma^{(3)}$.} after which the renormalized and bare beta functions may be calculated in a similar (though algebraically more complex) manner to the scalar case. One finds \cite{Montvay:1994cy}
\BE
\beta_0(g_0) = - a \frac{\dd g_0}{\dd a} = - \beta_0 g_0^3 - \beta_1 g_0^5  + O(g_0^7),
\EE
where the first few coefficients are
\BE
\beta_0 = \frac{N}{16 \pi^2} \cdot \frac{11}{3}, \quad \beta_1 = \Big(\frac{N}{16 \pi^2}\Big)^2 \cdot \frac{34}{3}.
\EE
We see that $g_* = 0$ is a UVFP, so as the continuum limit is approached at fixed $g_\ren$, the bare coupling approaches zero. In a sense, this result is a consistency check on the perturbative expansion. Thus, in simulations of pure gauge theory, one achieves the continuum limit by extrapolating $g_0 \to 0$ according to the perturbative results from lattice perturbation theory. We stress that although the UV theory approaches a gaussian fixed point, the IR physics (characterized by $g_\ren$) remains strongly-interacting.

\subsection{Fermions} The naive discretization of fermion fields, following what was done for scalars, leads to trouble. The euclidean continuum Dirac action in 4 dimensions is
\BE
S_D = \int \dd^4 x \; \overline\psi(x)(\slashed{\del}+m_0) \psi(x),
\EE
where $\psi(x)$ is a Grassmann-valued Dirac spinor (with 4 components). Its naive discretization with a symmetric difference operator is
\BE
S_N = a^4 \sum_x \overline\psi(x)\Bigg(\frac{1}{2a} \sum_\mu \gamma_\mu \big[ \psi(x+\mu) - \psi(x-\mu) \big] +m_0 \psi(x) \Bigg).
\EE
In momentum space, one computes
\BE
S_N = \frac{1}{V} \sum_p \overline\psi(p) \Big( m_0 + \frac{i}{a} \sum_\mu \gamma_\mu \sin p_\mu a \Big) \psi(p).
\EE
In the infinite volume limit, the naive fermion propagator $S(x-y)$ is then
\BE
S(x-y) = \int_{-\pi/a}^{\pi/a} \dd^d p \; \me^{i p (x-y)} \Big[ m_0 + \frac{i}{a} \sum_\mu \gamma_\mu \sin p_\mu a \Big]^{-1},
\EE
which may be written in integral form as
\BE
S(x-y) = \int_0^\infty \dd s \; \me^{-m_0 s} \int_{-\pi/a}^{\pi/a} \dd^d p \; \me^{i p (x-y)} \exp \Big[ - \frac{is}{a} \sum_\mu \gamma_\mu \sin p_\mu a \Big].
\EE
The continuum propagator $\int_p \me^{ipz}/(m_0 + i \slashed{p})$ should be obtained in the $a \to 0$ limit. Although an expansion of the sine function \textit{appears} to achieve this, $\sin p_\mu a$ in fact has zeros at all $2^d$ corners of the Brillouin zone. As $a \to 0$, one then finds $2^d$ saddle points of the integrand above, which means that in the continuum limit, $S(x-y)$ is a sum of $2^d$ copies of the desired propagator. This hiccup is called the \textit{doubling problem} for lattice fermions.

Various approaches to remedy the doubling problem have been forwarded over the years. In one approach, called \textit{Wilson fermions}, one adds a laplacian term to the naive fermion action, which shifts the bare mass in such a way as to guarantee that the doubler masses become infinitely heavy as $a \to 0$, thereby dropping out of the theory. But this solution comes at a cost \cite{Nielsen:1980rz}: for a massless theory, the Wilson term breaks chiral symmetry (invariance under $\psi = \me^{i\gamma_5\theta} \psi', \; \overline \psi = \overline \psi' \me^{i\gamma_5\theta}$), making the simulation of massless fermions a difficult task. Solutions to the doubling problem which allow for the retention of chiral symmetry (in some capacity) have therefore been sought over the years. One such approach is that of \textit{staggered fermions}.

Because we will report results of a lattice simulation using staggered fermions in four dimensions in chapter 2, we give a brief introduction to them here. The first step is to change variables from naive fermions in a peculiar way, the \textit{staggered transformation}:
\BE
\psi(x)  = \gamma_1^{x_1} \gamma_2^{x_2} \gamma_3^{x_3} \gamma_4^{x_4} \psi'(x), \quad \overline \psi(x) = \barpsi' (x) \gamma_4^{x_4} \gamma_3^{x_3} \gamma_2^{x_2} \gamma_1^{x_1}.
\EE
By repeatedly using the gamma matrix property $\{ \gamma_\mu, \gamma_\nu \} = 2 \delta_{\mu\nu} \mathbb{I}$, one can demonstrate that
\BE
\barpsi(x) \gamma_\mu \psi(x\pm\mu) = \eta_\mu(x) \barpsi'(x) \psi'(x\pm \mu), \quad \eta_\mu(x) := (-1)^{\sum_{\nu<\mu} x_\nu}.
\EE
This implies that the staggered transformation decouples the 4 Dirac components in the naive fermion action, leaving an action for 4 copies of the same kind of (1-component) fermion. One then defines the staggered action by retaining only one of the copies. Introducing the gauge field coupling to fermions in the standard way, one has
\BE \label{staggered_action}
S_\mrm{st} = a^4 \sum_x \overline\chi(x)\Bigg(\frac{1}{2a} \sum_\mu \eta_\mu(x) \big[ U_\mu(x)\chi(x+\mu) - U_\mu^\dag(x-\mu)\chi(x-\mu) \big] +m_0 \chi(x) \Bigg).
\EE
because only one of the Dirac components was kept, one expects intuitively that this action reduces the 16-fold degeneracy of the naive action to 4. To check this intuition one must perform a more detailed analysis \cite{Gattringer:2010zz}. What is found is that the staggered action describes 4 species, called ``tastes,'' of Dirac fermions, in terms of which the action resembles the Wilson fermion action, which has no doublers in the continuum limit. Furthermore, from eq. (\ref{staggered_action}) we see that the action possesses a \textit{remnant} chiral symmetry given by invariance under $\chi = \me^{i\eta_5(x)\theta} \chi', \; \overline \chi = \me^{i\eta_5(x)\theta} \overline \chi'$ when $m_0 = 0$.

The staggered transformation only reduces the doublers to 4, whereas some simulations want as few as 2 fermions (for up and down quarks), or 3 (to include the strange). To accommodate this situation, a practical but controversial procedure is adopted: take the square-root of the staggered determinant, and for the strange quark, take the fourth-root. There is some controversy in the literature about the validity of this procedure, however. See \cite{Gattringer:2010zz} and references therein.

\newpage

\chapter{Gradient flow and RG}

In this chapter we introduce the notion of \textit{gradient flow renormalization group} (GFRG) by comparing a type of diffusion known as \textit{gradient flow} with the block-spin transformations we saw in chapter 1. The comparison naturally leads to correlator scaling laws involving gradient-flowed observables that can be measured on the lattice. The comparison suggests a method for extracting scaling dimensions of operators from lattice simulations in a manner distinct from that of MCRG. In section 1, we describe gradient flow in the case of Yang-Mills theory and its primary application in lattice theory, namely, scale-setting. In section 2 we apply the GFRG method to scalar field theories in 2 and 3 dimensions, and in section 3 we apply it to a 4-dimensional gauge-fermion theory.

\section{Gradient flow}

The Yang-Mills gradient flow equation, in the context of lattice theory, first appeared in an exploration of the large-$N$ behavior of smeared Wilson loops in a paper by Narayanan and Neuberger in 2006 \cite{Narayanan:2006rf}.\footnote{It appears that the authors were inspired to choose this form by an analogy to the Langevin equation which generates quantum Yang-Mills theory under stochastic quantization \cite{Damgaard:1987rr}, a development of the 1980's. It was not at that point thought of as a smoothing transformation, although the concept of stochastic regularization was a clue. I also remark that the Yang-Mills flow equation (perhaps for the first time) appeared in the work of Atiyah and Bott in 1983 \cite{10.2307/37156}. It has since been used in the study of \textit{Ricci flow}, having appeared, for example, in \cite{Young:2008xk, STREETS2010454}, where it is one of two equations defining so-called Ricci-Yang-Mills flow. This flow refers to a smoothing evolution of the metric and connection on a principle bundle over a Riemannian manifold. \textit{Pure} Ricci flow was proposed in 1982 in the works of R. Hamilton \cite{hamilton1982}. Interestingly, however, the Ricci flow equations arose even earlier in the study of generalized nonlinear sigma models by D. Friedan in 1980 \cite{Friedan:1980jf}, where it was demonstrated that the RG flow of the model is a Ricci flow on the target space of the field theory, to lowest order in perturbation theory.} The lattice version of the equation was proposed independently by L\"uscher in 2009 \cite{Luscher:2009eq} in the context of so-called ``trivializing maps'' on field space. The idea behind these trivializing maps was to perform a transformation of the field variables $U_\mu(x)$ on the lattice in such a way that the jacobian exactly cancels the gauge field action, effectively mapping the theory to its strong-coupling (or high-temperature) limit; the hope was to improve the efficiency of the Hybrid Monte Carlo algorithm, which diminishes in the continuum limit $g_0 \to 0$ of lattice QCD. For our purposes, however, we will focus on the smoothing property of gradient flow, which will be demonstrated to provide an essential ingredient of a continuous RG transformation on the lattice. It should be noted that L\"uscher did speculate on the possibility of using trivializing maps in the context of RG \cite{Luscher:2009eq}, and this suggestion was followed up analytically in the works of Yamamura and others \cite{Kagimura:2015via,Yamamura:2015kva,Makino:2018rys}.

The continuum formulation of gradient flow for gauge theories runs as follows. Beginning with the initial gauge fields $A_\mu(x)$, one defines their \textit{flow} $B_\mu(x,t)$ to be the solution of the diffusion-type equation
\BE
\del_t B_\mu(x,t) = - \frac{\delta \hat S(A)}{\delta A_\mu(x)}\Big|_{A_\mu = B_\mu}, \quad B_\mu(x,0) = A_\mu(x),
\EE
where $\hat S$ can be called the \textit{flow action}. The parameter $t$ is called the \textit{flow time}, with dimensions of distance-squared. Typically, $\hat S$ is chosen to be the Yang-Mills action $S_\mrm{YM}$, in which case one obtains the Yang-Mills gradient flow,
\BE
\del_t B_\mu = D_\nu F_{\nu\mu},
\EE
where $F_{\mu\nu} = \del_\mu B_\nu - \del_\nu B_\mu + [B_\mu, B_\nu]$, and the covariant derivative in the adjoint representation is, for $X \in \mathfrak{g}$,
\BE
D_\mu X = \del_\mu X + [A_\mu, X].
\EE
L\"uscher and Weisz \cite{Luscher:2011bx} demonstrated perturbatively that the expectation values of observables at finite flow time required no \textit{further} renormalization above that of pure Yang-Mills theory, suggesting that such quantities will have well-defined continuum limits on the lattice.

A quantity of particular popularity is the Yang-Mills energy density at finite flow time,
\BE
E(t) := \smallfrac{1}{4} \langle \mrm{tr} \; F_{\mu\nu}^2(x,t) \rangle.
\EE
In \cite{Luscher:2010iy} it was demonstrated that $E(t)$ is renormalized if one computes it in bare perturbation theory and replaces the bare coupling $g_0$ by the $\MSbar$ coupling at scale $\mu^2  = 1/8t$, as expected from the general renormalizability of flowed observables mentioned above. It was then demonstrated that the lattice implementation of $E(t)$ can be of quite practical use in setting the scale. The ``theory scale'' $t_0$ defined through
\BE
t_0^2 E(t_0) = 0.3,
\EE
was demonstrated to scale to the continuum in roughly the same way as the Sommer parameter $r_0$. That is, by computing $t_0$ at several bare couplings for which $a$ was known already from Sommer parameter scale-setting (with $r_0=0.5$), it was demonstrated empirically that $t_0 / r_0^2$ is constant as $a \to 0$ under a simple (slightly-improved) discretization of $F_{\mu\nu}$, the so-called ``clover operator.'' What all this means is that one can approach the continuum by following the behavior of observables computed at time $t_0$ for each of their bare couplings, if the physical box size $aN$ is constant. See  \cite{Sommer:2014mea} for further discussion.

Perturbatively, the $\MSbar$-renormalized $t^2 E(t)$ is proportional to the renormalized coupling at tree level, suggesting that one can define an alternative renormalized coupling by
\BE
g^2_\mrm{GF}(t) := t^2 E(t).
\EE
Since the jacobian relating the two couplings is nonsingular to known orders in perturbation theory, this scheme change is expected to be valid. The computation of this quantity in finite volume has even led to a natural definition of a renormalized coupling that runs with the lattice size \cite{Fodor:2012td}, and which may be used in step-scaling analyses of the discrete beta function \cite{Fodor:2012td,Cheng:2014jba}. We note that this approach has been fruitful for several theories in the family of SU($N$) gauge theories with $N_f$ fermions. Recently, arguments based on Wilsonian RG have been offered in \cite{Hasenfratz:2019hpg} which suggest that the time derivative of $g^2_\mrm{GF}(t)$ can be used to estimate the renormalized beta function $\beta(g^2_\mrm{GF})$ in many gauge theories.

\section{Block-spin analogy}

In this section we will focus on scalar theories, so first we give a brief review of GF for scalar fields \cite{Monahan:2015lha,Monahan:2015fjf,Capponi:2016yjz,Aoki:2016ohw,Aoki:2016env,Aoki:2017bru,Fujikawa:2016qis}. The flowed fields will be denoted by $\phi_t(x)$. The general gradient flow equation for a one-component scalar field is
\BE
\del_t \phi_t(x) = - \frac{\delta \hat S(\phi)}{\delta \phi(x)} \Big|_{\phi = \phi_t}, \quad \phi_0(x) = \varphi(x).
\EE
If we choose a standard quartic action, for example,
\BE
\hat S(\phi) = \int \dd^d x \Big[ \frac{1}{2} \big(\del \phi(x)\big)^2 + \frac{m^2}{2} \phi^2(x) + \frac{\lambda}{4!} \phi^4(x) \Big],
\EE
then the flow equation reads
\BE \label{int_flow}
\del_t \phi_t(x) = \del^2 \phi_t(x) - m^2 \phi_t(x) - \frac{\lambda}{3!} \phi_t^3(x).
\EE
The utility of interacting flow for scalars was called into question by Suzuki and Fujikawa \cite{Fujikawa:2016qis}, who  determined that the finite-flow time observables are not entirely renormalized by a renormalization of the parameters in the bare action. This is intuitively clear from the presence of the cubic product $\phi^3(x)$ on the r.h.s. above, together with the lack of gauge symmetry which was crucial for the renormalizability proof of GF in Yang-Mills theory \cite{Luscher:2011bx}. The perturbative solution to the scalar GF equation involves local products of fields, which when self-contracted, lead to divergent tadpoles in flowed observables that are not eliminated by the standard renormalization procedure of $\phi^4$ theory. Fujikawa demonstrated, however, that suitably-modified definitions of the interacting flow can lead to a finite theory, ones involving derivative interactions in place of a point-vertex. We will revisit the notion of interacting flow in chapter 4 in the discussion of nonlinear RG's. In the remainder of this section, however, we will stick to noninteracting flows.

\subsection{GFRG transformation} If we specialize to the simplest kind of gradient flow, namely, massless free flow ($m^2=0,\lambda = 0$ in eq. (\ref{int_flow})),
\BE
\del_t \phi_t(x) = \del_x^2 \phi_t(x),
\EE
then we have a simple heat equation. The solution is given by the action of the heat kernel on the initial field,
\BE
\phi_t(x) = (K_t \varphi)(x) = \int \dd^d y K_t(x,y) \varphi(y) = \int \dd^d z K_t(z) \varphi(x+z).
\EE
In the last equality we have brought the solution to a suggestive form, using $K_t(x,y) = K_t(x-y)$. The (infinite-volume) heat kernel in $d$ dimensions is
\BE
K_t(z)  = \int_p \me^{ipz - p^2 t} = \frac{\me^{-z^2/4t}}{(4 \pi t)^{d/2}},
\EE
which rapidly decays when $z \gg \sqrt{4t}$. The solution is therefore reminiscent of the blocking transformation
\BE
\varphi_b(x_b) = (B_b \varphi)(x) = \frac{b^{\Delta_\phi}}{b^d} \sum_{\veps} \varphi(x + \veps),
\EE
when we identify $b \propto \sqrt{t}$, except that the averaging by the heat kernel depends continuously on its ``blocking parameter'' $t$. Importantly, we also do not have an analog of the rescaling factor $b^{\Delta_\phi}$. In this sense, free GF cannot of itself constitute an RG transformation.

We wish to define a smooth RG transformation based on the resemblance just noted. Since the field rescaling was an essential ingredient in block-spin RG which allowed the transformation to exhibit a fixed point, we propose that the analog of the blocked field $\varphi_b$ should be defined by
\BE \label{GFRG}
\mPhi_t(x_t) := b_t^{\Delta_\phi} \phi_t(x), \quad \mrm{with} \quad x_t = x / b_t,
\EE
where the exact form of $b_t$ is not yet determined, except that it must approach 1 as $t\to 0$ and it must be proportional to $\sqrt{t}$ for large enough $t$. This is because the mean-squared radius of the heat kernel is determined by
\BE
\langle z^2 \rangle = \int \dd^d z \; z^2 K_t(z) = 2dt,
\EE
which should correspond to the block-spin radius-squared (times $d$). In chapter 4 we will determine that, under Schwinger regularization (see eq. (\ref{effective_cutoff})) in the continuum, the rescaling factor is exactly
\BE \label{bt_form}
b_t = \sqrt{1 + 2 t \mLam_0^2},
\EE
where $\mLam_0$ is the bare cutoff. The function $b_t$ is expected to be regularization-dependent, however.

In the numerical implementation of GF, one must use the lattice heat equation. The continuum laplacian is replaced by its discretization, which we saw in chapter 1, so the GF equation is given by
\BE
\del_t \phi_t(x) = \sum_{\mu} \hat \del^*_\mu \hat \del_\mu \phi_t(x),
\EE
where $\hat \del, \; \hat \del^*$ are the forward and backward difference operators, respectively. In this case, the solution is
\BE
\phi_t(x) = \frac{1}{V} \sum_p \me^{ i p x - \hat p^2 t} \varphi(p),
\EE
where $\hat p_\mu = 2 \sin(p_\mu a /2) / a$. The lattice momenta are restricted to $p_\mu \in (-\pi/2a, \pi/2a]$, so we observe a monotonic increase in the suppression of high momentum modes, in a qualitatively similar way to the continuum gradient flow solutions. Thus we expect the lattice free flow equation to be equally capable of defining a continuous blocking transformation, which approaches the continuum formulation as $a \to 0$.

We also see in eq. (\ref{GFRG}) the introduction of a rescaled position $x_t = x / b_t$: the blocked field is defined on a rescaled space. In MCRG this leads to the necessity of considering lattices of different sizes when applying the method. In our case, the rescaled field must be said to live on a fictitious blocked lattice with non-integer spacing. We will avoid this subtlety in our analysis by always relating blocked observables to the bare observables and performing computations in the bare theory, as described below.

\subsection{Correlator ratios} The GFRG transformation proposed above leads to scaling relations among correlators which may be implemented in lattice simulations. Recall the correlator scaling formula for block-spin RG which relates the blocked and bare quantities,
\BE \label{corr_scaling2}
\langle \mcal{R}_a(\varphi_b; \hat z_b) \mcal{R}_a(\varphi_b; 0)  \rangle_{S_b} \sim b^{2\Delta_a} \langle \mcal{R}_a(\varphi; \hat z) \mcal{R}_a(\varphi; 0) \rangle_{S_0},
\EE
for $\hat z \gg b$. The scaling operators $\mcal{R}_a$ are generally polynomial in the field $\varphi$. Assuming that GFRG defines a valid RG transformation, the corresponding scaling formula reads
\BE \label{bare_ratio}
\langle \mcal{R}_a(\mPhi_{t}; \hat z_{t}) \mcal{R}_a(\mPhi_{t}; 0)  \rangle_{S_{t}} \sim b_{t}^{2\Delta_a} \langle \mcal{R}_a(\varphi; \hat z) \mcal{R}_a(\varphi; 0) \rangle_{S_0};
\EE
$S_t$ is the effective action generated by the GFRG transformation. The proper definition of this action will be described in chapter 4, but here we avoid it by use of the MCRG principle, which as described in chapter 1, allows one to compute observables in the blocked theory by computing blocked observables in the bare theory.

Now we consider eq. (\ref{bare_ratio}) at two times $t', \; t$ with $t' > t$, and take their ratio:
\BE
\frac{\langle \mcal{R}_a(\mPhi_{t'}; \hat z_{t'}) \mcal{R}_a(\mPhi_{t'}; 0)  \rangle_{S_{t'}}}{\langle \mcal{R}_a(\mPhi_{t}; \hat z_{t}) \mcal{R}_a(\mPhi_{t}; 0)  \rangle_{S_{t}}} \sim \Big(\frac{b_{t'}}{b_t}\Big)^{2\Delta_a}.
\EE
The quantities on the l.h.s. are defined on the lattice with points $\hat z_t = \hat z / b_t$. Using MCRG to write the expectations in the bare theory then yields a \textit{ratio formula},
\BE \label{ratioR}
\frac{\langle \mcal{R}_a(b_{t'}^{\Delta_\phi} \phi_{t'}; \hat z) \mcal{R}_a(b_{t'}^{\Delta_\phi} \phi_{t'}; 0)  \rangle_{S_0}}{\langle \mcal{R}_a(b_t^{\Delta_\phi} \phi_t; \hat z) \mcal{R}_a(b_t^{\Delta_\phi} \phi_t; 0)  \rangle_{S_0}} \sim \Big(\frac{b_{t'}}{b_t}\Big)^{2\Delta_a},
\EE
where now the position arguments refer to sites on the original lattice. Now, close to a fixed point, the correlator of any two operators $\mcal{O}_h, \; \mcal{O}_k$ may be expanded in correlators of scaling operators \cite{Amit:1984ms, Cardy:1996xt},
\BE \label{corr_expansion}
\langle \mcal{O}_h \mcal{O}_k \rangle = \sum_a c_{ha} c_{ka} \langle \mcal{R}_a \mcal{R}_a \rangle.
\EE
If it happens that one of the scaling operators, say $\mcal{R}_a$, dominates the sum, then one might expect that the ratio of correlators of $\mcal{O}_h, \; \mcal{O}_k$ can be used to measure $\Delta_a$. (At large enough distances, the leading operator always dominates the sum.) Letting $\mcal{O}_h$ be of order $n_h$ in $\varphi$ and $\ell_h$ in derivatives, we can factor out the rescalings $b_t^{\Delta_\phi}$ from each operator to obtain
\BE \label{ratioO}
\frac{\langle \mcal{O}_h( \phi_{t'}; \hat z) \mcal{O}_k(\phi_{t'}; 0)  \rangle_{S_0}}{\langle \mcal{O}_h(\phi_t; \hat z) \mcal{O}_k(\phi_t; 0)  \rangle_{S_0}} \sim \Big(\frac{b_{t'}}{b_t}\Big)^{2\Delta_a - n_{hk} \Delta_\phi - \ell_{hk}},
\EE
where $n_{hk} = n_h + n_k, \; \ell_{hk} = \ell_h + \ell_k$. The factors of $b_t^\ell$ arise because derivatives in the rescaled theory are related to those in the bare theory via $\hat \del_{\hat z_t} = b_t \hat \del_{\hat z}$. But when do we expect these ratio formulas to be valid? First, we need $\hat z \gg b_t$, so that the smeared operators do not overlap. Second, we need that the $\mcal{O}_h\mcal{O}_k$ correlator really \textit{is} dominated by $\mcal{R}_a$. Third, the scaling operators are only defined with respect to a fixed point, so we expect the formula above to be valid only in the vicinity of a fixed point, which means the RG transformation must be repeated enough times that proximity has been achieved; we interpret this as meaning that the flow time $t$ is large enough that the effective action $S_t$ is near the fixed point. Lastly, we remark that eq. (\ref{ratioR}) will be deduced without recourse to a block-spin analogy in chapter 4 in the framework of stochastic RG.

Notice that using $\mcal{O}_h = \mcal{O}_k = \phi$ in eq. (\ref{ratioO}) gives no information about $\Delta_\phi = d_\phi + \gamma_\phi$, since $\phi$ is the leading operator in the $\mathbb{Z}_2$-odd subspace. Generally, the ratios of the fundamental field cannot be used to extract an estimate for $\Delta_\phi$, and other methods are needed; there at least two options one may take.
\begin{itemize}
\item Option 1: If there exists an operator $\mcal{A}$ which is known \textit{a priori} to have zero anomalous dimension, with scaling dimension $\Delta_\mcal{A} = d_\mcal{A}$, then its ratio formula implies
\BE \label{Option1}
\frac{\langle \mcal{A}( \phi_{t'}; \hat z) \mcal{A}(\phi_{t'}; 0)  \rangle_{S_0}}{\langle \mcal{A}(\phi_t; \hat z) \mcal{A}(\phi_t; 0)  \rangle_{S_0}} \sim \Big(\frac{b_{t'}}{b_t}\Big)^{2(d_\mcal{A} - n_\mcal{A} \Delta_\phi - \ell_\mcal{A})}.
\EE
An example is the stress-energy tensor of a theory, or a conserved current such as the vector or remnant axial vector currents in gauge-fermion theories.

\item Option 2: In any theory, the operators fall into symmetry subspaces, e.g. $\mathbb{Z}_2$ in $\phi^4$ theory. In the domain of applicability of eq. (\ref{ratioO}), then, the mixed correlation functions $\langle \mcal{O}_i \mcal{O}_j \rangle$ in that subspace of operators all have a leading scaling behavior of $b_t^{2\Delta_a}$, and one can measure a family of exponents
\BE \label{Option2}
\delta_{ij} = 2\Delta_a - (n_i+n_j) \Delta_\phi - (\ell_i + \ell_j),
\EE
where $n_i + n_j$ is the total number of factors of $\phi$ on the l.h.s. and $\ell_i+\ell_j$ the total number of derivatives. From any pair $(i,j) \neq (h,k)$ one may then extract estimates of $\Delta_a,\; \Delta_\phi$, so long as neither correlator is $\langle \phi \phi \rangle$ itself, of course. One must have empirical or theoretical evidence that one operator \textit{does} dominate to make use of this method.
\end{itemize}

\subsection{Diagonalization method} In general, there will not be a dominant operator, and asymptotically large distances may not be accessible. And even if there is, one might instead want to extract the dimension of a subleading operator. One must then use eq. (\ref{ratioR}) directly, which requires a more involved approach. Expanding each $\mcal{R}_a$ in a basis of $\mcal{O}_k$'s, we obtain for the correlator of $\mcal{R}_a$'s
\BE
\langle \mcal{R}_a(b_{t}^{\Delta_\phi} \phi_{t}; \hat z) \mcal{R}_a(b_{t}^{\Delta_\phi} \phi_{t}; 0)  \rangle_{S_0} = \sum_{j,k} c_{aj} c_{ak} b_t^{(n_j+n_k) \Delta_\phi + \ell_j + \ell_k} \langle \mcal{O}_j(\phi_t; \hat z) \mcal{O}_k(\phi_t; 0)  \rangle_{S_0}.
\EE
We could use eq. (\ref{ratioR}) therefore if we knew the coefficients $c_{aj}$ by forming appropriate linear combinations of the correlators on the r.h.s, which are directly measured in lattice simulations. The $c_{aj}$ may be estimated numerically by recalling the consequence of conformal invariance on mixed scaling operator correlations \cite{Amit:1984ms, Cardy:1996xt},
\BE
\langle \mcal{R}_a(\hat z) \mcal{R}_b(0) \rangle = \delta_{ab} \frac{A_b}{\hat z^{2\Delta_b}},
\EE
valid exactly at the fixed point. What is suggested is to choose a basis of operators $\{\mcal{O}_k\}$ and compute the mixed correlations of the $\mcal{O}_k$, then forming the quantities
\BE
b_t^{(n_j+n_k) \Delta_\phi} \langle \mcal{O}_j(\phi_t; \hat z) \mcal{O}_k(\phi_t; 0)  \rangle_{S_0}
\EE
from an ansatz for $b_t$. As the full scaling dimension $\Delta_\phi$ is required for the GFRG transformation, one must here input a value for $\Delta_\phi$ if it is known. One then numerically diagonalizes the matrix of correlators at every distance $\hat z$ and time $t$ to obtain estimates for
\BE
\langle \mcal{R}_a(b_{t}^{\Delta_\phi} \phi_{t}; \hat z) \mcal{R}_a(b_{t}^{\Delta_\phi} \phi_{t}; 0)  \rangle_{S_0}.
\EE
Finally, one applies the ratio formula eq. (\ref{ratioR}) and measures $2 \Delta_a$ directly. The estimate for $\Delta_\phi$ obtained from this approach is merely a consistency check, while all other dimensions constitute genuine predictions.

\section{Scalar field theory}

We have applied the ratio formulas numerically in $\phi^4_d$ theory for $d=2, \; 3$ using the simulation action eq. (\ref{phi4_lat_action}), which reads
\BE
S(\varphi) = \sum_{x} \Bigg( - \beta \sum_{\mu} \varphi(x) \varphi(x+\mu)  + \varphi^2(x) + \lambda ( \varphi^2(x) - 1)^2 \Bigg),
\EE
where we drop hats for the lattice field out of convenience. We take the volume of the lattice $V = L^d$ to be cubic. To minimize the distance of the RG flow from the IRFP, the bare action must be tuned sufficiently well, meaning that given a value for $\lambda$, the neighbor coupling $\beta$ must be set as close to the critical value $\beta_c$ as possible, where $(\beta_c, \lambda)$ is a point on the system's critical ($a=0$) surface. A popular method for determining $\beta_c$ in spin systems is via the Binder cumulant.

\subsection{Tuning to the critical surface} The order parameter of $\mathbb{Z}_2$ symmetry breaking in any spin model is the magnetization $m$, which is defined on a particular configuration $\varphi$ by
\BE
m = \frac{1}{V}  \sum_x \varphi(x),
\EE
and is equal to the zero mode of the field. In the ordered phase where spins are aligned, $\langle m \rangle \neq0$, while in the disordered phase one has $\langle m \rangle =0$. One can study the probability distribution of the magnetization through an analysis of the finite volume zero mode effective action obtained by integrating out all modes in the box with $p \neq 0$ \cite{ZinnJustin:2002ru}. The moments of the magnetization distribution exhibit universal properties in the infinite volume limit. One such observable of particular practicality is the \textit{Binder cumulant}, which is defined by
\BE
U_L(\beta) = \frac{3}{2} - \frac{1}{2} \frac{\langle m^4 \rangle}{\langle m^2 \rangle^2}.
\EE
In the lower temperature limit, $U_L \to 1$, while in the high temperature limit, $U_L \to 0$. It was argued by Binder long ago \cite{Binder:1981sa} that, at the critical value $\beta_c$, the cumulant has a universal value $U_* \neq 0$ as $L \to \infty$, universal in the sense that it is independent of $\lambda$ and shared by all systems in the Ising universality class. The approach of $U_L(\beta_c)$ to $U_*$ is determined by \textit{corrections to scaling} of the form \cite{Amit:1984ms, Cardy:1996xt, ZinnJustin:2002ru}
\BE \label{corrections_to_scaling}
U_L(\beta_c) = U_* + c_1(\lambda) L^{-\omega} + O( L^{-2\omega}, L^{-\omega'}),
\EE
where $\omega = -y _4 > 0$ is the exponent corresponding to the leading irrelevant RG eigenvalue of the system, and $\omega'$ stands for the next-to-leading irrelevant exponent. The universal values $U_*$ depend only on the dimension of the system for spin systems with only 1 internal degree of freedom.  They are known to very high precision \cite{Salas:1999qh, Kaupuzs:2016hpl, Hasenbusch:1999mw}: in 2d, $U_* = 0.9160386(24)$, while in 3d, $U_* = 0.69832(13)$.

\begin{figure}[h!]
\centering
\begin{minipage}{.48\textwidth}
\includegraphics[width=1.05\textwidth]{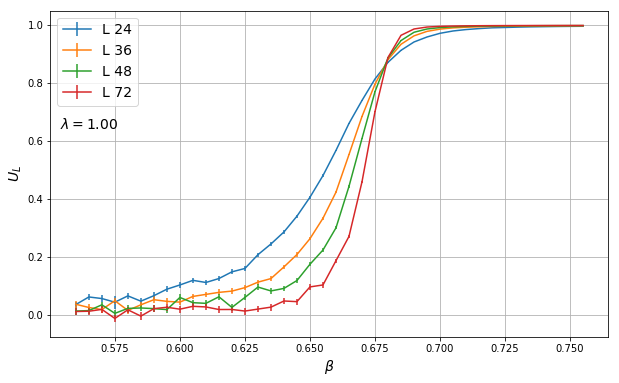}
\end{minipage}\hfill
\begin{minipage}{.48\textwidth}
\includegraphics[width=1.0\textwidth]{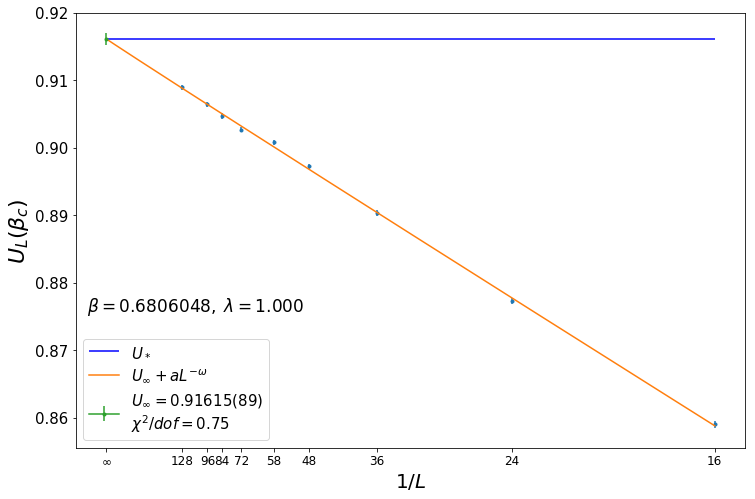}
\end{minipage}
\caption{\small{(Left) Binder cumulant on multiple volumes over a wide range of $\beta$ in 2d. The curves for different $L$ nearly intersect close to the critical value $\beta_c$. (Right) Extrapolation of the Binder cumulant computed at $\beta_c$ to infinite volume in 2d according to eq. (\ref{corrections_to_scaling}).\label{fig:binderplots}}}
\end{figure}

In figure \ref{fig:binderplots} we plot the behavior of the cumulant $U_L$ in 2d as a function of $\beta$ on several lattices. We see that there exists a region where the different volumes nearly intersect each other. The cumulant is analytic in $\beta - \beta_c$, and therefore this behavior is expected to occur in the vicinity of $\beta_c$ according to eq. (\ref{corrections_to_scaling}), up to $O(L^{-\omega})$ deviations. Furthermore, exactly at the critical point, an infinite volume extrapolation of the cumulant should yield the universal value $U_*$. Very precise estimates exist in the literature for the critical couplings $\beta_c$ at various interaction parameter values $\lambda$. Using these values, we have checked that our system is well-tuned by performing infinite volume extrapolations as suggested by eq. (\ref{corrections_to_scaling}) at leading order. This is depicted for our simulation in $d=2$ in figure \ref{fig:binderplots}. In both dimensions we obtain good fits consistent with the universal value at $L=\infty$ within $1\sigma$; the fit results are exhibited in table \ref{binder_fitresults}.

\begin{table}[h!]
\begin{center}
 \begin{tabular}{||c c c c c c c||} 
 \hline
$d$ & $\lambda$ & $\beta_c$ & $U_\infty$ & $\omega$ & $a$ & $\chi^2/\mrm{dof}$\\ 
 \hline\hline
2 & 1.00 & 0.6806048 & 0.91615(89) & 0.989(38) & -0.890(86) & 0.75 \\ 
 \hline
3 & 1.100 & 0.3750966 & 0.6971(20) &  0.845(10)  & 0.036(44) & 0.43 \\ 
\hline
\end{tabular}
\caption{Estimates of the universal Binder cumulant and related fit parameters from our simulation in dimensions $2$ and $3$, performed at the critical $\beta$ values from refs. \cite{Kaupuzs:2016hpl}, \cite{Hasenbusch:1999mw}, respectively.}
\label{binder_fitresults}
\end{center}
\end{table}

\subsection{Simulation details} We simulated $\phi^4_d$ theory using Markov Chain Monte Carlo (MCMC) methods. In what follows we report the details for 3 dimensions, for simplicity. The MC chain of field configurations was generating using a mixed update algorithm consisting of Metropolis updates for the size of $\phi$ and Wolff cluster updates for the sign of $\phi$ \cite{PhysRevLett.62.361}. One Metropolis update involved picking $V = L^d$ random sites in sequence and for each pick updating the spin length according to
\BE
\phi' = \phi + r (u - 0.5),
\EE
where $u$ is a random number uniformly distributed in the interval $[0,1]$, with probability $\me^{-\Delta S}$, with $\Delta S = S' - S$ being the change in the action due to the proposed spin update. The number $r$ is the maximum radial update length, which was chosen to be $r = 2.00$ in both dimensions. One cluster update consisted of the attempted construction of a cluster, which picks a site at random and adds neighboring spins to the cluster with probability $\me^{-2\beta \sigma_i \sigma_j}$, where $\sigma_i = \mrm{sgn}(\phi_i)$. Once built, the signs of all spins in the cluster are flipped. The Twister PRNG was used to generate all random variables \cite{10.1145/272991.272995}.

The ratio of radial updates to cluster updates was chosen to be that which led to the best extrapolation of the Binder cumulant to infinite volume for given sample size at criticality. In 3d, we chose 10 cluster updates per radial update, yielding $\tau_\mrm{int} \approx 10$, with variations of order 1 between different observables. Measurements in the full simulation were then carried out every 5 MC sweeps, where one sweep was defined to be 50 cluster updates and 5 radial updates.
The autocorrelation was estimated in two ways: (1) errors $\eps_b$ were computed on binned data for various bin sizes, and the integrated time is estimated as $\tau_\mrm{int} \approx \eps_{*}^2 / 2 \eps_0^2$, where $\eps_*$ was the error on the binning plateau \cite{Amit:1984ms}, and (2) a direct computation of the integrated autocorrelation time on sequential subsets of the MC chain, repeated for every subset and averaged together. We checked that $\tau_\mrm{int}$ for $\phi^2$ and $\phi(0) \phi(z)$ at maximum distance $L/2$ were comparable in both cases.

To multiply the statistics, the MC chain was split at 10k sweeps (the thermalization cut) into 10 branches. After a few sweeps, the data from separate chains was checked to be essentially uncorrelated. On each branch, almost 1M sweeps were carried out (for every volume except the two largest, 64 and 72, which had 150k sweeps per chain), yielding a total of $\approx$ 10M MC sweeps, and thus 2M measurements.  To saturate the errors, the data was then binned with bin size $b=10$, yielding about 20k independent statistical samples per branch. We simulated on volumes $L=24,36,48,56,64,72$.
Lastly, the numerical integration of the gradient flow was performed using a fourth order Runge-Kutta integration scheme.

\subsection{Ratios and exponents}

The ratio formulas proposed above have been tested by measuring the mixed correlation functions in the odd-operator subspace with basis $\{\phi, \phi^3\}$ and even subspace with basis $\{\phi^2, \phi^4\}$, each one containing the leading two operators according to canonical dimension. Corresponding to each of these operators is a scaling dimension $\Delta_i, \; i = 1,2,3,4$, although those are the dimensions of the \textit{scaling} operators that are dominated by the corresponding monomial operators $\phi^i$. The most precise estimates we know of from lattice simulations, except $\Delta_3$, are given in table \ref{preciseDeltas} \cite{Hasenbusch:1999mw}. The $\Delta_3$ dimension is predicted to be $\Delta_3 = 2 + \Delta_1$ \cite{Rychkov:2015naa}. We are unaware of any direct numerical determinations of $\Delta_3$ apart from the conformal bootstrap \cite{Poland:2018epd}, so we use the prediction just mentioned. Preliminary results of this section were reported in \cite{Carosso:2018rep,Carosso:2019tan}. In what follows, we describe the analysis in the context of $\phi^4_3$. At the end, we briefly report preliminary results for $\phi^4_2$.

\begin{table}[h!]
\begin{center}
 \begin{tabular}{||c c c||} 
 \hline
$\Delta$ & $d=2$ & $d=3$ \\ 
 \hline\hline
$\Delta_1$ & 0.125 & 0.51790(20) \\ 
 \hline
$\Delta_2$ & 1 & 1.41169(76) \\ 
\hline
$\Delta_3$ & 2.125 & $\sim$ 2.5 \\
\hline
$\Delta_4$ & 2 & 3.845(11) \\
\hline
\end{tabular}
\caption{Exact scaling dimensions in 2d, known from the solution of the 2d Ising model. In 3d, we report the most precise estimates of scaling dimensions from the lattice \cite{Hasenbusch:1999mw}.}
\label{preciseDeltas}
\end{center}
\end{table}

\begin{figure}[h!]
\centering
\begin{minipage}{.48\textwidth}
\includegraphics[width=0.98\textwidth]{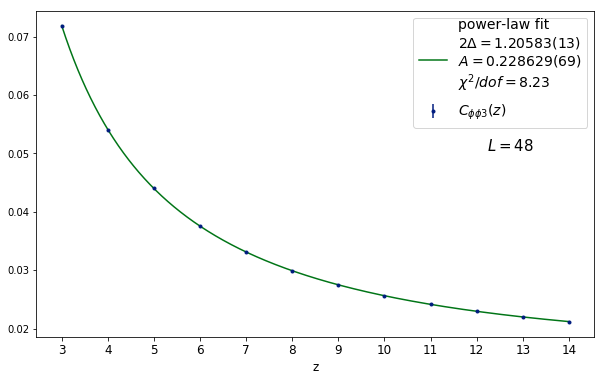}
\end{minipage}\hfill
\begin{minipage}{.48\textwidth}
\includegraphics[width=1.0\textwidth]{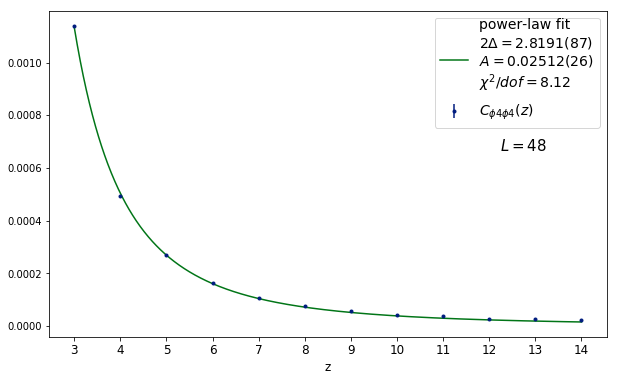}
\end{minipage}
\caption{\small{Power law exponential fits of the $\phi\phi^3$ and $\phi^4\phi^4$ correlators that exhibit the dominance of the leading operators $\phi$ and $\phi^2$. \label{fig:powerlawfits}}}
\end{figure}

At criticality, the point-point correlation functions are expected to exhibit power law decay of the form $C(z)=A/z^{2\Delta}$ in infinite volume. Since $\Delta_1, \; \Delta_2$ are the leading dimensions in their respective subspaces, they are expected to govern the leading power law behaviors. In figure \ref{fig:powerlawfits} we plot the $\phi \phi^3$ and $\phi^4 \phi^4$ correlators together with their fits to a periodic power-law $C(z) + C(L-z)$ with $C(z) =  A/z^{2\Delta}$.
We observed power laws that clearly indicate the dominance of the leading operators in each subspace, with exponents (reported in the plots) close to the expected $2\Delta_1 \approx 1.0358, \; 2\Delta_2 \approx 2.82$. For $\phi\phi^3$, the subleading power law has exponent $2\Delta' \approx 5$; thus, for both correlators we observe a clear dominance by the leading operator.

\begin{figure}[h!]
\centering
\begin{minipage}{.48\textwidth}
\includegraphics[width=1.0\textwidth]{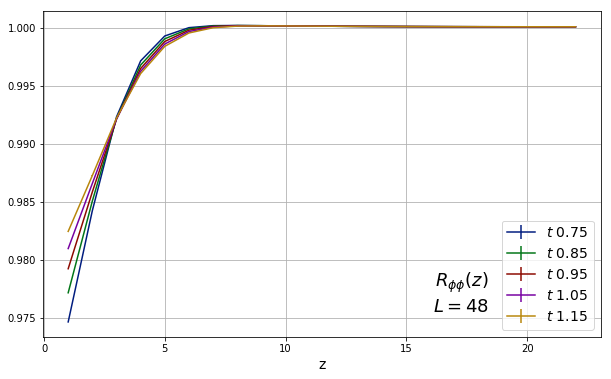}
\end{minipage}\hfill
\begin{minipage}{.48\textwidth}
\includegraphics[width=1.0\textwidth]{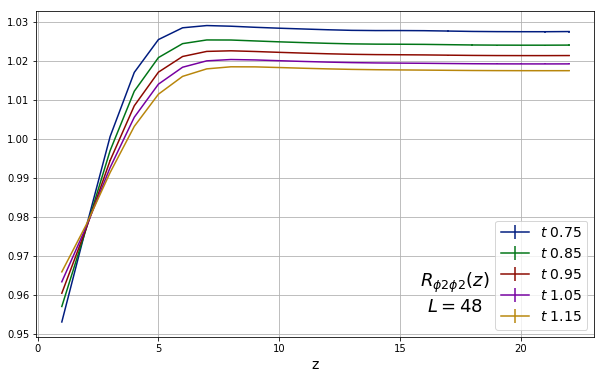}
\end{minipage}
\caption{\small{Ratios of flowed correlators versus distance. Plateaus form at large distances where scaling sets in and the ratio formulas become valid.\label{fig:ratios}}}
\end{figure}

In figure \ref{fig:ratios}, we plot the ratios of correlators at several flow times as functions of distance on the lattice. We observe the short-distance region where the smeared spins overlap as dips in the ratios. At larger distances, plateaus form where the ratios approach their asymptotic forms, although there appears to be slight residual $z$-dependence. For the $\langle\phi\phi\rangle$ correlator, the plateau moves extremely little with flow time, as predicted by eq. (\ref{ratioO}). For all other correlators there is notable movement. The residual $z$-dependence could come from a number of sources. First, we expect that even with a dominant operator, there will always be subleading corrections due to the leading irrelevant operators. If we keep the first subleading term in eq. (\ref{corr_expansion}) and compute the ratio in eq. (\ref{ratioO}), assuming power law correlations, we can derive the expected form of these corrections (in infinite volume). Denoting ratios by $R(z,t)$, we find
\BE
R(z; t) \sim b_t^{2\Delta_a} \Big( 1 + (b_t^{2(\Delta_b - \Delta_a)} - 1) \frac{c_1}{z^{2(\Delta_b - \Delta_a)}} + b_t^{2(\Delta_b - \Delta_a)} \frac{c_1^2}{z^{4(\Delta_b - \Delta_a)}} \Big).
\EE
In 3 dimensions, $\Delta_b - \Delta_a \gtrsim 2$ in both subspaces, so we expect these corrections to be small at large distances. We were unable to extract estimates for these subleading terms from fits. A second source of $z$-dependence is the fact that the finite volume heat kernel has nontrivial behavior in $z$. However, we expect such corrections to be multiplied by factors of $O(\me^{-n^2 L^2 / 4 t}) \; \forall n>0$. See appendix A for details about the finite volume heat kernel.

We therefore have attempted to extract $\Delta_1$ from applying eqs. (\ref{ratioR}), (\ref{Option2}) in the odd subspace, and  separately $\Delta_1, \; \Delta_2$ from the even subspace, according to Option 2 outlined in section 2. In applying the ratio formula, we compare flow times separated by $\eps = 0.05$ and fit using the form (inspired by eq. (\ref{bt_form}))
\BE
R_{\phi^i \phi^j}(z;t) \sim b_\eps(t)^{\delta_{ij}} = \Big(1 + \frac{\eps}{c + t}\Big)^{\delta_{ij}/2},
\EE
where $\delta_{ij} = 2\Delta_a - (n_i + n_j) \Delta_1$ is expected for the operators we use, $a$ being the index of the dominant operator in the correlator $\langle \mcal{O}_i \mcal{O}_j \rangle$. We remark that this form allows one to attempt fitting at arbitrarily small $t$ values, but that the scaling form is expected only for larger times. The correlators at nearby flow times are statistically highly correlated due to the smoothing effect of GF, making the estimation of errors by classical means a risky task,
 much like the high correlation of correlator values at successive distances in spectrum measurements. We thus adopt a jackknifing procedure whereby a number of sub-ensembles are generated from the whole ensemble by removing chunks of $J$ samples in sequence, then the correlator ratios are computed and the fits performed, and the best fit parameters are collected together into an ensemble \cite{DeGrand:2006zz}. To account for the noise observed at large distances we have included in our jackknife ensemble fits from every $z$ value from regions where a stable plateau is identifiable in the ratio plots. On each volume we chose a ratio $N/J \approx 40$, where $N$ is the total number of samples. The final estimates on a given volume are then obtained by computing the means and covariance of the ensemble of best fit parameters.

\begin{figure}
\centering
\begin{minipage}{.48\textwidth}
\includegraphics[width=1.0\textwidth]{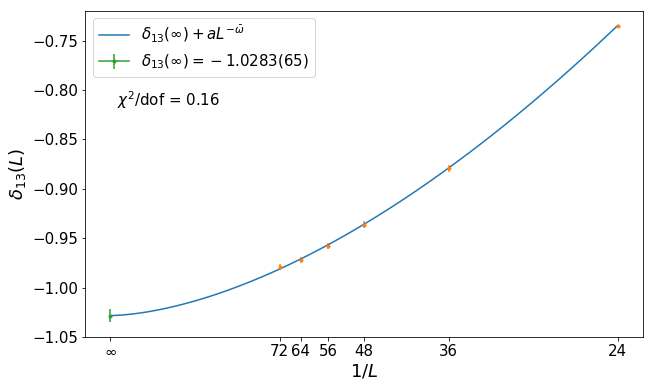}
\end{minipage}\hfill
\begin{minipage}{.48\textwidth}
\includegraphics[width=1.0\textwidth]{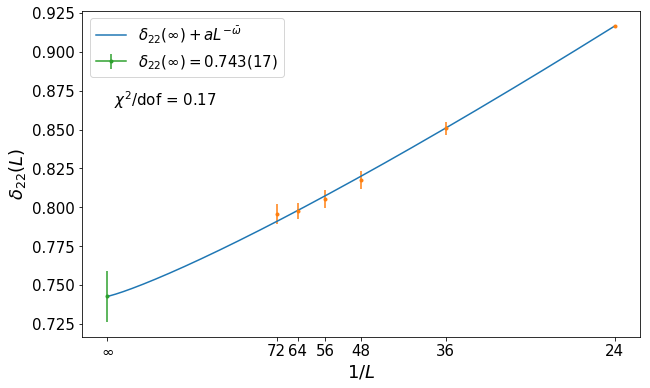}
\end{minipage}
\caption{\small{Infinite volume extrapolation of the $\delta_{13}$ and $\delta_{22}$ exponents according to eq. (\ref{inf_vol_ansatz}).}}
\label{fig:delta22_extrap}
\end{figure}

The results show a notable dependence on volume. We therefore extrapolate to infinite volume using a leading correction-to-scaling ansatz,
\BE \label{inf_vol_ansatz}
\delta_{ij}(L; a, \bar \omega) = \delta_{ij} + a L^{-\bar \omega}.
\EE
In \fig{fig:delta22_extrap}, the examples of $\delta_{13}$ and $\delta_{22}$ are displayed. In the odd subspace, one can extract $\Delta_1$ from a single $\delta_{ij}$. In the even subspace, one can extract $\Delta_1, \; \Delta_2$ from any \textit{pair} of $\delta_{ij}$'s. Best fit parameters of the infinite volume extrapolations are reported in table \ref{delta_fits}, and the corresponding scaling dimensions are reported in table \ref{MyDeltas}.

\begin{table}[h!]
\begin{center}
\begin{tabular}{||c c | c c c c||} 
 \hline
$(i,j)$ & $\delta_{ij}\text{\small{(table \ref{preciseDeltas})}}$ & $\delta_{ij}$ & $\bar \omega$ & $a$ & $\chi^2/\mrm{dof}$ \\ 
 \hline\hline
$(1,3)$ & -1.03580(40) & -1.0283(65) & 1.667(80) & 59(14) & 0.16 \\ 
 \hline
$(3,3)$ &  -2.07160(80) & -2.055(18) & 1.73(11) & 154(51) & 0.30 \\ 
\hline
$(2,2)$ & 0.7518(17) & 0.743(17) & 1.17(18) & 7.2(3.5) & 0.17 \\
\hline
$(2,4)$ &  -0.2840(19) & -0.308(38) & 0.96(20) & 5.8(2.9) & 0.11 \\
\hline
$(4,4)$ &  -1.3198(22) & -1.310(29) & 1.56(16) & 84(38) & 0.20 \\
\hline
\end{tabular}
\caption{\small{Fit results of the infinite volume extrapolations of $\delta_{ij}$ from mixed correlators $\langle \phi^i \phi^j \rangle$ using eqs. (\ref{ratioO}), (\ref{Option2}) and fitting to eq. (\ref{inf_vol_ansatz}). The expected values from table \ref{preciseDeltas} are tabulated for comparison.}}
\label{delta_fits}
\end{center}
\end{table}

\begin{table}[h!]
\begin{center}
 \begin{tabular}{||c c c||} 
 \hline
$(i,j), (h,k)$ & $\Delta_1$ & $\Delta_2$ \\ 
 \hline\hline
$(1,3)$ & 0.5141(32) & ---  \\ 
 \hline
$(3,3)$ & 0.5138(45) & ---  \\ 
\hline
$(2,2),(2,4)$ & 0.525(21) & 1.422(46)  \\
\hline
$(2,2),(4,4)$ & 0.5132(84) & 1.398(22) \\
\hline
$(2,4),(4,4)$ & 0.501(24) & 1.349(88)  \\
\hline\hline
table \ref{preciseDeltas} & 0.51790(20) & 1.41169(76) \\
\hline
\end{tabular}
\caption{\small{Estimates of leading scaling dimensions in 3d from mixed correlators $\langle \phi^i \phi^j \rangle$ (or pairs thereof) using eqs. (\ref{ratioO}), (\ref{Option2}). Values are obtained from the results reported in table \ref{delta_fits}.}}
\label{MyDeltas}
\end{center}
\end{table}

\begin{figure}[h!]
\centering
\begin{minipage}{.48\textwidth}
\includegraphics[width=1.0\textwidth]{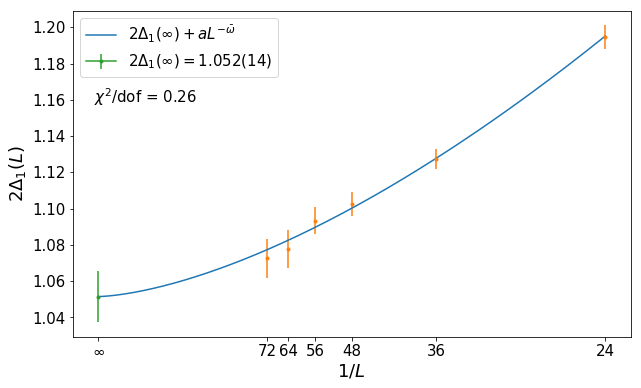}
\end{minipage}\hfill
\begin{minipage}{.48\textwidth}
\includegraphics[width=1.0\textwidth]{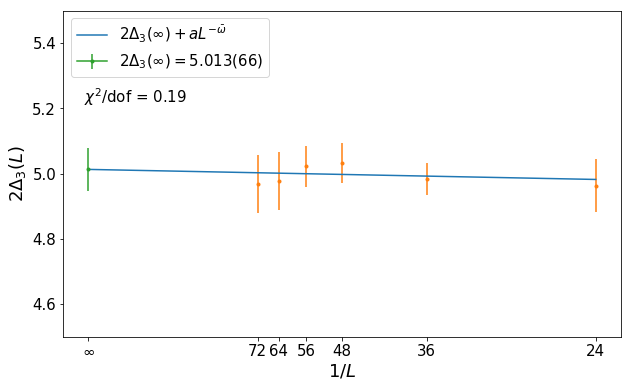}
\end{minipage}
\caption{\small{Infinite volume extrapolations of $2\Delta_1$ and $2\Delta_3$ from the diagonalization method using correlator ratios from a reduced set of distances, $z \in [10,14]$. \label{fig:diag_D1D3}}}
\end{figure}

Next, we report the results of the diagonalization procedure based on eq. (\ref{ratioR}). The leading diagonalized correlators with dimensions $\Delta_1$ and $\Delta_2$ had a clean signal at all distances past $z = 10$ on every lattice, in the sense that their plateaus exhibited no notable noise. The subleading correlators, however, tended to exhibit wild fluctuations past certain distances ($z \gsim 15$), where the signal from the subleading operators becomes small. At shorter distances ($z \lsim 10$), there was a systematic tendency for exponents to be underestimated.
Repeating the same data analysis used to extract the $\delta_{ij}$ and their infinite volume extrapolations, but now extracting directly estimates of $2\Delta_a$, and using the first value of $\Delta_1$ from table \ref{preciseDeltas} as the necessary input dimension in eq. (\ref{ratioR}), we obtained estimates for $\Delta_1$ and $\Delta_2$. The value for $\Delta_1$ is slightly displaced from the input value, but the value for $\Delta_2$ is consistent with those extracted in table \ref{MyDeltas}. 
For the $\mcal R_3$ scaling operator, distances beyond $z \approx 15$ were left out of our analysis because of a poor signal. In figure \ref{fig:diag_D1D3}, we plot the extrapolations for $\Delta_1$ and $\Delta_3$ using the same limited $z$-range $z \in [10,14]$. The value $\Delta_3 = 2.506(33)$ is consistent with the prediction of $2 + \Delta_1 \approx 2.518$ from \cite{Rychkov:2015naa, Hasenbusch:1999mw}. The signal for $\Delta_4$ was generally quite poor. The data was not clean enough to perform an infinite volume extrapolation. The most reasonable estimates on each volume were obtained from distances $z < 14$. A crude estimate obtained from averaging results from every volume in the range $z \in [10, 13]$, for example, yields $2\Delta_4 = 7.98(65)$, while adding one more distance, so $z \in [10,14]$, yields $2\Delta_4 = 8.31(99)$. The expected value is $2\Delta_4 \approx 7.69$. We take this as a good sign, but without the infinite volume extrapolation the result is not as precise as the lower dimensions $\Delta_1, \; \Delta_2, \; \Delta_3$.

\begin{figure}[h!]
\centering
\begin{minipage}{.48\textwidth}
\includegraphics[width=1.0\textwidth]{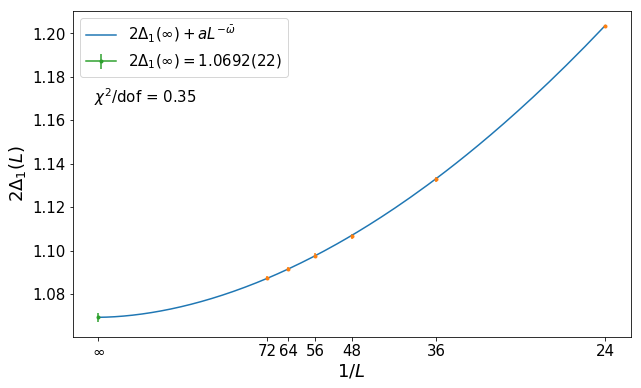}
\end{minipage}\hfill
\begin{minipage}{.48\textwidth}
\includegraphics[width=1.0\textwidth]{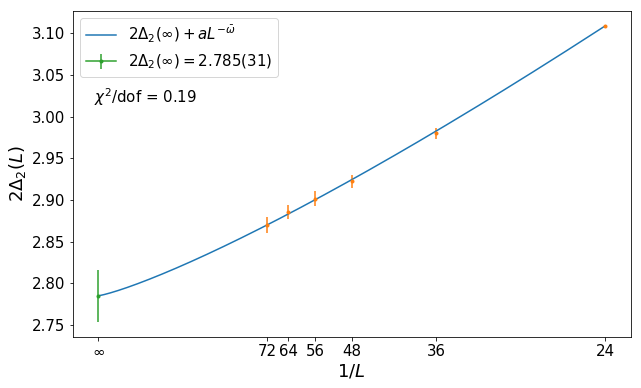}
\end{minipage}
\caption{\small{Infinite volume extrapolations of $2\Delta_1$ and $2\Delta_2$ from the diagonalization method using correlator ratios at all distances past $z=10$. \label{fig:diag_D1D2}}}
\end{figure}

\newpage
In 2 dimensions, we have carried out a preliminary analysis to obtain estimates of $\delta_{ij}$ in an identical manner as above, although with about a fifth of the statistics, so far. The results are reported in table \ref{delta2d_fits}. In 2d, the canonical dimension of $\phi$ is zero, and therefore all scaling dimensions are purely anomalous. Ratios for $\phi\phi$ again exhibited minimal variation with time, while higher operator ratios exhibited significant movement. The plateaus were generally less flat than they were in 3d, possibly due to larger contributions from subleading operators. Nonetheless, suggestive estimates were obtained for the $\delta_{ij}$. In the odd-subspace, the results deviate from the exact values by many standard deviations, indicating perhaps the stronger presence of the subleading operator and the necessity of a diagonalization analysis. This has not yet been carried out.

\begin{table}[h!]
\begin{center}
\begin{tabular}{||c c | c c c c||} 
 \hline
$(i,j)$ & $\delta_{ij}\text{\small{(table \ref{preciseDeltas})}}$ & $\delta_{ij}$ & $\bar \omega$ & $a$ & $\chi^2/\mrm{dof}$ \\ 
 \hline\hline
$(1,3)$ & -0.25 & -0.2616(14) & 2.48(21) & 127(83) & 0.21 \\ 
 \hline
$(3,3)$ &  -0.50 & -0.5279(28) & 2.35(18) & 161(91) & 0.55 \\ 
\hline
$(2,2)$ & 1.50 & 1.538(20) & 1.92(35) & 90(93) & 0.35 \\
\hline
$(2,4)$ &  1.25 & 1.299(31) & 1.79(31) & 103(93) & 0.30 \\
\hline
$(4,4)$ &  1.00 & 1.061(60) & 1.62(25) & 129(93) & 0.39 \\
\hline
\end{tabular}
\caption{\small{(Preliminary) Fit results of the infinite volume extrapolations of the $\delta_{ij}$ in 2 dimensions. The expected values from table \ref{preciseDeltas} are tabulated for comparison.}}
\label{delta2d_fits}
\end{center}
\end{table}

Lastly, we note that the exponents we have measured above, of course, are not as precisely determined as they are from finite-size scaling (FSS) analyses, and neither method is nearly as precise as the conformal bootstrap predictions \cite{Poland:2018epd} for scalar field theories. We note that the FSS results from \cite{Hasenbusch:1999mw}, however, had roughly 10-100 times the statistics we have. An advantage of GFRG methods is that they are expected to be applicable in a much broader class of lattice field theories, including gauge-fermion systems in 4 dimensions. In such systems, the nonperturbative determination of anomalous dimensions is a lively and ongoing research program, and the conformal bootstrap has only recently made progress in these systems \cite{Poland:2018epd,Li:2020bnb}. In the next section, we describe an application of GFRG to one such system.

\section{12-flavor SU(3) gauge theory}

A model of central interest in the beyond Standard Model lattice community is the nearly-conformal $N_f$-flavor SU(3) gauge theory and its generalizations. It is a candidate for explaining the electroweak symmetry breaking which produces the Higgs boson, arising from new strong interactions at a higher energy scale \cite{Appelquist:2016viq}. Its motivation lies in the fact that for a range of $N_f$ values, called the ``conformal window,'' the theory may contain a light scalar identified with the Higgs. The perturbative beta function for SU($N$) gauge-fermion systems is given to 2-loop order by \cite{Montvay:1994cy}
\BE
\beta(g_0) = - \beta_0 g_0^3 - \beta_1 g_0^5  + O(g_0^7),
\EE
where
\BE
\beta_0 = \frac{1}{16 \pi^2}\Big( \frac{11 N}{3} - \frac{2 N_f}{3} \Big), \quad \beta_1 = \frac{1}{(16 \pi^2)^2} \Big( \frac{34 N^2}{3} - \frac{10 N N_f}{3} - \frac{(N^2-1)N_f}{N} \Big).
\EE
As $N_f$ is increased from 0 the theory eventually develops an interacting infrared fixed point (IRFP), whose coupling strength decreases with $N_f$: the Caswell-Banks-Zaks fixed point \cite{Caswell:1974gg,Banks:1981nn}. Above $N_f = 16.5$, the IRFP merges with the gaussian fixed point and the theory loses asymptotic freedom. In the conformal regime, the IRFP is characterized by a set of scaling dimensions of local operators. The values of these scaling dimensions are highly nontrivial, especially as $N_f$ is lowered from $16.5$ and one begins to leave the class of weakly-coupled IRFPs. Much work has been done to determine the anomalous dimensions, both analytically \cite{Gracey:2018oym, DiPietro:2020jne} and on the lattice \cite{Cheng:2013bca, Giedt:2015alr}, and lattice simulations in particular have focused on the determination of the fermion mass anomalous dimension for reasons of phenomenology as well as practicality. 

We expect that the ratio formula eq. (\ref{ratioO}) is applicable in generic field theories, since blocking may be defined in any theory. In fact, the \textit{first} application of the correlator ratio method outlined in this chapter was to a $N_f = 12,$ SU(3) lattice gauge theory using staggered fermions \cite{Carosso:2018bmz}. We computed ratios of the pion, axial vector, vector, and baryon correlators, and extracted estimates for the mass anomalous dimension of the fermions and the leading baryon. Because staggered fermions exhibit a conserved current from a remnant chiral symmetry, we used its correlator ratio to estimate the fermion field anomalous dimension, and to eliminate its contribution to other ratios. Before reporting results, we will describe our prescription for the flow and the method of measurement of flowed hadronic observables.

\subsection{Flow definitions} The first question to address is what type of gradient flow should be used in a gauge theory. If we wish to keep the effective action in the same universality class of the bare theory, the flow should preserve the symmetries of the theory. In this case, the flow equation must maintain gauge invariance. The simplest gauge-invariant action is the Yang-Mills action, so we expect YM gradient flow to be sufficient in the continuum. It is clear from a perturbative analysis of the equations that the flowed fields have the desired damping of high-momentum modes of the gauge field $A_\mu(x)$ \cite{Luscher:2010iy}. We also note that the requirement of gauge invariance necessitates a nonlinear gradient flow, a feature we did not see in the scalar case. 

When translating to the lattice, one must decide on the discretization of the gauge action. From a Wilsonian RG perspective, it is expected that any discretization is fine, in principle, so long as it reduces to the YM flow in the continuum limit. The simplest lattice discretization of the Yang-Mills gradient flow is called \textit{Wilson flow}. Beginning with bare links $U_\mu(x)$, their flow $V_\mu(x,t)$ is determined by
\BE \label{wilsonflow}
\del_t V_\mu(x,t) = - g_0^2 \del_{x,\mu} S_{W}(V_t) \; V_\mu(x,t),
\EE
where $S_{W}(U)$ is the Wilson action, eq. (\ref{Waction}), and $\del_{x,\mu} = T^a \del_{x,\mu}^a$ is a Lie algebra-valued derivative. It is defined on functions $f(U)$ on the group $G$, where $U = \{U_\nu(y)\}$ is the set of gauge links on the lattice, by
\BE
\del_{x,\mu}^a f(U) := \frac{\dd}{\dd s} f \big( \me^{sX_\mu(x)} U \big) \Big|_{s = 0}, \quad \mrm{where} \quad X_\mu(x) := \delta_{\mu\nu} \delta(x,y) T^a.
\EE
In words, then, the derivative operation first replaces $U_\mu(x)$ in $f(U)$ by $\me^{s T^a} U_\mu(x)$, differentiates with respect to $s$, and sets $s = 0$. It's then clear that $\del_{x,a} f(U) \in \mathfrak{g}$. It is nothing but the definition of the tangent vectors to the gauge group $G$ at element $U$. As the simplest discretization, the Wilson flow has significant lattice artifacts, and therefore alternative flow definitions have been given \cite{Ramos:2015baa}, but we will not discuss these.

The flow of the fermions $\psi(x)$ can be defined with a simple gauge-covariant diffusion, i.e., gauged heat equation \cite{Luscher:2013cpa}. The covariant derivative in the fundamental representation is $D_\mu(A) = \del_\mu + A_\mu$. We denote flowed fermion fields by $\chi_t(x)$. The diffusion equation in the continuum is then
\BE \label{fermionflowcontinuum}
\del_t \chi_t(x) = D^2(B) \chi_t(x).
\EE
The simplest discretization of this flow would be
\BE \label{fermionflow}
\del_t \chi_t(x) = \Delta(V_t) \chi_t(x),
\EE
where $\Delta(U) = \sum_\mu \nabla^*_\mu \nabla_\mu$, and the covariant difference operators are
\BE
\nabla_\mu \chi(x) = U_\mu(x) \chi(x+\mu) - \chi(x), \quad \nabla^*_\mu \chi(x) = \chi(x)  - U_\mu^\dag(x-\mu) \chi(x-\mu).
\EE
But the choice of flow is to some extent arbitrary, so long as it serves to damp high modes while preserving the symmetries of the field. Thus, one could alternatively define the flow by the square of the Dirac operator $\slashed{D}^2(U)$,
\BE
\del_t \chi_t(x) = \slashed{D}^2(V_t) \chi_t(x).
\EE
In fact, the two kinds of second derivative are related as
\BE
\slashed{D}^2 = D^2 + \sigma_{\mu\nu} F_{\mu\nu}, \quad \sigma_{\mu\nu} = \smallfrac{1}{2} [\gamma_\mu,\gamma_\nu].
\EE
We can think of the difference between these flows as follows. The flow generates an effective action which typically contains all possible terms consistent with the symmetries of the theory, so terms like $c(t) \bar \chi \sigma_{\mu\nu} F_{\mu\nu} \chi$ would be present for both choices of flow. They would only differ in their dependence on $t$, on the precise form of the coefficient $c(t)$. But since the dynamics of the theory is controlled by its IRFP, differences in the exact details of $c(t)$ will become less important as $t$ increases and the effective action approaches the fixed point action.

\subsection{Flowed observables}

For gauge fields and scalar fields, the way to compute observables at finite flow time is straight-forward. One simply evaluates the operators within expectation values on the flowed fields. For fermions, however, the problem is more subtle, because fermion fields are Grassmann-valued and therefore they are not directly manipulated and measured in lattice simulations. One therefore must do some work for any given observable to understand how it should be measured.

The simplest fermionic observables are the mesonic operators. A flowed meson operator has the general bilinear form
\BE
P_t(x) = \overline \chi_t(x) \Gamma \chi_t(x)
\EE
for some gamma matrix $\Gamma$ (or staggered equivalent thereof), as these are the operators which can have the same quantum numbers as the mesons out in nature. Their correlators are then defined by
\BE \label{fullcorr}
C_t(x,y) = \langle P_t(x) P_t(y) \rangle.
\EE
It is also convenient to define \textit{partially-flowed} correlators by
\BE \label{partialcorr}
\tilde C_t(x,y) = \langle P_t(x) P_0(y) \rangle,
\EE
as these are simpler to measure and differ from the fully-flowed correlators by terms of $O(\sqrt{t} / |x-y|)$, as we argue in the next section. The flowed baryon operators are similarly defined. For the simplest staggered baryon, the flowed operator is given by
\BE
B_t(x) = \eps_{abc} \chi_t^a(x) \chi_t^b(x) \chi_t^c(x),
\EE
where $a,b,c=1,\dots,3$ are the color indices. Their correlations are given just as in eqs. (\ref{fullcorr}, \;\ref{partialcorr}). 

To understand how such correlators are measured, first we compute their contractions. Letting $S(x,y) = \slashed{D}^{-1}(x,y)$ be the inverse Dirac operator, one finds \cite{Luscher:2013cpa}
\begin{align}
\contraction{}{\psi(x)}{}{ \bar{\psi}(y)} \psi(x)\bar{\psi}(y) &= S(x,y), \nonumber \\
\contraction{}{\chi_t(x)}{}{ \bar{\psi}(y)} \chi_t(x) \bar{\psi}(y) &= \sum_v \! K(t,x;0,v) S(v,y), \nonumber \\
\contraction{}{\psi(x) }{}{\bar{\chi}_t(y)}\psi(x) \bar{\chi}_t(y) &= \sum_v \! S(x,v) K(t,y;0,v)^\dag, \nonumber \\
\contraction{}{\chi_t(x)}{}{\bar{\chi}_t(y)} \chi_t(x)\bar{\chi}_t(y) &= \sum_{vw} \! K(t,x;0,v) S(v,w) K(t,y;0,w)^\dag,
\end{align}
where $K(t,x;s,y)$ is the gauge covariant Green function solution of the fermion flow equation, and therefore depends on the gauge field in a nontrivial way. One formally writes the solution as 
\BE
K(t,s) = \exp \int_s^t \dd \sigma \Delta(V_\sigma),
\EE
where $V_\sigma$ solves the flow equation (\ref{wilsonflow}). Using the contractions above, one integrates over the fermions in the flowed expectation values to obtain expressions in terms of the gauged heat kernel and the gauge fields. For example, integrating over the fermions in eq. (\ref{partialcorr}) gives
\begin{align}
- \Gamma_{\alpha\beta}\Gamma_{\gamma\delta} & \sum_{v}\! K(t,y;0,v) S(v,x)_{\delta\alpha} \cdot \sum_{u}\! S(x,u)_{\beta\gamma} K(t,y;0,u)^\dag \nonumber \\
& = - \int_{vu} \! \mrm{tr}\big[ K(t,y;0,v) S(v,x) \Gamma \; S(x,u) K(t,y;0,u)^\dag \Gamma \big].
\end{align}
In the case of pions (for Wilson fermions, say), the gamma matrix is $\Gamma = \gamma_5$, and from $\gamma_5$-hermiticity, $\gamma_5 S(x,y) \gamma_5 = S^\dag(y,x)$, we have
\BE
\langle P(x) P_t(y) \rangle = - \sum_{vu} \! \mrm{tr}\big[ K(t,y;0,v) S(v,x) \; \big( K(t,y;0,u) S(u,x) \big)^\dag \big].
\EE
Now, on the lattice, one computes $S(x,y)$ by placing a point source $\eta_y$ at site $y$ defined by $\eta_y(x) = \delta(x,y)$, and numerically inverts the Dirac operator on the source. For point-point correlators, then, the quantity
\BE
(K_t S \eta_y)(x) = \sum_{v} \! K(t,x;0,v) S(v,y),
\EE
is simply the solution of eq. (\ref{fermionflow}) with $\psi_t \to S$ and initial condition $S(v,y)$, with $y$ held fixed. Denoting the inversion of $\slashed{D}$ on the point source by $(S \eta_y)(x)$, the pion correlator on a single gauge configuration takes the form
\BE
- \sum_y | (K_t S\eta_x)(y) |^2,
\EE
which is numerically implementable. Thus, to measure the partially-flowed pion correlator, one inverts the Dirac operator on a point source and integrates the gauged heat equation with initial condition being the vector field $S\eta_y$.

Observables that are fully-flowed, such as eq. (\ref{fullcorr}), are much harder to measure, because one must integrate instead the \textit{adjoint} heat equation \cite{Luscher:2013cpa}, and the computational cost increases drastically. Even some local observables, like the chiral condensate $\bar \chi_t (x) \chi_t(x)$, require adjoint flow. None of the observables used below required the computation of adjoint flow, however.

\subsection{Super ratios}

As we saw above, it is numerically advantageous to compute expectation values with only a single flowed operator in the correlator. But the original ratio formula eq. (\ref{ratioO}) requires both operators to be flowed. If such partially-flowed correlators are approximately equal to the full correlators at large distances, then we expect a modified ratio formula (for $i = j$)
\begin{equation}
\frac{ \langle \op(0) \op_t(z) \rangle}{ \langle \op(0) \op(z) \rangle} \propto t^{\ell_\op / 2 + \gamma_\op/2 - n_\op \eta/4} + O(\sqrt{t}/z) \label{eq:partialratio},
\end{equation}
where the dependence on $t$ is the square root of the dependence in eq. (\ref{ratioO}), at large times. Intuitively, this should hold for distances much larger than the smearing radius, $z \gg \sqrt{t}$. Notice also that we switch to an emphasis on anomalous dimensions rather than full scaling dimensions in this section, as is customary in four dimensions.

To motivate this form of the ratio formula, let us consider the case of $\langle \phi_t(z) \phi_t(0) \rangle$. From $\phi_t(y) = (K_t \varphi)(y)$, where $K_t$ is the heat kernel, we obtain
\BE
\langle \phi_t(z) \phi_t(0) \rangle = \int_y K_t(y) \langle \phi_t(z) \varphi(y) \rangle \equiv \int_y K_t(y) G_t(z - y).
\EE
Now we expand $G_t$ about $z$ using
\BE
|z-y| = |z| \Big( 1 + \frac{y^2}{z^2} - 2 \frac{y \cdot z}{z^2} \Big)^{1/2} = |z|\Big(1 + \frac{y^2}{2 z^2} - \frac{y \cdot z}{z^2} - \frac{(y \cdot z)^2}{2(z^2)^2} + O(y^3) \Big),
\EE
where $x^2 = |x|^2$. The expansion of the correlator above is then
\BE
\int_y K_t(y) G_t(z - y) = \int_y K_t(y) \Big[ G_t(z) +  \Big(\frac{y^2}{2 z^2} - \frac{y \cdot z}{z^2} - \frac{(y \cdot z)^2}{2(z^2)^2} + O(y^3)\Big) z \cdot \nabla_z G_t(z) + O(\nabla_z^2) \Big].
\EE
Now, from the moments of the heat kernel $\langle 1 \rangle = 1$, $\langle y \rangle = 0$, and $\langle y_i y_j \rangle = 2t \delta_{ij}$, we obtain
\BE
\int_y K_t(y) G_t(z - y) = G_t(z) + (d-1) \frac{t}{z^2} z \cdot \nabla_z G_t(z) + O(t^2 / z^4, \nabla^2_z).
\EE
At large distances $z \gg \sqrt{t}$, the partially flowed correlators are then approximately equal to the fully-flowed correlators, and we expect eq. (\ref{eq:partialratio}) to be valid.

We can use eq. (\ref{eq:partialratio}) to determine the field anomalous exponent $\eta$ along the lines of Option 1 outlined above at eq. (\ref{Option1}). Once $\eta$ is determined, any other anomalous dimension can be predicted.  Alternatively, we may construct a \textit{super ratio} of the form
\begin{eqnarray}
\Rop_{\op}(t,x_0) &=& \frac{ \langle \op(0) \op_t(x_0) \rangle}{ \langle \op(0) \op(x_0) \rangle} \label{eq:ratio_full}
\Bigg( \frac{ \langle \symop(0) \symop(x_0) \rangle}{ \langle \symop(0) \symop_t(x_0) \rangle} \Bigg)^{n_\op/n_\symop}  \\ 
&=& b^{\Delta_\op - (n_\op / n_\symop) d_\symop}  \nn \\
&\propto& t^{\gamma_\op/2 + \delta/2}, \quad x_0 \gg a\sqrt{t},  \nn
\end{eqnarray}
which cancels the anomalous dimension $\eta$ directly, leaving only the desired anomalous dimension $\gamma_\op$ and some possible residual dependence on the canonical dimensions of $\op$ and $\symop$ through $\delta \equiv d_\op - (n_\op / n_\symop) d_\symop$. If the operators contain no derivatives then $\delta = 0$; this will be the case for all operators we consider in our numerical study.

Eq.~\ref{eq:ratio_full} is valid only on the critical $m=0$ surface and at sufficiently large flow times such that the linear basin of attraction of the IR-stable fixed point has been reached.  Otherwise, we expect the predicted $\gamma_\mcal{O}$ from eq.~\ref{eq:ratio_full} to show additional dependence on $t$ coming from irrelevant operators.  In practice, the flow time $t$ which can be reached is limited by the finite lattice volume.

\subsection{Finite volume corrections} To correct for finite volume, a different approach was used in this system than was later used in the scalar case.


The correlator scaling formula under a blocking transformation in a finite volume reads
\BE
C(z/b;g',L/b) = b^{\Delta} C(z;g,L),
\EE
where $g'$ is the coupling of the blocked theory after blocking by $b$. Now consider the same formula on a larger volume $sL$ at distance $sz$, under a rescaling by $sb$:
\BE
C(sz/sb;\tilde g',sL/sb) = (sb)^{\Delta} C(sz;\tilde g,sL),
\EE
where a possibly different coupling $\tilde g$ is used on the larger lattice. The two scaling formulas above imply the two ratios
\BE
R_b(g;L) = b^\Delta, \quad R_{sb}(\tilde g; sL) = (sb)^\Delta,
\EE
from which it follows that
\BE
R_{sb}(\tilde g; sL) = s^\Delta R_b(g;L).
\EE
Repeating the argument above on a volume $L'$ leads to
\BE
R_{sb}(\tilde{\bar g}; sL') = s^\Delta R_b(\bar g; L').
\EE
Letting $L' = sL$ and taking the difference of the previous two equations implies
\BE
R_{sb}(\tilde{\bar{g}}; s^2 L) - R_{sb}(\tilde g; sL) = s^\Delta \big[ R_b(\bar g;sL) - R_b(g;L) \big].
\EE
If the blocking steps above are performed sufficiently close to the IRFP, the effective couplings $g, \bar g$ are close to their fixed point values. Expanding each ratio about $g_*$, we find
\BE\label{eq:finite_vol_corr}
R_{sb}(g_*; s^2 L) - R_{sb}(g_*; sL) = s^\Delta \big[ R_b(g_*; sL) - R_b(g_*; L) \big] + O(g_i - g_*).
\EE
Eq. (\ref{eq:finite_vol_corr}) predicts the ratio $\Rop(g)$ on volume $s^2 L$ in terms of ratios on smaller volumes, plus a correction term $O(g - g_*)$.  We will absorb the latter term as a $g$ dependent correction and assume that the ratio on $s^2 L$ volumes approximates infinite volume.  Assuming that conformal symmetry is broken only by the finite number of spatial lattice points $L$, we expect finite volume corrections to depend only on the dimensionless ratio $b/L$, and thus on the flow time as $\sqrt{t}/L$.

\subsection{Simulation details}

We carried out a pilot study of SU(3) gauge theory with $N_f = 12$ degenerate fermions in the fundamental representation.  We used a set of gauge configurations that were originally generated for finite-size study of this system~ \cite{Cheng:2013xha} using a plaquette gauge action and  nHYP-smeared staggered fermions \cite{Hasenfratz:2001hp,Hasenfratz:2007rf}.  Further details on the lattice action can be found in Refs.~\cite{Cheng:2011ic,Cheng:2013eu,Cheng:2013bca,Cheng:2013xha}. We considered five values of the bare gauge coupling $\beta=4.0,5.0,5.5,5.75$ and $6.0$, analyzing 46 and 31 configurations on lattice volumes of $24^3\times 48$ and $32^3\times 64$, respectively.  The fermion mass was set to $m = 0.0025$, small enough that we expect the breaking of scale invariance to be dominated by the finite spatial extent $L$.

We considered only fermionic operators, and used the axial charge $A^4$  for our conserved operator $\mathcal{A}$. Since staggered fermions have a remnant U(1) symmetry, it is straightforward to construct a conserved axial charge operator with $Z_A=1$ \cite{Aoki:1999av}. We used on-site staggered operators for the pseudoscalar, vector, and nucleon, and a 1-link operator for the axial charge states. Our individual correlators were consistent with simple exponential decay, although we cannot rule out a functional dependence that includes a Yukawa-like power law correction \cite{Ishikawa:2013tua}.

We considered 10 flow time values between $1.0 \le t/a^2 \le 7.0$ (note that the flow range is $\sqrt{8t}$ in four dimensions.) The strong correlations in GF lead to very small statistical errors in the flow-time dependence.  

\subsection{Analysis}

\begin{figure}
\centering
\includegraphics[width=0.6\textwidth]{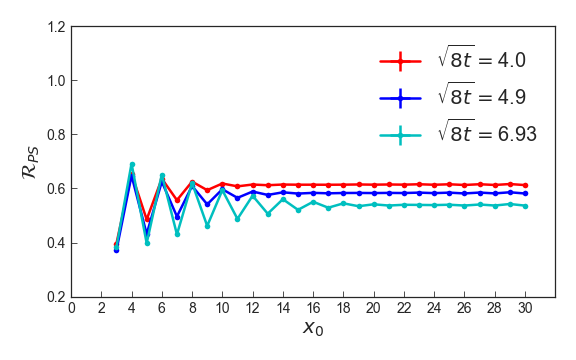}
\caption{\small{Dependence of the correlator ratio $R_P$ on source-sink separation $x_0$ and flow scale $\sqrt{8t}$.  For each value of $\sqrt{8t}$, a stable plateau in $R_P$ is seen for $x_0 \gtrsim 2\sqrt{8t}$.  The results shown here are on $32^3\times 64$ volumes at $\beta=5.75$. \label{fig:corr-flow}}}
\end{figure}

In the following, we work in lattice units.  The ratio given in eq. (\ref{eq:ratio_full}) should be independent of $x_0$ at large $x_0$, as long as the operator $\op$ has well defined quantum numbers. At distances comparable to the flow range, $x_0 \lesssim \sqrt{8t}$, the flowed operators overlap and the ratios could have non-trivial and non-universal structure. Since we  used staggered fermions where the action has oscillating phase factors, in the small $x_0$ region we observed significant oscillation, as shown in \fig{fig:corr-flow} for the $\gamma_5$ pseudoscalar operator that does not have a partner in the channel. The width of the oscillation is about $2\sqrt{8 t}$, after which a stable plateau develops.  The decrease in the value of the plateau as the flow time increases predicts the anomalous dimension of the pseudoscalar operator.

We worked directly with the ratio $\Rop(t)$ of eq. (\ref{eq:ratio_full}), and did not attempt to extrapolate the fermion field exponent $\eta$ (obtained from using $\symop$ in eq. (\ref{Option1})) to the infrared limit, as it showed much stronger finite-volume and bare coupling dependence than the full operator ratios. At fixed $t$ and $\beta$ we typically found $\eta \lesssim 0.1$.  

As a consistency check we considered the vector operator, but found large systematic effects due to oscillation; although we cannot quote a precise extrapolated value, we generally found the associated anomalous dimension consistent with zero as expected.

We predicted the anomalous dimension as a function of $t$ by comparing the ratios at consecutive $(t_1,t_2)$ flow time values
\begin{equation}
\gamma_\op (\beta, \bar{t}, L) = 
  \frac{ \rm{log}(\Rop_\op(t_1,\beta,L) / \Rop_\op(t_2,\beta,L)) } { \rm{log}(\sqrt{t_1}/\sqrt{t_2}) }
\end{equation}
where $\bar{t}=(t_1+t_2)/2$. The mass anomalous dimension is predicted by considering the pseudoscalar operator, recalling that $\gamma_m = -\gamma_{S}  = - \gamma_{PS} $. We estimated the finite volume corrections by eq. (\ref{eq:finite_vol_corr}), estimating $\gamma_m$ iteratively. We had numerical data on $24^3\times 48$ and $32^3\times 64$ volumes so $s=32/24$, and eq. (\ref{eq:finite_vol_corr}) increased the effective volume to $42.66$.

\begin{figure}
\centering
\begin{minipage}{.45\textwidth}
\includegraphics[width=1.1\textwidth]{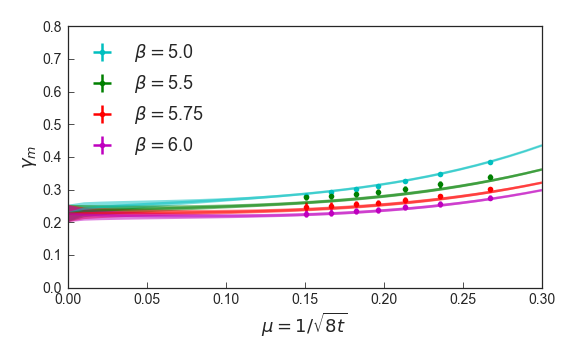}
\caption{\small{Extrapolation of the mass anomalous dimension $\gamma_m$ to the infrared limit, as described in the text. \label{fig:gamma-M}}}
\end{minipage}\hfill
\begin{minipage}{.45\textwidth}
\includegraphics[width=1.1\textwidth]{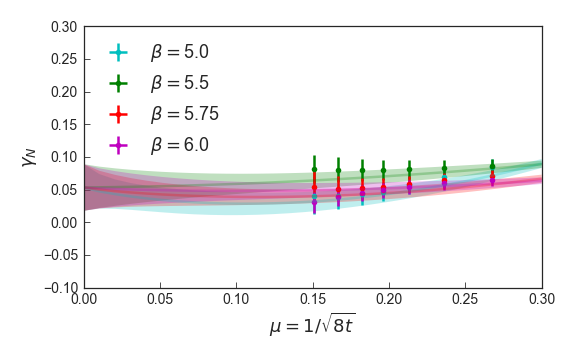}
\caption{\small{Extrapolation of the mass anomalous dimension $\gamma_N$ to the infrared limit, as described in the text. \label{fig:gamma-N}}}
\end{minipage}
\end{figure}

In \fig{fig:gamma-M} we show the infinite volume estimated $\gamma_m$ as a function of $\mu \equiv 1/\sqrt{ 8 \bar{t}}$. 
There is significant dependence on the bare gauge coupling $\beta$ and also on the flow time $t$, as expected in a slowly running system. We extrapolated to the $t \to \infty$ limit as
\begin{equation}
\gamma_m(\beta,t) = \gamma_0 + c_\beta t^{\alpha_1} + d_\beta t^{\alpha_2}
\label{eq:extrapolation}
\end{equation}
motivated by the expectation that the correction terms should be due to the slowly evolving irrelevant couplings,  associated with higher-dimensional operators that can mix with the operator of interest. Based on Refs.~\cite{Cheng:2013xha,Cheng:2013eu,Cheng:2013bca} we expect the FP to be closest to the $\beta=5.5-6.0$ range, so that the dependence on $\beta$ should be weakest in this range.

We performed a combined fit versus $\beta$ and $t$ using common $\gamma_0$, $\alpha_1$ and $\alpha_2$, but allowing $\beta$ dependent coefficients $c_\beta$ and $d_\beta$. The central fit, as shown in \fig{fig:gamma-M}, omits $\beta=4.0$ and discards the smallest and two largest $t$ values, predicting $\gamma_m=0.23$. The other exponents obtained were $\alpha_1 = -0.25(14)$ and $\alpha_2 = -2.37(29)$; these likely include some remaining finite-volume effects and thus should not correspond directly to irrelevant operator dimensions.

We varied the analysis by dropping small/large $t$ values, and also including or discarding $\beta=4.0$ and $\beta=6.0$ from the fit; from these variations we estimated a systematic error of $0.04$ on $\gamma_m$. As an additional cross-check on our finite volume correction procedure, we performed an alternative analysis in which a global fit to $\Rop_\op(t)$ were carried out assuming power-law dependence on the dimensionless ratio $\sqrt{8t}/L$. This gave a central value of 0.27. We conservatively took the difference in central values as an estimate of our finite-volume extrapolation systematic, giving the final prediction
\begin{equation}
\gamma_m = 0.23(6)
\end{equation}
combining the systematic errors in quadrature. 

A significant advantage of this technique is that more complicated composite operators can be dealt with in a straightforward way.  To demonstrate this, we considered the nucleon operator with our method. The nucleon showed more significant oscillations in the ratio $\Rop_N$, continuing into the plateau region; we accounted for the oscillations by averaging over adjacent pairs of $x_0$ values to obtain $\Rop_N$. The oscillations at large $x_0$ may be due to the coupling of the staggered nucleon operator to other wrong-parity states; numerically, the coupling is small in the ratio. We defined the nucleon anomalous dimension with an additional negative sign, $\gamma_N \equiv \Delta_N - d_N$, to match the convention of refs. \cite{Pica:2016rmv,Gracey:2018oym}. Repeating the full analysis as described yields \fig{fig:gamma-N} and predicts
\begin{equation}
\gamma_N = 0.05(5)
\end{equation}
where the finite-volume systematic error is estimated to be 0.03 and the remaining combined systematic and statistical error is 0.04.

\section{Afterword}

We have demonstrated that gradient flow (GF) can be used to extract estimates of scaling dimensions by testing the correlator ratio method in both relatively well-understood theories, $\phi^4_3, \; \phi^4_2$, and a relatively complicated theory, the 12-flavor SU(3) gauge-fermion system. The method is entirely nonperturbative. Furthermore, our method avoids the costly procedure of ensemble matching that is required in most MCRG studies \cite{Hasenfratz:2009kz}. Now, in the scalar theories we worked with a well-tuned system, and in the gauge-fermion case, we worked effectively at zero fermion mass. An important avenue for future work will be to consider the effects of deviations from criticality in the scalar system. Another question to address is whether the method may be extended to systems without IRFP's, such as QCD. We also expect the method should be fully applicable to other conformal theories than those we have considered, such as $\mathcal{N}=4$ super-Yang-Mills \cite{Schaich:2015daa}, and it has already been applied by Bergner, et al. to adjoint QCD with 1, 3/2, and 2 flavors \cite{Bergner:2019kub, Bergner:2019dim}. Lastly, we plan to apply the method to compute anomalous dimensions of electron bound states in 3-dimensional noncompact QED with $N_f$ flavors, a system with interesting and controversial infrared properties.	

\newpage

\chapter{Functional RG}

In this chapter we will introduce the \textit{functional} or \textit{exact renormalization group} program (FRG), the goal of which is to systematically define and solve \textit{functional} PDE's which describe the evolution of field-theoretic quantities of interest under continuous RG transformations. Examples of such quantities are the flowing effective action or the flowing 1PI generator. The RG equations we encountered in chapter 1 were differential equations in the couplings for the observables of a theory, be they renormalized or bare ones. In contrast, the functional RG equations are PDE's in the field variables, which track the evolution of the flowing action \textit{as a whole}. By expanding these functionals in powers of the field, one typically obtains an infinite hierarchy of coupled (non-functional) PDE's for the coefficient functions multiplying the fields. An important difference between FRG and the perturbative RG methods we encountered in chapter 1 is that FRG allows for nonperturbative approaches to the study of RG, which do not rely on Callan-Symanzik-type equations or perturbative renormalizability. But of course, in practice, the method must implement its own approximation strategies in order to solve the functional PDE's, which are often highly nonlinear. We will see that FRG can be thought of as a continuous, or smoothed-out, implementation of block-spin RG

To get started, we introduce a functional tensor notation which proves to be convenient when working with functional PDE's. Then, as a warm-up to the general program of FRG, we will describe in detail a version of \textit{smooth} high-mode elimination RG, to be compared with the typical textbook example of \textit{sharp} high mode elimination, in the framework of perturbation theory, and we will derive the perturbative Wilson-Fisher fixed point (for the second time in this thesis). Along the way, we will introduce some new generating functionals, and we will develop an understanding of what is meant by \textit{effective field theory}. This presentation will also mirror the one given later in the context of stochastic RG. We then will describe the general derivation of FRG equations, and the phenomena of RG fixed points in this formalism. We will compute the gaussian fixed point action following the analysis of \cite{Wilson:1973jj}. To close the chapter, we will briefly survey various applications of FRG which have emerged over the years. 

\section{Notation}

In this subsection a functional (index-free) tensor notation is introduced, to be used extensively in this chapter and the next. It is based on that of \cite{Kopietz:2010zz} and certain conventions from differential geometry. The notation often renders quite simple the expression of otherwise cumbersome functional equations by avoiding explicit position and momentum integrations and arguments, when it is appropriate to do so. We will develop the notation by recasting the generating functionals we defined in chapter 1 in a new form.

The first bit of notation was introduced back in chapter 1:
\BE
J \circ \varphi := \int \dd^d x J(x) \varphi(x).
\EE
Sometimes this is written alternatively as $(J,\varphi)$. We can think of this notation as expressing the contraction of two vectors $J$ and $\varphi$ whose components are $J(x)$ and $\varphi(x)$, since they have one ``index." On a lattice, the integral would be replaced by a summation. Now, with the notion of a functional vector comes the notion of functional tensor products, and thus functional tensors. For example, $J \otimes \phi$ is a rank-2 tensor, and the ``dot product'' above may be written in yet another way:
\BE
J \circ \varphi = \mrm{tr}[ J \otimes \varphi ].
\EE
$n$-point functions may be written in terms of functional tensor products. For example, the connected functions may be written as\footnote{For fermions one will need to be careful about ordering in this notation. See\cite{Kopietz:2010zz} for one approach.}
\BE
\langle \varphi \otimes \cdots \otimes \varphi \rangle^\con = \frac{\delta}{\delta J} \otimes \cdots \otimes \frac{\delta}{\delta J} \; W(J) \Big|_{J=0} = W^{(n)},
\EE
and one can say $W^{(n)}$ is rank-$n$. We can further introduce a multilinear notation for contracting tensors against vectors, e.g.\footnote{This notation is commonly used in differential geometry as it allows for the expression of tensorial quantities in a coordinate-free manner. See \cite{Nakahara:2003nw} for examples.}
\BE
W^{(n)} (J, \dots, J) := \prod_{i=1}^n \int \dd^d x_i J(x_i) \cdot W^{(n)}(x_1, \dots, x_n),
\EE
so that the expansion of the generator in $J$ is simply written as
\BE
W(J) = \sum_{n=0}^\infty \frac{1}{n!} W^{(n)} (J, \dots, J).
\EE
Whether the arguments of $W^{(n)}$ refer to position, momentum, or functional vectors should be clear from context.

 Another instance of multilinear notation is in the relationship between the 1PI vertices $\Gamma^{(n)}$ and the $W^{(n)}$. We noted in chapter 1 that $\Gamma^{(4)} = - W^{(4)} / W^{(2)} \cdots W^{(2)}$, that is, the 4-point vertex function is a full-propagator-amputated connected 4-point function (for a $\mathbb{Z}_2$-symmetric theory). Such a relation can be expressed nicely in multilinear notation as
\BE
W^{(4)}(\chi, \dots, \chi) = - \Gamma^{(4)}(W^{(2)} \chi, \dots, W^{(2)} \chi),
\EE
for an arbitrary functional vector $\chi$. We leave the determination of the corresponding relation for $W^{(6)}$ as an exercise for the curious reader. Lastly, we remark that rank-2 tensors are functional matrices, and we may speak of their inverses as usual. For example, the inverse propagator and the 2-point vertex are related by $[W^{(2)}]^{-1} \Gamma^{(2)} = \mathbb{I}$, where the functional identity has the Dirac delta as its components.  With all this new notation, we make our lives easier in many computations that come up in functional RG, as we will observe below.

\section{High-mode elimination} 

\textit{High-mode elimination} refers to the quintessential form of RG, namely, the systematic elimination of high-momentum (or short-distance) modes $\varphi(p)$ from a field theory, while preserving its low-momentum (or long-distance) structure. In a lattice system, this can be achieved nonperturbatively with block-spin RG or the GFRG of chapter 2. The former was a discrete transformation, while the latter was a continuous transformation. If one wants to analyze the structure of the effective action obtained by these transformations, it is usually simpler to work in the continuum and to use a continuous RG transformation. The framework of high-mode elimination enables one to study such effective actions, and it is perhaps most straight-forward to use perturbation theory, although the ultimate goal of functional RG is to apply nonperturbative methods to study the same problem. Now, there are many ways to carry out this program; we will adopt a method which combines elements of the analyses of Zinn-Justin \cite{ZinnJustin:2002ru}, Peskin and Schroeder \cite{Peskin:1995ev}, Igarashi et al. \cite{Igarashi:2009tj}, and Kopietz et al. \cite{Kopietz:2010zz}, and which utilizes perturbation theory. 

\subsection{The bare theory} The initial motivation of high-mode elimination RG is to mimic the discrete blocking transformations of real space RG in the context of continuum field theory. One begins with a bare theory at cutoff $\mLam_0$ and with partition function and action, respectively,
\BE
Z = \int \mathscr{D} \varphi \; \me^{-S_0(\varphi)}, \quad S_0(\varphi) = \frac{1}{2} ( \varphi, M_0 \varphi) + V_0(\varphi).
\EE
If one regularizes with a \textit{sharp} cutoff on the momentum integrals in $S_0$, then one is working with a pseudo-lattice model of infinite spatial volume and with non-lattice kinetic terms (e.g. $p^2$ vs $\sin^2 p a$). Furthermore, the sharpness of the cutoff will lead to nonanalyticity in position space, in general, and is therefore unfavorable for some purposes. There are other means of implementing a cutoff in the continuum, however. For a scalar field theory, a common choice is to define a \textit{smooth cutoff function} $K_0(p)$ such that $K_0(0) = 1$ and $K_0(p) \ll 1$ for $p \gg \mLam_0$, and implement it by a modified free propagator,\footnote{The quantization of field theories with smooth cutoffs is described in \cite{Namsrai:1986md} under the name of ``nonlocal QFT,'' and has been achieved with reasonable rigor. Here, we work in euclidean spacetime and concern ourselves with statistical field theories. It would be interesting, however, to attempt to apply the procedures below in quantum field theory, proper. It seems possible, for example, to obtain effective nonrelativistic field theories in such a manner.}
\BE
M_0 = \mDelt_0^{-1} := K_0^{-1} \mDelt^{-1}, \quad \mrm{where} \quad \mDelt^{-1}(p) = p^2 + m_0^2,
\EE
for example. By implementing a cutoff this way, the theory may be regularized in perturbation theory, since every internal line will correspond to a factor of
\BE
\mDelt_0 (p) = \frac{K_0(p)}{p^2 + m_0^2}.
\EE
A convenient choice we shall adopt for the rest of this work is Schwinger regularization,
\BE
K_0(p) = \me^{-p^2 / \mLam_0^2} = \me^{-p^2 a_0^2},
\EE
where the inverse cutoff $a_0 = \mLam_0^{-1}$ has been defined. Notice that such a $K_0$ is nothing but a momentum space heat kernel $f_t(p)$ at ``time'' $t = a_0^2$.

In general, the regulator should be chosen to preserve the symmetry of the fields appearing in the theory. In this case, Schwinger regularization suffices, but one must take greater care when dealing with gauge theories, or theories with constraints like spin models.

\subsection{The low-mode action} We want to define a low-mode, or \textit{effective} action, corresponding to the bare theory in such a way that the cutoff of the effective theory is lower, $\mLam < \mLam_0$, and the low-momentum (or long-distance) observables $\langle \mcal{O}(\varphi) \rangle_{S_0}$ are unaffected. We begin by deriving a peculiar functional identity:
\begin{align}
\mcal{N} \int \mathscr{D} \varphi & \exp\Big[-\frac{1}{2} ( \varphi, [A + B]^{-1} \varphi) - V(\varphi) \Big] \nonumber \\
& = \int \mathscr{D} \phi_1 \mathscr{D} \phi_2 \exp\Big[- \frac{1}{2} ( \phi_1,  A^{-1} \phi_1) - \frac{1}{2} ( \phi_2,  B^{-1} \phi_2) - V(\phi_1 + \phi_2) \Big],
\end{align}
where $A$ and $B$ are invertible matrices, and $\mcal{N}$ is a constant. To prove it, begin with the r.h.s., and redefine the $\phi_2$ field via $\varphi = \phi_1 + \phi_2$. The quadratic part of the action becomes
\BE
\frac{1}{2} ( \phi_1,  [A^{-1} + B^{-1}] \phi_1) - ( \phi_1,  B^{-1} \varphi) + \frac{1}{2} ( \varphi,  B^{-1} \varphi).
\EE
The integral over $\phi_1$ is then gaussian (being an instance of Wick's theorem) evaluating to
\BE
\det\Big[ \frac{2 \pi}{A^{-1}+B^{-1}} \Big] \exp \Big[ \frac{1}{2} \big( B^{-1} \varphi, [A^{-1}+B^{-1}]^{-1} B^{-1} \varphi \big) \Big],
\EE
which identifies the constant $\mcal{N}$. The matrix in the quadratic part of the $\varphi$ action is therefore
\BE
B^{-1} - (B^{-1})^\top [A^{-1}+B^{-1}]^{-1} B^{-1} = [A + B]^{-1}
\EE
(notice that the matrices need not commute, we just need $B^{-1}$ symmetric).

Next, we write the quadratic term $M_0$ in the bare action $S_0$ as
\BE
K^{-1}_0 \mDelt^{-1} = [K_0 - K_{\mLam} +  K_{\mLam}]^{-1}\mDelt^{-1} = [\mDelt(K_0 - K_{\mLam}) +  \mDelt K_{\mLam}]^{-1},
\EE
thereby identifying $A$ and $B$. We use the identity above and afterward relabel $\phi_1 \to \varphi$, $\phi_2 \to \phi$. The integral over $\varphi$ we denote by
\BE \label{lowmode_ampconn}
\me^{-A_\mLam(\phi)} = \int \mathscr{D} \varphi \exp\Big[-\frac{1}{2} ( \varphi, \mDelt_{\mLam\mLam_0}^{-1} \varphi) - V(\varphi + \phi) \Big],
\EE
where $\mDelt_{\mLam\mLam_0} = (K_0 - K_{\mLam})\mDelt$ is the so-called \textit{high-mode propagator}, because $\mDelt_{\mLam\mLam_0}(p) \to 0$ rapidly for $p \ll \mLam$, while $\mDelt_{\mLam\mLam_0}(p) \approx K_0(p) \mDelt(p)$ for $\mLam \ll p$. Complimentarily, we may regard $\mLam$ as a sliding infrared cutoff for the bare theory. We have therefore avoided the nonanalytic division of sharp high-mode elimination techniques $\phi = \phi_< + \phi_>$. The $\phi$ field in the argument of $V$ is sometimes called a \textit{background field} with respect to the $\varphi$ action. For our purposes, however, we will show that the functional $A_\mLam(\phi)$ is in fact the generating functional of free-propagator-amputated connected $n$-point functions.

To understand this, we derive another functional identity \cite{Kopietz:2010zz}. For arbitrary action $S(\chi) = \frac{1}{2} ( \chi, M \chi) + V(\chi)$, define
\BE
\me^{-A(\eta)} := \frac{1}{Z_0} \int \mathscr{D} \chi \exp \Big[-\frac{1}{2} ( \chi, M \chi) - V(\chi + \eta) \Big].
\EE
Let $\chi = \chi' - \eta$. By foiling out the quadratic term (for symmetric $M$), 
\BE
\frac{1}{2} ( \chi, M \chi) = \frac{1}{2} ( \chi', M \chi') - ( \chi', M \eta) + \frac{1}{2} ( \eta, M \eta),
\EE
we observe that the $\chi'$-integral is just the generating functional of disconnected functions $Z(J)|_{J = M\eta}$, and is therefore the exponential of the connected generator $W(J)|_{J=M\eta}$. Hence
\BE \label{MAW}
A(\eta) =  \frac{1}{2} ( \eta, M \eta) - W(M\eta).
\EE
Letting $M = \mDelt^{-1}$ and taking two functional derivatives, we find
\BE
A^{(2)} = \mDelt^{-1}  - \mDelt^{-1} W^{(2)} \mDelt^{-1}.
\EE
The second term is the connected 2-point function with external free propagators divided out, or ``amputated.'' Differentiating $n>2$ times, one finds a relation best expressed in tensor notation,
\BE
A^{(n)}(\eta, \dots , \eta) = - W^{(n)}(\mDelt^{-1}\eta, \dots , \mDelt^{-1}\eta).
\EE

We collect together all the information we have just uncovered, starting with the bare theory on the l.h.s., in the form
\BE
\int \mathscr{D} \varphi \; \me^{-S_0(\varphi)} = Z = \int \mathscr{D} \phi \; \me^{- S_\mLam(\phi) },
\EE
where the \textit{low-mode effective action} $S_\mLam$ has been defined:
\BE \label{lowmode_action}
S_\mLam(\phi) := \frac{1}{2} ( \phi, \mDelt_{\mLam}^{-1} \phi) + A_\mLam(\phi)
\EE
Thus, we have an exact functional expression for the low-mode effective action. Because it is given in terms of the generator $A_\mLam$, we also know how to systematically compute it in perturbation theory: compute the amputated-connected $n$-point functions determined by a theory with (high-mode) action
\BE
S_{\mLam \mLam_0} (\varphi) = \frac{1}{2} ( \varphi, \mDelt_{\mLam\mLam_0}^{-1} \varphi) + V_0(\varphi).
\EE
Furthermore, every transformation we performed was \textit{passive}, and therefore the observables $\langle \mcal{O}(\varphi) \rangle_{S_0}$ are unchanged, except that $\mcal{O}(\varphi) \to \mcal{O}'(\phi)$ are not generally of the same functional form. If $\mcal{O}(\varphi)$ is some polynomial, then $\mcal{O}'(\phi)$ will generally be a different polynomial. This is an instance of what is referred to as \textit{operator mixing}, a phenomenon we have seen already in the context of scaling operators. Lastly, notice that for any nontrivial theory (i.e. interacting, $V_0 \neq 0$), $A_\mLam(\phi)$ contains nonvanishing ``coefficients'' $A^{(n)} \neq 0 \; \forall n$, each of which has the same symmetries as the bare action. It follows that the effective action contains every possible term consistent with the symmetry of the bare theory, indeed an infinite number of terms. This generic feature is an integral aspect of the phenomenological approach to effective field theory, which we will discuss below.

\subsection{Effective couplings}

We now consider as an example the case of $\phi^4_d$ theory with bare action in momentum space given by
\BE
S_0(\varphi) = \frac{1}{2} \int_p K_0^{-1}(p)[p^2 + m_0^2] \varphi(p) \varphi(-p) + \frac{\lambda_0}{4!} \int_{\bo p} \tilde \delta(p_\mrm{tot}) \varphi(p_1) \varphi(p_2) \varphi(p_3) \varphi(p_4).
\EE
Here we use for convenience the notations
\BE
\int_p = \int_{\mathbb{R}^d} \frac{\dd^d p}{(2\pi)^d}, \quad \bo p = p_1 \oplus \cdots \oplus p_n, \quad p_\mrm{tot} = \sum_{i=1}^n p_i, \quad \mrm{and} \quad \tilde \delta(p) = (2\pi)^d \delta(p).
\EE
The value of $n$ should be clear from context; in the quartic term above it is 4, for example. We write the functional expansion of $S_\mLam(\phi)$ as usual,
\BE
S_\mLam(\phi) = \sum_{n=0}^\infty \frac{1}{n!} S^{(n)}_\mLam(\phi, \dots , \phi).
\EE
The effective couplings at scale $\mLam$ are determined by expanding the functions $S^{(n)}_\mLam(\bo p)$ about $\bo p = \bo 0$, and writing $P_i$, $i=1,...,dn$ for the $i^\mrm{th}$ component of the direct sum $\bo p = p_1 \oplus \cdots \oplus p_n$:	
\BE
S^{(n)}_\mLam(\bo p) = \sum_{m=0}^\infty \frac{1}{m!} g_{\mLam, i_1\cdots i_m}^{(n,m)} P_{i_1} \cdots P_{i_m}, \quad g_{\mLam, i_1\cdots i_m}^{(n,m)} = \frac{\del^m S^{(n)}_\mLam(\bo p)}{\del P_{i_1} \cdots \del P_{i_m}}\Big|_{\bo p = 0}.
\EE
Since the bare theory is rotationally invariant and $\mathbb{Z}_2$-symmetric, the only non-vanishing couplings have $n, \;m$ even. We will only focus on a few of the most important couplings, out of simplicity. 

The quadratic part of the low-mode action is given by the sum
\BE
S^{(2)}_\mLam = \mDelt^{-1}_\mLam + A^{(2)}_\mLam.
\EE
Now, the high mode action has a free propagator $\mDelt_{\mLam_0\mLam}$. In perturbation theory in the bare coupling $\lambda_0$, the amputated 2-point function is then, to first order,
\BE
A^{(2)}_\mLam(p) = \mDelt^{-1}_{\mLam_0\mLam}(p) - \mDelt^{-1}_{\mLam_0\mLam}(p) W^{(2)}_{\mLam_0\mLam}(p) \mDelt^{-1}_{\mLam_0\mLam}(p) = \frac{\lambda_0}{2} I^d_{\mLam_0\mLam}(m_0^2) + O(\lambda_0^2),
\EE
where the snail loop is
\BE
I^d_{\mLam_0\mLam}(m_0^2) = \int_\ell \mDelt_{\mLam_0\mLam}(\ell) = \int_\ell \frac{\delta K_{\mLam_0 \mLam} (\ell)}{\ell^2 + m_0^2}, \quad \delta K_{\mLam_0 \mLam} (\ell) = K_0(\ell) - K_\mLam(\ell).
\EE
The leading behavior of $K_\mLam (p) S^{(2)}_\mLam(p)$ in the momenta is just $p^2 + O(\lambda_0^2)$, and therefore the kinetic term coefficient has no 1-loop contribution. Denoting that coefficient by $c_\mLam$, we have $c_\mLam = 1 + O(\lambda_0^2)$. The momentum-independent part of the function $S^{(2)}_\mLam(p)$ defines the effective mass term in $S_\mLam$:
\BE \label{eff_mass}
m_\mLam^2 = g_\mLam^{(2,0)} = m_0^2 + \frac{\lambda_0}{2} I^d_{\mLam_0\mLam}(m_0^2) + O(\lambda_0^2).
\EE
The next-most important coupling is the quartic coupling, which comes from the amputated 4-point function at zero external momenta. A standard perturbative calculation gives
\BE \label{eff_coupling}
\lambda_\mLam = g_\mLam^{(4,0)} = \lambda_0 - \frac{\lambda_0^2}{2} \sum_{i = 1}^3 C^d_{\mLam_0 \mLam}( p_{\sigma_i}, m_0)|_{\bo p = 0} + O(\lambda_0^3),
\EE
where
\BE
C^d_{\mLam_0 \mLam}(p, m_0) = \int_\ell \frac{\delta K_{\mLam_0 \mLam} (\ell)}{\ell^2 + m_0^2} \frac{\delta K_{\mLam_0 \mLam} (\ell + p)}{(\ell + p)^2 + m_0^2},
\EE
and where the $p_{\sigma_i}$ are the sums $p_k + p_j$ for each distinct pairing $(k,j)$ of external momenta, without overcounting by momentum conservation $p_1 + p_2 + p_3 + p_4 = 0$. Other couplings, such as the sextic coupling $g_6$, may also be computed; if there is no such coupling in the bare action, then the low-mode couplings will be a function only of the bare mass and quartic couplings.

Momentum-dependent vertices may also be computed simply by expanding the $A_\mLam^{(n)}(\bo p)$ in $\bo p$ near zero, which may be done for nonzero bare mass. To get a feel for what these ``higher'' effective couplings look like, we consider the example of $g_6$. One finds
\BE
g^{(6,0)}_\mLam = A^{(6)}_\mLam(\bo p)|_{\bo p = 0} = \lambda_0^2 \sum_\sigma \mDelt_{\mLam\mLam_0}(p_\sigma) - \frac{1}{2} \lambda_0^3 \sum_{\sigma'}D^d_{\mLam\mLam_0}(p_{\sigma'}, m_0) \Big|_{\bo p=0},
\EE
where the 1-loop integral is
\BE
D^d_{\mLam\mLam_0}(p, k, m_0) = \int_\ell \frac{\delta K_{\mLam_0 \mLam} (\ell)}{\ell^2 + m_0^2} \frac{\delta K_{\mLam_0 \mLam} (\ell + p)}{(\ell + p)^2 + m_0^2} \frac{\delta K_{\mLam_0 \mLam} (\ell + p + k)}{(\ell + p + k)^2 + m_0^2}.
\EE
Since the high-mode propagator vanishes as $p\to 0$, the leading behavior of the effective 6-point coupling is determined by the 1-loop term, and power-counting implies it is of order $O(\lambda_0^3 / \mLam^3)$.

We remark that in the low-momentum sector $p \ll \mLam$ of the action, diagrams that are not 1PI are suppressed by terms of order $p^2/\mLam^2$, which explains why they do not contribute to the effective (zero-momentum) couplings above. This is because every non-1PI diagram contains internal propagators with no loop momenta, namely, the ones which connect 1PI pieces. Such propagators are proportional to $\delta K_{\mLam_0\mLam}(p) \approx p^2 / \mLam^2$, which means that they are highly suppressed. In sum,
\BE
A^{(n)}_{\mLam}(\bo p) \xrightarrow[\bo p \to 0]{} \Gamma^{(n)}_\mLam (\bo p) \Big|_{\bo p = 0}.
\EE
We conclude that the low-momentum sector of the effective action coincides with that of the quantum effective action, which in a sense gives further justification to the name.

\subsection{RG} A renormalization group analysis concerns itself with the scale dependence of the theory and, if there be any, the fixed points in the space of possible actions. Now, we like to compare actions by comparing their coefficients. But the presence of a cutoff function in the action makes the comparison of coefficients at different scales ambiguous.
This is most clearly understood by reverting back to the sharp cutoff approach: comparing two actions $S_\mLam, \; S_{\mLam'}$ would involve comparing integrals with different limits, and worse, as $\mLam \to 0$, it would seem that all momentum integrals vanish. But, for any $S_\mLam$, by rescaling $p = \bar p \mLam$, we can normalize the integration limits to $\bar p \in [0,1]$. We can then compare two actions at different scales (almost) unambiguously. In our smooth cutoff approach, $p = \bar p \mLam$ corresponds to conventionally using $K(\bar p) = \me^{-\bar p^2}$ as the cutoff function.


The second ambiguity relates to the field normalization. The kinetic term in the effective action has coefficient $c_\mLam$, which is not equal to the bare coefficient $c_0$. For any field theory, the normalization of the fields is to some extent arbitrary, however. This freedom is reflected in the ability to always normalize one coupling in the action to 1. The convention in field theory is to normalize the kinetic term. Now, after performing the momentum redefinition $p = \bar p \mLam$ described above, the kinetic and mass terms in $S_\mLam$ have the form
\BE
\frac{1}{2} \mLam^d \int_{\bar p} K(\bar p)^{-1}[c_\mLam \mLam^2 \bar p^2 + m_\mLam^2] \phi(\bar p \mLam) \phi(-\bar p \mLam).
\EE
One then defines the dimensionless \textit{rescaled} effective field $\mPhi$ by
\BE
\mPhi(\bar p) := c_\mLam^{1/2}\mLam^{-d_\phi} \phi(\bar p \mLam),
\EE
after which the quadratic terms take the form
\BE
\frac{1}{2} \int_{\bar p} K(\bar p)^{-1}[\bar p^2 + \mLam^{-2} m_\mLam^2 / c_\mLam] \mPhi(\bar p) \mPhi(-\bar p).
\EE
This suggests that the natural effective mass coupling to consider when performing an RG transformation is the dimensionless rescaled parameter
\BE
r_\mLam := m_\mLam^2 \mLam^{-2} / c_\mLam = b^2_\mLam \hat m^2_\mLam / c_\mLam,
\EE
where the scale change parameter $b_\mLam := \mLam_0 / \mLam$ has been introduced, and hats denote removal of scale with $\mLam_0$, as we did in lattice theory. Similarly, the momentum and field redefinition lead to a natural redefinition of the effective quartic coupling,
\BE
u_\mLam :=  \mLam^{d-4}  \lambda_\mLam c_\mLam^{-2}= b_\mLam^{4-d} \hat \lambda_\mLam c_\mLam^{-2}.
\EE
The resulting terms in the effective action are then
\BE
S_\mLam(\phi) \supset \frac{1}{2} \int_{\bar p} K(\bar p)^{-1}[\bar p^2 + r_\mLam] \mPhi(\bar p) \mPhi(-\bar p) + \frac{u_\mLam}{4!} \int_{\bar{\bo p}} \delta(\bar{p}_\mrm{tot}) \mPhi(\bar p_1) \mPhi(\bar p_2) \mPhi(\bar p_3) \mPhi(\bar p_4).
\EE
It is then unambiguous to compare this effective action at different scales $\mLam$.

To study the possible fixed point behavior of this RG transformation, we derive ODE's for the flowing couplings $r_\mLam, u_\mLam$ perturbatively. One differentiates the effective couplings, eqs. (\ref{eff_mass}, \ref{eff_coupling}), with respect to $\mLam$, replaces bare couplings by effective couplings using their perturbative relationships, and looks for stationary points of the resulting system of ODE's.
To begin, note that
\begin{align}
\mLam \frac{\dd m^2_ \mLam}{\dd \mLam} & = \frac{\lambda_0}{2} \mLam \frac{\dd}{\dd \mLam} I^d_{\mLam_0\mLam}(m_0^2) + O(\lambda_0^2),  \nonumber \\
\mLam \frac{\dd \lambda_ \mLam}{\dd \mLam} & = -\frac{3\lambda_0^2}{2} \mLam \frac{\dd}{\dd \mLam} C^d_{\mLam_0\mLam}(0, m_0^2) + O(\lambda_0^3).
\end{align}
Closed-form expressions for these integrals exist, but they are algebraically cumbersome. It is simplest to first compute the derivatives and then perform asymptotic expansions for $\mLam_0$ large and $\hat m_0^2$ small.\footnote{Closed-form expressions for both integrals exist in fact for any $ 2 < d \leq 4$. See \cite{Igarashi:2009tj} for examples in 4 dimensions. We also note that integrals that come up in smooth high-mode elimination allow one to understand the relationship between dimensional regularization and cutoff field theory. See \cite{Kleppe:1991ru} for some discussion of this.} One finds, keeping a few subleading terms,
\begin{align}
\mLam \frac{\dd}{\dd \mLam} I^3_{\mLam_0\mLam}(m_0^2) & = \Omega_3 \Big[ - \frac{\sqrt{\pi}}{2} \mLam + \sqrt{\pi} m_0^2 / \mLam + O(m_0^3 / \mLam^2) \Big] \nonumber \\
& = \Omega_3 \mLam_0 \Big[ - \frac{\sqrt{\pi}}{2} b_\mLam^{-1} + \sqrt{\pi} \hat m_0^2 b_\mLam + O(\hat m_0^3 b_\mLam^2) \Big],
\end{align}
and
\begin{align}
\mLam \frac{\dd}{\dd \mLam} C^3_{\mLam_0\mLam}(0, m_0^2) & = \Omega_3 \Big[ (\sqrt{2} -2) \sqrt{\pi} \Big( \frac{1}{\mLam} + \frac{8 m_0^2}{\mLam^3} \Big) + \frac{\sqrt{\pi} \mLam}{\mLam_0^2 } + O(m_0^3/ \mLam^4, m_0^2 / \mLam \mLam_0^2) \Big] \nonumber \\
& = \Omega_3 \mLam_0^{-1} \Big[ (\sqrt{2} -2) \sqrt{\pi} \Big( b_\mLam + 8 \hat m_0^2 b_\mLam^3 \Big) + \sqrt{\pi}b_\mLam^{-1} + O(\hat m_0^3b_\mLam^4, \hat m_0^2 b_\mLam) \Big].
\end{align}
The next step is to replace the bare couplings by their perturbation series in $\lambda_\mLam, \; m_\mLam^2$, by inverting eqs. (\ref{eff_mass}, \ref{eff_coupling}), and then write $\lambda_\mLam, \; m_\mLam^2$ in terms of $u_\mLam, \; r_\mLam$. Noting that $\mLam \del_\mLam = - b \del_b$, we then compute the flow of the rescaled couplings,
\begin{align}
b \frac{\dd r_\mLam}{\dd b} & = 2 r_\mLam + b^2 \mLam \frac{\dd \hat m^2_ \mLam}{\dd \mLam}, \nonumber  \\
b \frac{\dd u_\mLam}{\dd b} & = u_\mLam + b \mLam \frac{\dd \hat \lambda_ \mLam}{\dd \mLam}.
\end{align}
As $b \to \infty$, these equations asymptotically approach the system of ODE's
\begin{align}
b \frac{\dd r}{\dd b} & = 2 r + \alpha_1 u + O(r u, u^2), \nonumber \\
b \frac{\dd u}{\dd b} & = u - \alpha_2 u^2 + O(r u^2, u^3),
\end{align}
where the coefficients $\alpha_i > 0$ are
\BE
\alpha_1 = \frac{1}{8 \pi^{3/2}}, \quad \alpha_2 = \frac{3(2-\sqrt{2})}{4 \pi^{3/2}}.
\EE
We immediately observe the existence of two fixed point solutions at this order in perturbation theory. The first is the gaussian fixed point $r_* = u_* = 0$, while the second is the famous \textit{Wilson-Fisher fixed point} (WFFP)\footnote{The WFFP is usually found within the epsilon expansion, but here we have worked explicitly in $d=3$.}
\BE
u_* \approx 1/ \alpha_2, \quad r_* \approx - \alpha_1 u_* / 2.
\EE
The sign of the beta function $\beta(u)$ is opposite that of the discussion in the lattice theory chapter because increasing $b$ corresponds to decreasing $\mu$.

It is worthwhile to pause for a moment and summarize what has happened. We began with the bare theory $S_0$ involving field modes up to $\mLam_0$. We then performed a particular change of variables which yielded an effective theory $S_\mLam$ involving field modes up to $\mLam < \mLam_0$. A passive change of momenta and field variables, determined by removing canonical mass dimensions with the scale $\mLam$, together with a further rescaling for the field in order to normalize the kinetic term, led to a dimensionless, rescaled effective theory. The flow of this theory as $\mLam$ decreased was analyzed by studying the leading effective couplings, and it was found that as $\mLam \to 0$, the system of ODE's for these couplings had a fixed point. This is just the kind of IRFP discussed in chapter 1 in the context of block-spin RG. We note that, because the transformations involved were passive, the rescaled and unrescaled effective actions are numerically equal, so a fixed point of one is a fixed point of the other. The correlations of the rescaled theory, for example $\langle \mPhi \cdots \mPhi \rangle_\mLam$, approach the correlations of the fixed point theory as $\mLam \to 0$, whereas the correlations of the unrescaled theory $\langle \phi \cdots \phi \rangle_\mLam$ asymptotically approach a scaling determined by $c_\mLam^{1/2} \mLam^{d_\phi}$, the wave function renormalization. We emphasize the distinction between rescaled and unrescaled variables because it is quite important in lattice simulations, as we will see in chapter 4 (and as we already saw in chapter 2 with GFRG).

\subsection{Effective field theory} The fundamental assumption of effective field theory (EFT) is that \textit{all theories we currently work with and will continue to work with, up to the possible exception of a quantum theory of gravity, have a limited range of applicability, in terms of distance or energy scales}. This is certainly true of every real-world theory that has been tested to date. This means that every theory we formulate should contain within it a parameter which functions as a (possibly unknown) \textit{cutoff} above which the theory becomes invalid. The bare theory we considered above was not intended to describe any physics above $\mLam_0$, for example. Now, for condensed matter systems such as ferromagnets, the cutoff not only sets the cutoff scale, but has a literal manifestation: the atomic spacing. In quantum field theory, on the other hand, all or nearly all theories under consideration take place in a continuum. But these theories possess a cutoff, nevertheless, and therefore must correspond to some kind of smooth cutoff, qualitatively similar to what was used above. In practice, examples of such cutoffs are the masses of particularly heavy particles, or symmetry-breaking scales like that of chiral perturbation theory.\footnote{Another interesting example of how a natural smooth cutoff can arise is in the interpretation of nonlocal quantization given in \cite{Namsrai:1986md}, where the cutoff function $K_0$ arises from an underlying stochastic spacetime.} The general question of how to formulate such theories based on the data we have at low energies therefore becomes of central interest in particle physics.

We noted above that the effective action $S_\mLam(\phi)$ contains all terms consistent with the symmetries of the bare theory. The action therefore contained all possible ``nonrenormalizable'' (NR) interactions once the scale was lowered even slightly from $\mLam_0$. The action $S_\mLam$ at scale $\mLam$ is an effective theory; it describes the same physics as $S_{0}$, but with lowered cutoff $\mLam$. Now, in the real world, we might not know what $S_0$ is, according to the fundamental assumption stated above. The best we can do at first is to write down some effective action with cutoff $\mLam$ as its regularization, and go perform scattering experiments, say, at energy $p$; some of these measurements are used to set the renormalized couplings. But we know that this action typically contains all sorts of interactions, including the NR ones, so we have to also set those by experiment too. The reason our effective theory remains predictive is the fact that, to any given order in $p^2/\mLam^2$, only a \textit{finite} number of NR interactions must be set (see \cite{Lepage:1989hf} for a detailed explanation). And once set, we can produce predictions for all \textit{other} processes to that order. As we approach energies closer to $\mLam$, the number of NR interactions we must set will proliferate, and the theory will break down. In many cases, therefore, we can \textit{estimate} the breakdown scale by measuring the strength of the NR interactions.


 In some cases we know the bare theory, in others we do not. We know that QCD is the high-energy theory (or ``UV completion'') whose low-energy interactions involve mesons and nucleons.  We have to use the methods of EFT to describe the low-energy processes of QCD, however, because perturbation theory breaks down at low energies as the gauge coupling becomes strong.\footnote{This ``low-energy'' scale corresponds to $\mLam$ in the discussion above; it arises from the dynamics of QCD. In this context, $\mLam_0$ refers instead to whatever the cutoff of QCD might be.} This EFT is called \textit{chiral perturbation theory}. It is widely expected that the Standard Model itself is an effective theory, and experimental measurement of the NR-interactions allows for predictions of its breakdown scale. Lastly, we remark that even quantum gravity can be treated in an effective manner, because whatever its correct description might be at very high energies (the \textit{Planck scale} $M_\mrm{pl}$), it is still sensible to use the (nonrenormalizable) theory obtained by direct quantization of General Relativity, expanded about a flat spacetime metric, for processes at scales far, far below $M_\mrm{pl}$. For an introduction to EFT in QED and the Standard Model, we refer the reader to \cite{Lepage:1989hf}, to \cite{Petrov:2016azi} for a more systematic account (including gravity, chiral perturbation theory, and non-relativistic EFT's), and for a rigorous exposition of the existence of effective scalar field theory in 4 dimensions, to \cite{Ball:1993zy}.

\section{Exact RG equations}

We now turn to the derivation of the so-called \textit{exact} RG equations which are studied in the enterprise of functional RG (FRG). We will discuss the nature of fixed points of such transformations, finding a close parallel with the block-spin analysis of chapter 1, and a few examples will be given along the way. Before plunging into these derivations, we first describe some of the early history of FRG.

\begin{figure}
\centering
\includegraphics[width=0.8\textwidth]{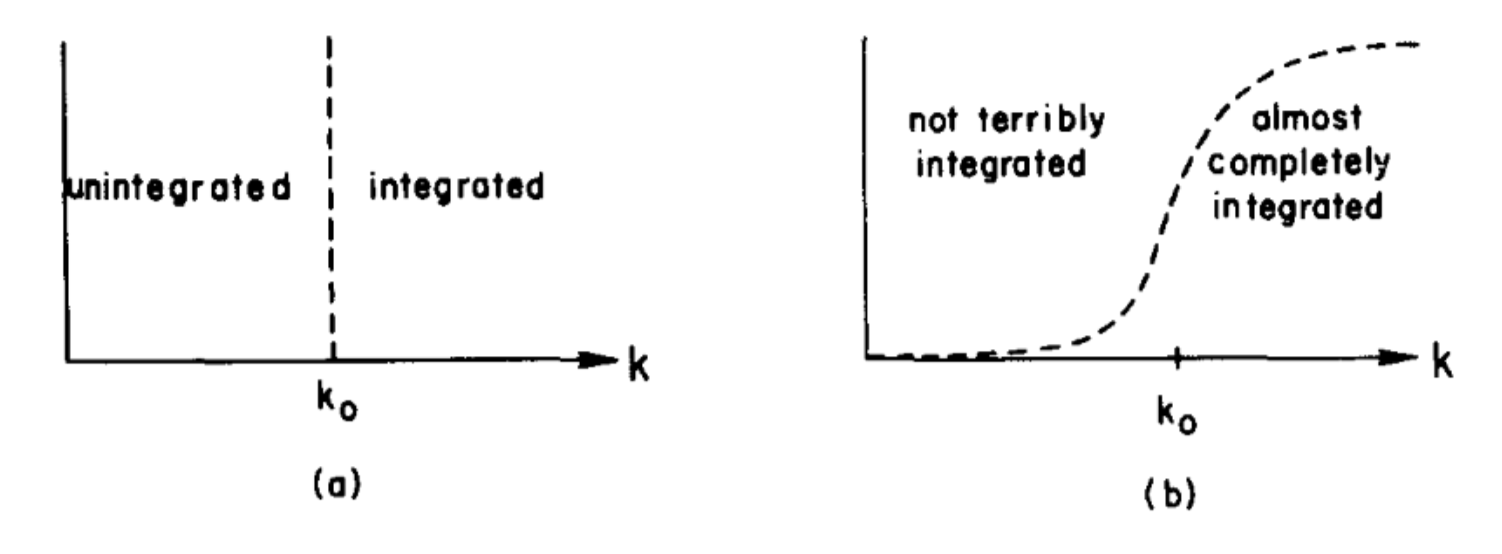}
\caption{\small{Wilson and Kogut's qualitative depiction of the difference between sharp and smooth high-mode elimination under exact RG transformations. Adapted from \cite{Wilson:1973jj}. \label{fig:WK_plot}}}
\end{figure}

On 2 June 1971, Wilson's paper \cite{Wilson:1971dh} was received, in which his approximate RG recursion formula for blocking transformations was introduced. In October, his paper with Fisher \cite{Wilson:1971dc} on the epsilon expansion was submitted. In this paper they described the recursion formula in $4-\eps$ dimensions.
By 27 October of 1972, Wegner and Houghton \cite{Wegner:1972ih} derived a differential equation for the ``blocked'' Hamiltonian using a sharp cutoff, which implied continuous versions of the recursion formulas of Wilson and Fisher. The paper was published in July 1973, the same month that Wilson and Kogut's review \cite{Wilson:1973jj} of the epsilon-expansion was received. Deep in their grand review, on the $74^\mrm{th}$ page, was presented a differential equation for the blocked Hamiltonian, which, like Wegner's, involved a sharp cutoff for the bare theory, but unlike Wegner's, utilized a \textit{smooth suppression} of high modes, rather than a sharp elimination, in a manner similar to what we saw under smooth high-mode elimination above. They distinguished low modes from high modes by referring to the former as ``not terribly integrated,'' and the latter as ``almost completely integrated''; see figure \ref{fig:WK_plot}.

In the review, Wilson notes that he presented the exact RG equations at a conference at Irvine in 1970 \cite{Wilson:1973jj}. It is possible to imagine Wilson having put them away while pursuing the more tractable approach provided by the epsilon-expansion, as he goes on to state that ``these equations are very complicated so they will not be discussed in great detail.'' Nevertheless, the works of Wilson, Kogut, Wegner, and Houghton constituted the first instances of functional RG equations, which differ in kind from the recursion formulas and epsilon-expansion by tracking the evolution of the effective action, \textit{as a whole}, rather than a small number of couplings. Thus we see that FRG was developed essentially simultaneously with modern RG theory. In the ensuing decade, the subject was advanced somewhat slowly, as the Callan-Symanzik approach combined with epsilon expansion (and/or dimensional regularization) proved its worth, having been used to demonstrate the asymptotic freedom of Yang-Mills theories and QCD \cite{Gross:1973id, Politzer:1973fx,tHooft:1998qmr}. 

Although a small number of researchers continued to work on FRG during this time, it is fair to say that FRG remained somewhat stagnant. Its revival did not come until the late 80's and early 90's after the works of Polchinski \cite{Polchinski:1983gv}, Hasenfratz and Hasenfratz \cite{Hasenfratz:1985dm}, and Wetterich \cite{Ringwald:1989dz,Wetterich:1989xg} recalled the work of the early days and proposed new methods and applications of the FRG.

\subsection{Wegner's approach} The following derivation is based on that of Wegner \cite{Wegner:1974} and that of Rosten \cite{Rosten:2010vm}. We have seen that, because the RG transformations are passive, the partition functions ``at different scales'' are equal. Let us imagine we have transformed down to scale $\mLam$, and then perform a small 
transformation such that the new scale is $\mLam - \delta \mLam$. Labeling the partition functions with respect to the  effective cutoffs in their actions, we can characterize the invariance condition as
\BE
\frac{\dd Z_\mLam}{\dd \mLam} = \lim_{\delta \mLam \to 0} \frac{1}{\delta \mLam} \big( Z_\mLam - Z_{\mLam - \delta \mLam} \Big) = 0,
\EE
The second term in the limit definition is
\BE
Z_{\mLam - \delta \mLam} = \int \mathscr{D} \phi \;  \me^{-S_{\mLam - \delta \mLam}(\phi)}.
\EE
We imagine that the effective action at $\mLam-\delta\mLam$ was obtained by a transformation of field variables $\phi'$, a continuous analog of a blocking transformation,
\BE \label{cov}
\phi = \phi' - \delta \tau \Psi_\mLam(\phi') + O(\delta \tau^2),
\EE
where $\phi'$ was the field at scale $\mLam$, and $\delta\tau = - \delta \mLam / \mLam = \delta \mLam^{-1}$ (obtained from $\tau = \ln \mLam_0 / \mLam$), so that decreasing $\mLam$ corresponds to increasing $\tau$. For example, the new fields might have been obtained from the old ones by a local smoothing transformation which damps high modes,\footnote{We will see that this transformation is in fact not sufficient as an RG transformation. The form of the transformation corresponding to the high-mode elimination RG we considered earlier is given in eq. (\ref{highmode_psi}).}
\BE
\phi(p) = \phi'(p) - \delta \tau p^2 \phi'(p) + O(\delta \tau^2).
\EE
The functional vector $\Psi_\mLam$ is sometimes called the ``flow-vector.'' Introducing a blackboard bold gradient symbol for the functional derivative, the notation\footnote{The \LaTeX \; command for this symbol is available upon request.}
\BE
\Bnab_\phi = \fdelphi
\EE
will be used in what follows, often omitting the $\phi$ subscript when it is not too confusing. The expansion of the action in $\delta \tau$ is then
\BE
S_{\mLam - \delta \mLam}(\phi' - \delta \tau \Psi_\mLam(\phi')) = S_\mLam(\phi') +  \delta \tau \mLam \del_\mLam S_\mLam (\phi') - \delta \tau \Psi_\mLam(\phi') \circ \Bnab S_\mLam (\phi') +O(\delta \tau^2 ).	
\EE
In general, the measure will change as well:
\BE
\mathscr{D} \phi = \mathscr{D} \phi' \det\Big[\mathbb{I} - \delta \tau \Bnab \otimes \Psi_\mLam(\phi') \Big] = \mathscr{D} \phi' \Big[ 1 - \delta \tau \; \Bnab \circ \Psi_\mLam(\phi') + O(\delta \tau^2) \Big].
\EE
From the invariance of the partition function above, and because the Boltzmann factor is positive definite, we equate the integrand to zero. This yields the Wegner flow equation, dropping primes,
\BE
- \mLam \del_\mLam S_\mLam (\phi) = - \Psi_\mLam(\phi) \circ \Bnab S_\mLam (\phi) + \Bnab \circ \Psi_\mLam(\phi).
\EE
Wegner suggested that the flow equation \textit{must} be nonlinear in $S_\mLam$ in order to constitute an RG transformation, meaning that $\Psi_\mLam$ must depend on $S_\mLam$. We will attempt to give an explanation for  \textit{why} in what follows. We  then choose the form (which can be called ``diffusive'' for reasons discussed later on)
\BE \label{diffusive_ERGE}
\Psi_\mLam (\phi) = \frac{1}{2} C_\mLam \Bnab S_\mLam(\phi) - B_\mLam(\phi),
\EE
where $C_\mLam$ is some appropriately chosen (positive) cutoff function, which is a functional matrix, and $B_\mLam$ is a functional vector. In Rosten's analysis, $B_\mLam = C_\mLam \Bnab \hat S_\mLam$ where $\hat S_\mLam$ is the ``seed action.'' The resulting flow equation is (abusing matrix notation slightly by factoring out $C_\mLam$)
\BE \label{ERGES}
- \mLam \del_\mLam S_\mLam = \frac{1}{2} C_\mLam \Big[ - \Bnab S_\mLam \circ \Bnab S_\mLam + \Bnab^2 S_\mLam \Big] + B_\mLam \circ \Bnab S_\mLam - \Bnab \circ B_\mLam.
\EE
This equation will generally be called the ``exact RG equation'' (ERGE). Most of the ERGE's for effective actions considered in the literature so far have been of this form. The most common choices for scalar theories are $B_\mLam = 0$ or $B_\mLam \propto \phi$. When $B_\mLam$ is a polynomial of higher order in $\phi$, the RG transformation is called \textit{nonlinear}. Wilson and Bell considered an example of such a flow in 1974, but very few others have been considered analytically (spin models with the constraint $|\phi| = 1$, however, induce a nonlinearity, which is implemented numerically by projections). We shall look at nonlinear RG's in chapter 4.

We can understand intuitively what the ERGE is doing by considering the Euler approximation of the PDE,
\BE
S_{\mLam - \delta \mLam} = S_\mLam + \frac{\delta \mLam}{\mLam} \Bigg( \frac{1}{2} C_\mLam \Big[ - \Bnab S_\mLam \circ \Bnab S_\mLam + \Bnab^2 S_\mLam \Big] + B_\mLam \circ \Bnab S_\mLam - \Bnab \circ B_\mLam \Bigg).
\EE
If the action $S_\mLam$ is more than quadratic in $\phi$, the $(\Bnab S_\mLam)^2$ term generates higher-polynomial terms, the $\Bnab^2 S_\mLam$ term generates lower-polynomial terms, and depending on the choice of $B_\mLam$, the last two terms can either generate new terms and/or modify existing terms. For example, with a $\phi^2 + \phi^4$ type action, the effective action after one step would have modified quadratic and quartic couplings but also a sextic term is generated. This kind of generation and mixing of terms in the action is what one expects of generic RG transformations. If $\Psi_\mLam$ had been chosen to be independent of $S_\mLam$, then this behavior would not have occurred (at least for linear $B_\mLam$); the effective action might have terms generated by $B_\mLam$, but there would be no feedback of $S_\mLam$ on itself. We take this as justifying Wegner's instinct.

In terms of the Boltzmann weight $\rho_\mLam = \me^{-S_\mLam}/Z_\mLam$, the flow equation can be written as
\BE
\mLam \del_\mLam \rho_\mLam = \Bnab \circ ( \Psi_\mLam \rho_\mLam ).
\EE
The flow vector $\Psi_\mLam$ typically depends on $\rho_\mLam$, so this form is not very useful. For the choice of diffusive $\Psi_\mLam$ above, it becomes
\BE
- \mLam \del_\mLam \rho_\mLam = \frac{1}{2} C_\mLam \Bnab^2 \rho_\mLam + \Bnab \circ ( B_\mLam \rho_\mLam).
\EE
This equation is of the form of a Fokker-Planck equation with diffusion matrix $C_\mLam$ and drift $- B_\mLam$, an observation which forms the basis of stochastic renormalization group, which we explore in the last chapter. We also see the reason for calling the choice eq. (\ref{diffusive_ERGE}) ``diffusive'': the Fokker-Planck equation describes the evolution of a diffusion process.

In the section on high-mode elimination, we observed the importance of considering rescaled effective degrees of freedom when searching for RG fixed points. To implement the effect of rescaling on the level of the flow equations, we can change variables in the ERGE above. From the definitions $p = \bar p \mLam$ and $\mPhi(\bar p) = \mLam^{-d_\phi} \phi(\bar p \mLam)$, we can replace functional derivatives using
\BE
\frac{\delta}{\delta \phi(p)} = \int_{\bar k} \frac{\delta \mPhi(\bar k)}{\delta \phi(p)} \frac{\delta}{\delta \mPhi(\bar k)} = \mLam^{-d - d_\phi} \int_{k} \frac{\delta \phi(k)}{\delta \phi(p)} \frac{\delta}{\delta \mPhi(\bar k)} =  \mLam^{-d - d_\phi}  \frac{\delta}{\delta \mPhi(\bar p)}.
\EE
The second-derivative term then becomes
\BE
\frac{1}{2} \Bnab_\mPhi \circ \overline C_\mLam \Bnab_\mPhi \rho_\mLam,
\EE
where $\overline C_\mLam(\bar p) = \mLam^2 C_\mLam(\bar p \mLam)$. The seed term similarly becomes
\BE
\Bnab_\mPhi \circ ( \overline B_\mLam \rho_\mLam),
\EE
with $\overline B_\mLam(\mPhi; \bar p) = \mLam^{-d_\phi} B_\mLam(\mLam^{d_\phi} \mPhi(\bar p); \bar p \mLam)$. The distribution $\bar \rho_\mLam(\mPhi)$ of the rescaled fields is equal to $\rho_\mLam(\phi)$ up to an overall normalization, but because they have different dependence on $\mLam$ at fixed field argument, the $\mLam$-derivative on the l.h.s. of the ERGE changes. Write the distribution as $\rho_\mLam(\phi) = \rho(\mLam, \phi)$. Then the relation of the two distributions is
\BE
\rho(\mLam, \phi) = \bar \rho(\mLam, \mPhi) \Big|_{\mPhi(\bar p) = \mLam^{-d_\phi} \phi(\bar p \mLam)},
\EE
and the derivative becomes
\begin{align}
\frac{\del}{\del \mLam} \rho (\mLam, \phi) = & \frac{\del}{\del \mLam} \bar \rho (\mLam,\mPhi) + \int_{\bar p} \frac{\delta \bar \rho (\mLam, \mPhi)}{\delta \mPhi(\bar p)} \frac{\del}{\del \mLam} \big[ \mLam^{-d_\phi} \phi(\bar p \mLam) \big] \nonumber \\
= & \frac{\del}{\del \mLam} \bar \rho (\mLam,\mPhi) - \mLam^{-1} \int_{\bar p} \big(d_\phi - \bar p \cdot \nabla_{\bar p}\big) \mPhi(\bar p) \frac{\delta \bar \rho (\mLam,\mPhi)}{\delta \mPhi(\bar p)}.
\end{align}
The operator
\BE
(D \mPhi) \circ \Bnab = \int_{\bar p} \big(d_\phi - \bar p \cdot \nabla_{\bar p}\big) \mPhi(\bar p) \frac{\delta}{\delta \mPhi(\bar p)}
\EE
is a representation of the dilatation generator on functionals. If the field rescaling involves the full wave function renormalization, $\zeta_\mLam = \mLam^{d_\phi} c_\mLam^{1/2}$, then
\BE
d_\phi \longrightarrow \Delta_\phi(\mLam) = \mLam \frac{\dd \ln \zeta_\mLam}{ \dd \mLam},
\EE
in $D$, while the cutoff function and drift terms become
\BE
\overline C_\mLam(\bar p) = \mLam^{-d} \zeta_\mLam^2 C_\mLam(\bar p \mLam), \quad \overline B_\mLam(\mPhi;\bar p) = \zeta_\mLam B_\mLam(\zeta_\mLam^{-1} \mPhi; \bar p \mLam).
\EE
The rescaled flow equation for $\bar \rho_\mLam$ is then
\BE
- \mLam \del_\mLam \bar \rho_\mLam = \frac{1}{2} \overline C_\mLam \Bnab^2 \bar \rho_\mLam + \Bnab \circ ( \overline B_\mLam \bar \rho_\mLam ) - D\mPhi \circ \Bnab \bar \rho_\mLam.
\EE
The dilatation generator can be thought of as a redefinition of $B_\mLam$, but we will keep them separate in our presentation. It has been suggested that $C_\mLam$ and $B_\mLam$ should be chosen such that $\overline C_\mLam$ and $\overline B_\mLam$ are independent of $\mLam$. I do not regard this as an essential ingredient of RG transformations, but only one of practical benefit. Solving this equation directly is generally difficult, but can be done exactly in a few simple cases. However, it is interesting that the equation obtained by setting  $\del_\mLam \bar \rho_\mLam = 0$, in principle, determines the RG fixed point action. Apart from the approximations made in FRG (described at the end of this chapter), we remark that one can attempt applying a \textit{functional} method of characteristics \cite{Dahmen:1972jx}; we remark that this approach can reproduce the solution to many linear RG's. 

By direct differentiation of the formula for the effective low-mode action, eq. (\ref{lowmode_ampconn}), we can obtain the ERGE for high-mode elimination as performed in section 2. One finds that it has the form of eq. (\ref{ERGES}) with diffusion and drift
\BE
C_\mLam(p) = \frac{2p^2 K_\mLam(p)}{p^2 + m_0^2}, \quad B_\mLam(p) = 2 p^2,
\EE
and therefore the flow vector is
\BE \label{highmode_psi}
\Psi_\mLam(p) = \frac{2p^2 K_\mLam(p)}{p^2 + m_0^2} \frac{\delta S_\mLam(\phi)}{\delta \phi(p)} - 2 p^2 \phi(p),
\EE
which determines the appropriate type of change of variables eq. (\ref{cov}) in this case, although we did not need to perform it in this way to study the effective action.

\subsection{Constraint functionals} Wilson and Kogut (WK) did not follow the approach above to arrive at their functional RG equation \cite{Wilson:1973jj}. They began, rather, with an analogy to the Green function solution of partial differential equations, generalizing the notion to functional equations. They noted that the functional\footnote{WK use an unconventional field $\phi$ with mass dimension $d/2$, i.e., their kinetic term coefficient is dimensionful. But they work in units such that $\mLam_0 = 1$ and integrals have a sharp cutoff.}
\BE
G_\tau(\phi,\varphi) = \mcal{N}_\tau \exp \Big[ - \frac{1}{2} \int_p \mLam_0^2 \frac{\big(\phi(p) - \me^{-\alpha_\tau(p)} \varphi(p) \big)\big(\phi(-p) - \me^{-\alpha_\tau(-p)} \varphi(-p)\big)}{1 - \me^{-2 \alpha_\tau(p)}} \Big],
\EE
where $\alpha_\tau(p) = p^2 (\me^{2\tau} - 1) + \beta(\tau)$ for some as-yet undetermined $\beta(\tau)$, is the Green functional of the PDE
\BE
\fdelAB{\rho_\tau (\phi)}{\alpha_\tau(p)} = \frac{\delta}{\delta \phi(p)} \Big[ \frac{\delta \rho_\tau(\phi)}{\delta \phi(-p)} + \phi(p) \rho_\tau(\phi) \Big]
\EE
subject to the initial condition $\rho_0(\phi) = \delta(\phi - \varphi)$, when the normalization $\mcal{N}_\tau$ is chosen appropriately. For arbitrary initial condition $\rho_0(\varphi)$, the solution would then be
\BE
\rho_\tau(\phi) = \int \mathscr{D} \varphi \; G_\tau(\phi,\varphi) \; \rho_0(\varphi).
\EE
In words, the distribution of fields $\phi$ is a gaussian-smearing of the initial distribution $\rho_0$, such that the mean value of the field $\phi(p)$ is $\me^{-\alpha_\tau(p)} \varphi(p)$, within a variance determined by the denominator in the exponent. Thus, the Green function can be thought of as imposing a statistical constraint which suppresses the high modes of $\varphi$ in a smooth fashion.\footnote{RG transformations which use a delta function rather than $G_t$ are sometimes used. Traditional spin-blocking RG is an example. But one must be careful if using delta functions in the continuum, as we discuss in chapter 4.} If the Green functional is such that
\BE
\int\mathscr{D} \phi \; G_\tau(\phi,\varphi) = \text{independent of $\varphi$},
\EE
then one can insert the r.h.s. into the partition function, integrate over $\varphi$, and obtain $\rho_\tau(\phi)$ as the new Boltzmann weight. Such a procedure seems different from that of Wegner, on the face of it, but we will see below that WK's route is a special case of Wegner's. From the relation
\BE
\frac{\del \rho_\tau(\phi)}{\del \tau} = \int_p \frac{\dd \alpha_\tau(p)}{\dd \tau} \fdelAB{\rho_\tau (\phi)}{\alpha_\tau(p)},
\EE
one obtains the ERGE
\BE
\frac{\del \rho_\tau(\phi)}{\del \tau} = \int_p \dot \alpha_\tau(p) \frac{\delta}{\delta \phi(p)} \Big[ \frac{\delta \rho_\tau(\phi)}{\delta \phi(-p)} + \phi(p) \rho_\tau(\phi) \Big].
\EE
By comparison with Wegner's formalism in the previous section, we see that WK's exact RG is a special case of diffusive FRG with\footnote{The ``cutoff function'' $C_t$ is not truly a cutoff function for Wilson and Kogut. They use a sharp cutoff on all momentum integrals, $|p| \in [0,\mLam_0]$.}
\BE
B_\tau(\phi;p) =  \dot \alpha_\tau(p) \phi(p), \quad C_\tau(p) = \dot \alpha_\tau(p).
\EE
Inspection of $\alpha_\tau(p)$ suggests that the effective scale is $\mLam = \mLam_0 \me^{-\tau}$, which implies the rescaled variables
\BE
p = \mLam_0 \me^{-\tau} \bar p, \quad \mPhi(\bar p) = \mLam_0^{\frac{d}{2}} \me^{\frac{d}{2} \tau} \phi(p).
\EE
WK insist that the rescaling factor $\zeta_\tau = \me^{\frac{d}{2} \tau}$ should be determined by demanding that the ERGE is independent of $\tau$ in rescaled variables. Since $\dot \alpha_\tau(p) = 2 \bar p^2 + \dot \beta(\tau)$, the rescaled ERGE is
\BE
\frac{\del \rho_\tau(\mPhi)}{\del \tau} = \int_{\bar p} \big(\smallfrac{d}{2} + \bar p \cdot \nabla_{\bar p}\big) \mPhi(\bar p) \frac{\delta \rho_\tau(\mPhi)}{\delta \mPhi(\bar p)} + \int_{\bar p} \big(2 \bar p^2 + \dot \beta(\tau)\big) \frac{\delta}{\delta \mPhi(\bar p)} \Big[ \frac{\delta \rho_\tau(\mPhi)}{\delta \mPhi(-\bar p)} + \mPhi(\bar p) \rho_\tau(\mPhi) \Big].
\EE
The function $\beta(\tau)$ is determined by choosing a normalization condition for the kinetic term for all $\tau$. In the gaussian model, it will turn out that $\beta(\tau) = \tau$ is appropriate. We will describe the gaussian fixed point of WK's ERGE in the next section. 

\subsection{Fixed points}

In terms of $\tau = \ln \mLam_0 / \mLam$, a fixed point solution is an action $S_*$ for which $\del_\tau S_* = 0$, which typically must occur in the limit $\tau \to \infty$. Dropping bars for dimless quantities, the ERGE eq. (\ref{ERGES}) implies that a fixed point action satisfies
\BE
0 = - \frac{1}{2} C_* \Big[ \Bnab S_* \circ \Bnab S_* - \Bnab^2 S_* \Big] + B_* \circ \Bnab S_* - \Bnab \circ B_* - D \mPhi \circ \Bnab S_*,
\EE
assuming there is some limit of $C_\tau$ and $B_\tau$ as $\tau \to \infty$.

We are often interested in the behavior of actions that are slightly deformed from the fixed point, in order to study the various asymptotic behaviors of these deformations. We may perturb about the fixed point by letting
\BE
S_\tau = S_* + \mcal{E}_\tau, \quad B_\tau = B_* + \mcal{F}_\tau, \quad C_\tau = C_* + \mcal{G}_\tau,
\EE
with $\mcal{E}_\tau, \; \mcal{F}_\tau, \; \mcal{G}_\tau$ small for large $\tau$, and linearizing the flowing equation:
\begin{align}
\del_\tau \mcal{E}_\tau + D \mPhi \circ \Bnab \mcal{E}_\tau = - \frac{1}{2} C_* \Big[ & 2 \Bnab S_* \circ \Bnab \mcal{E}_\tau - \Bnab^2 \mcal{E}_\tau \Big] - \frac{1}{2} \mcal{G}_\tau \Big[ \Bnab S_* \circ \Bnab S_* - \Bnab^2 S_* \Big] \\
& - \Bnab \circ \mcal{F}_\tau + B_* \circ \Bnab \mcal{E}_\tau + \mcal{F}_\tau \circ \Bnab S_*.
\end{align}
Assuming $\mcal{G}_\tau$ decays with time, we drop the $\mcal{G}_\tau$ term in what follows. Furthermore, we see that assuming $B_\tau$ is independent of $\tau$ further simplifies the equation by dropping $\mcal{F}_\tau$. It is then clear how WK's demand of a $\tau$-independent ERGE can simplify analyses. In this simplified (but typical) case, then, the linearized flow equation becomes
\BE
\del_\tau \mcal{E}_\tau + D \mPhi \circ \Bnab \mcal{E}_\tau = - \frac{1}{2} C_* \Big[2 \Bnab S_* \circ \Bnab - \Bnab^2 \Big] \mcal{E}_\tau + B_* \circ \Bnab \mcal{E}_\tau.
\EE
Variables separate, $\mcal{E}_\tau = f(\tau) \mcal R(\phi)$, and if we let $f(\tau) = f(0) \me^{y \tau}$, then we have an eigenvalue equation for $\mcal{R}$,
\BE
y \mcal R = - \frac{1}{2} C_* \Big[ 2 \Bnab S_* \circ \Bnab - \Bnab^2 \Big] \mcal R + B_* \circ \Bnab \mcal{R} - D \mPhi \circ \Bnab \mcal{R},
\EE
which generally will have a spectrum $\{ y_a \}$ (which will be discrete under certain assumptions \cite{Rosten:2010vm}). Therefore, the perturbed action near a fixed point can be written as
\BE
S_\tau(\phi) = S_*(\phi) + \sum_{i} \alpha_i \; \me^{y_a \tau} \mcal R_a(\phi).
\EE
The $\mcal{R}_a$ are called \textit{scaling operators}, and the $y_a$ are their RG eigenvalues. We observe three distinct types of behavior for such perturbations. For the operator $\mcal{R}_a$, we have
\begin{itemize}
\item $y_a < 0$: the perturbation decays with time exponentially, and is called \textit{irrelevant},
\item $y_a = 0$: the perturbation is independent of time, and is called \textit{exactly marginal},
\item $y_a > 0$: the perturbation increases exponentially, and is called \textit{relevant}.
\end{itemize}
Thus we recover the same kind of behavior for the perturbations about a fixed point that were observed using the discrete block-spin theory of chapter 1, except that the RG flow is parameterized by $\tau$ rather than $b$. It will be useful in what follows to use the continuous analog of the scale factor, $b_\tau = \me^\tau = \mLam_0 / \mLam$.

The expectation values of scaling operators behave in a simple way under RG transformations in the vicinity of the fixed point. Suppose $S_\tau \to S_{\tau+\eps}$ close to $S_*$ under the RG flow.\footnote{The following derivation is adapted to FRG from the approach described in \cite{ZinnJustin:2007zz}.} Then deform the initial action via the scaling operator $\mcal{R}_a$,
\BE
S_\tau(\theta) := S_\tau + \theta \mcal{R}_a,
\EE
where $\theta$ is a smooth parameter. The flowing action $S_\tau$ may be written as a linear combination of the scaling operators. Thus $\theta$ can be viewed as a deformation of the associated coupling. This means that, under an RG transformation $\tau \to \tau + \eps$, the coupling changes simply: $\theta \to b_\eps^{y_a} \theta$, where $b_\eps = b_{\tau + \eps} / b_\tau$. Now, from the general relation
\BE
\langle \mcal{R}_a \rangle_{S_\tau} = - \frac{\dd}{\dd \theta} \Big( \ln \int_\mPhi \me^{-S_\tau(\theta)} \Big) \Big|_{\theta = 0},
\EE
we may derive $\langle \mcal{R}_a \rangle_{S_{\tau+\eps}} = b_\eps^{-y_a} \langle \mcal{R}_a \rangle_{S_\tau}$. The scaling operators above are volume integrals of local scaling operators $\mcal{R}_a(\bar x)$, where $\bar x = \hat x b_\tau^{-1}$ and $\hat x$ is a dimensionless distance at the bare scale ($x = a_0 \hat x$). Hence
\BE \label{scalingops}
\langle \mcal{R}_a(\bar x) \rangle_{S_{\tau+\eps}} = b_\eps^{\Delta_a} \langle \mcal{R}_a (b_\eps \bar x) \rangle_{S_\tau}, \quad \text{with} \quad \Delta_a := d - y_a
\EE
being the \textit{scaling dimension} of the operator. By letting $S_\tau \to S_\tau(\theta)$ in this scaling formula, we can derive scaling laws for higher $n$-point functions of $\mcal{R}_a$ by further differentiation. Thus, when we say that a scaling operator changes as $\mcal{R}_a \to b_\eps^{\Delta_a} \mcal{R}_a$, it is true either as a term in the effective action (at constant coupling), or within expectation values at different RG scales. In particular, it implies the correlator scaling laws for scaling operators that we described in block-spin RG and which formed the basis of the GFRG method of chapter 2.

The example of the GFP of Wilson and Kogut's ERGE will now be discussed. In terms of the action $S_\tau$, their ERGE becomes
\BE
\del_\tau S_\tau = - D \mPhi \circ \Bnab S_\tau - \dot \alpha \Big[ \Bnab S_\tau \circ \Bnab S_\tau - \Bnab^2 S_\tau - \mPhi \circ \Bnab S_\tau \Big].
\EE
If the bare action is gaussian, the effective action will also still be gaussian. Writing
\BE
S_\tau(\mPhi) = \frac{1}{2} u_\tau \mPhi \circ \mPhi,
\EE
where $u_\tau = u(\tau,\bar p)$, and noting that integration by parts (discarding boundaries) implies
\BE
\int_{\bar p} \bar p \cdot \nabla_{\bar p} \mPhi( \bar p) \; u_\tau(\bar p) \mPhi(-\bar p) = - \frac{1}{2} \int_{\bar p} \big[ d + \bar p \cdot \nabla_{\bar p} u_\tau(\bar p) \big] \mPhi(\bar p) \mPhi(-\bar p),
\EE
then determines a PDE for the 2-point term $u_\tau( \bar p)$:
\BE
\del_\tau u_\tau = - \bar p \cdot \nabla_{\bar p} u_\tau - 2 \dot \alpha \big[1 - u_\tau \big] u_\tau,
\EE
For $\beta(\tau) = a \tau + b$, then $\dot \alpha = 2 \bar p^2 + a$. We solve the equation by the method of characteristics. First we find the integral curves of the vector field on $\mathbb{R}^{1+d}$,
\BE
X = \del_\tau + \bar p \cdot \nabla_{\bar p},
\EE
namely, the curves $(\tau_\lambda, \bar p_\lambda)$ parameterized by $\lambda$ and determined by the ODE's
\BE
\frac{\dd \tau}{\dd \lambda} = 1, \quad \frac{\dd \bar p}{\dd \lambda} = \bar p \quad \Rightarrow \quad \tau_\lambda = \lambda, \quad \bar p_\lambda = \bar p \me^{\lambda},
\EE
with initial condition $\tau_{0} = 0$ --- hence we can just use $\tau$ as the parameter along every curve. The first order PDE above then transports $u(\tau,p)$ along the integrals curves of $X$ via
\BE
\frac{\dd u}{\dd \lambda} = 2 (2\bar p^2_\lambda + a) u(\lambda) \big( 1 - u(\lambda) \big),
\EE
whose solution $u(\lambda) = u(\tau_\lambda, \bar p_\lambda)$ is the value of $u$ at a point along the curve determined by $\lambda$ and the initial conditions for $\tau_\lambda, \; \bar p_\lambda$. The solution is then
\BE
u(\lambda) = \frac{u(0)}{u(0) + (1 - u(0)) \exp B(\lambda)},
\EE
where $u(0) = u(\tau_0, \bar p_0) = u(0,\bar p)$, and 
\BE
B(\lambda) = - 2 \int_0^\lambda \dd \lambda' \big[ 2 \bar p_{\lambda'}^2 + a \big] = -2 \bar p^2 ( \me^{2\lambda} - 1) - 2 a \lambda.
\EE
The initial condition for $u$ is the quadratic part $S^{(2)}_0(\bar p) = \omega(\bar p)$ of bare action. Geometrically, it is the value of $u$ along the $\tau=0$ axis. If we want the solution at a point $(\tau,\bar k)$, we use the fact that $\bar k = \bar p_\tau = \bar p \me^\tau$ can be taken as the value of the momentum at parameter value $\tau$ along a curve starting from the $\tau$-axis, where the momenta are $\bar p_0 = \bar p$. Writing $\tau=\lambda$, we have
\begin{align}
u(\tau,\bar k) & = \frac{\omega(\bar p)}{\omega(\bar p) + (1 - \omega(\bar p)) \exp [ -2 \bar p^2 ( \me^{2\tau} - 1) - 2 a \tau]} \nonumber \\
& = \frac{\omega(\me^{-\tau} \bar k)}{\omega(\me^{-\tau} \bar k) + (1 - \omega(\me^{-\tau} \bar k)) \exp [ -2 \bar k^2 (1 - \me^{-2\tau}) - 2 a \tau]}.
\end{align}
If the initial condition is the standard kinetic term, $\omega(\bar p) = z \bar p^2$, we find
\BE
u(\tau,\bar k) = \frac{z \bar p^2}{z \bar p^2 + (1 - z \bar p^2 \me^{-2\tau}) \exp [ -2 \bar p^2 (1 - \me^{-2\tau}) - 2 (a-1) \tau]}.
\EE
To obtain a nonzero and nonuniform Boltzmann distribution as $\tau \to \infty$, we see that we must choose $a=1$ in $\beta(\tau)$. The limit is then
\BE
u_*(\bar p) = \frac{z \bar p^2}{z \bar p^2 + \me^{ - 2 \bar p^2}}.
\EE
Thus, the ERGE of Wilson and Kogut indeed has a gaussian fixed point, and in fact it possesses a \textit{line} of fixed points parameterized by $z$ such that the Boltzmann factor is bounded above. One may alternatively solve the fixed point equation directly, in which case $z$ arises as an integration constant.


\section{Various implementations} 

I close this chapter with a brief summary of various methods and applications of the formalism of functional RG that have arisen after its initial formulation in the 1970's. We will not give  in-depth accounts of them and refer instead to other sources and reviews for the interested reader. Also note that the items below are, of course, not necessarily mutually exclusive.
\begin{itemize}

\item \textit{Polchinski equation.} In 1983 Polchinski wrote down an RG equation inspired by that of Wilson and Kogut \cite{Polchinski:1983gv}. Rather than using his ERGE to study fixed points and RG flows in the abstract, he used it to prove perturbative renormalizability in a novel and simpler way than usual. It also laid the groundwork for precise formulations of effective field theory. His proof has since been made quite rigorous \cite{Keller:1990ej,Ball:1993zy}, and versions of it have been carried out in other systems, like QED4 \cite{Bonini:1993kt}.

\item \textit{Derivative expansion.} The simplest truncation strategy in FRG is the \textit{local potential approximation} (LPA) \cite{Hasenfratz:1985dm,Golner:1985fg,Bagnuls:2000ae}. It proceeds by fixing the kinetic term as $(\del \phi)^2$ and ignoring all other momentum dependence in the effective action, yielding an ERGE for the potential $V(\phi)$. 
The LPA is a first-order approximation to the more general \textit{derivative expansion}, in which the effective action is expanded in powers of momenta. A drawback to the most basic approach (the LPA) is that the $\eta$ exponent is zero, because there is no need to correct for would-be changes of the kinetic term, and therefore will only be expected to be accurate in systems like $\phi^4_3$ where $\eta$ is small, but this defect lessens with higher orders \cite{Bagnuls:2000ae}.

\item \textit{Average effective action.} The concept of a constraint functional was revived in the works of Wetterich in the early 1990's. He introduced a quantity called the \textit{average effective action} $\Gamma_k$, which corresponds to the high-mode action discussed above \cite{Wetterich:1989xg}. This action satisfies an ERGE that is often simpler in form than that of the flowing effective action $S_\mLam$, as it deals directly with 1PI functions \cite{Wetterich:1992yh}, whereas in general the contributions $S_\mLam$ are merely connected. Wetterich's original application of this formalism was to the evolution of the effective potential in the broken-symmetry phase of scalar field theories, but has since found numerous applications. 

\item \textit{Condensed matter.} Functional RG has been adapted to fermionic models in condensed matter theory since the early 2000's in order to study long-distance properties of such systems. By ``long distance,'' one here means \textit{close to the Fermi surface}. An important difference with respect to RG as presented in this thesis is that the rescaling step in the RG transformation should rescale momenta relative to the Fermi surface. See \cite{Kopietz:2010zz} for an exposition.

\item \textit{Asymptotic safety.} In the realm of quantum gravity it was proposed by Weinberg long ago \cite{Weinberg:1980gg} that a possible solution to the puzzle of the high-energy limit of quantized General Relativity would be the existence of a UVFP for the gravitational interaction. Perturbative methods are typically assumed to be untrustworthy at high energies in this theory, so it is natural to attempt to apply the nonperturbative methods of FRG to quantum gravity in search of fixed points. See \cite{Eichhorn:2020mte,Reichert:2020mja} for reviews.

\end{itemize}

\newpage

\chapter{Stochastic RG}

In this chapter, we will demonstrate the equivalence of certain kinds of FRG transformations with a class of stochastic (Markov) processes on field space.\footnote{See \cite{Pavliotis:2014} for a mathematician's introduction to stochastic processes, \cite{ZinnJustin:2002ru} for a physicist's introduction, and \cite{Damgaard:1987rr, ZinnJustin:2002ru} for an introduction to their field-theoretical generalization in the context of stochastic quantization. The essential aspects are reviewed in Appendix B.} It has been noted before \cite{Gaite:2000jv, Pawlowski:2017rhn, ZinnJustin:2007zz} that the functional RG equations for effective actions, when written in terms of effective Boltzmann weights, are of the form of a Fokker-Planck (FP) equation, whose solution is therefore a probability distribution over effective fields. Taking this observation seriously, and recalling that Fokker-Planck distributions can be thought of as being generated by a Langevin equation on the degrees of freedom appearing in the FP distribution, one may ask what kinds of Langevin equation generate the FRG effective actions. In what follows, we will define an RG transformation by a particularly simple (linear) choice of Langevin equation, and show by direct calculation that the transition functions resemble the constraint functionals found in the literature of FRG. The effective action for the specific case of $\phi^4$ theory in three dimensions will then be discussed, and the existence of a nontrivial IR fixed point will be checked to 1-loop order in perturbation theory. It will therefore become apparent that although the stationary distribution of the FP equation would be expected to be gaussian, a simple rescaling of variables allows for an interacting fixed point solution.\footnote{This is not surprising from the FRG perspective, of course, but it may be unexpected from the standpoint of stochastic processes, where the stationary distributions of the Fokker-Planck equation are expected to involve the potential whose gradient appears as the drift term in the Langevin equation \cite{Pavliotis:2014}.}

In chapter 2 we described the relationship of gradient flow (GF) with RG, and at various points mentioned that an alternative approach to the theory, not based on a block-spin analogy, was possible. Before describing the approach, we mention that other analytic work has been done \cite{Kagimura:2015via,Yamamura:2015kva,Makino:2018rys, Abe:2018zdc, Sonoda:2019ibh} connecting GF to the framework of functional RG. In particular, it was noted by Abe and Fukuma that certain definitions of a GF effective action lead to a kind of Langevin equation \cite{Abe:2018zdc} (though different from what we propose here), and later by Sonoda, that the connected $n$-point functions of a particular FRG effective theory are equal to the GF observables up to proportionality \cite{Sonoda:2019ibh}. The relationship between stochastic processes with ``colored noise'' and RG has also been explored recently in \cite{Ziegler:2019jgf}.

The equivalence we discuss here is a formulation of the Monte Carlo Renormalization Group (MCRG) principle for FRG. Recall that the kind of MCRG discussed by Swendsen \cite{Swendsen:1979gn} in the 1980's provided a prescription for computing observables in an effective theory by computing \textit{blocked} observables in a bare theory, that is, without having to know the effective action. A similar property will be found for the stochastic RG transformation, namely, that effective observables may be computed from the stochastic observables generated by the Langevin equation, whose initial condition is the bare field. The MCRG property will be valid for both lattice and continuum theories alike, thereby suggesting the possibility of computing general observables in an effective theory on the lattice by integrating a Langevin equation on top of the ensemble generated in the MCMC simulation of the corresponding bare theory. 

The relationship to gradient flow will then follow from an observation made by Wilson and Kogut \cite{Wilson:1973jj}, and recently connected to gradient flow by Sonoda and Suzuki \cite{Sonoda:2019ibh}. In the context of the stochastic RG transformation, it follows from the MCRG equivalence that the connected expectation values of an FRG effective theory are equal to gradient-flowed expectations up to additive corrections that depend on the choice of Langevin equation, and which decay exponentially at large distances. This relationship implies that the measurement of gradient-flowed quantities is sufficient for the determination of long-distance critical properties of the theory, in much the same way as spin-blocked observables at large distances. This avoids the necessity of performing a full Langevin equation simulation if one only cares about long-distance properties.

A virtue of the characterization of FRG in terms of stochastic processes is that the observables of the effective theory satisfy differential equations involving the generator of the Markov process, allowing one to study the flow of the observables directly, without knowledge of the effective action. An analysis of these equations for discrete, small time steps leads to the stochastic RG instantiation of usual RG scaling laws of correlations of the fundamental field, as well as of composite operators built from it. In particular, by virtue of the stochastic MCRG equivalence, one is led to correlator ratio formulas of the sort described in chapter 2, implying a method for measuring scaling dimensions of operators close to a critical fixed point. Thus, the results of chapter 2 may be regarded as a consequence of the stochastic RG idea.

What follows is an exposition of stochastic RG based on the contents of \cite{Carosso:2019qpb}, but expanded upon in various places.

\section{Stochastic processes and FRG}

Here we discuss the general framework for stochastic RG. The RG transformation will be defined by a Langevin equation on the degrees of freedom of a field theory. The simplicity of the equation will allow for an explicit calculation of the probability distribution which it generates, and the functional form of the distribution will entail an equivalence to conventional FRG transformations. A brief consideration of the observables generated by the stochastic process will lead to the MCRG equivalence between the effective theory and the stochastic observables. Lastly, we will comment on the pitfalls of a seemingly simpler definition of the effective theory.

\subsection{The Langevin equation}

We will define an RG transformation by a stochastic process $\phi_t$ on field space over $\mathbb{R}^d$, determined by a Langevin equation (LE) of the form
\BE \label{LE_momspace}
\del_t \phi_t(p) = -\omega(p) \phi_t(p) + \eta_t(p), \quad \phi_0(p) = \varphi(p),
\EE
where $\omega(p)$ is positive for $\|p\|>0$ and $\omega(0) \geq 0$, e.g. $\omega(p) = p^2$, where $p^2 := \| p \|^2$.\footnote{Of course, the realm of stochastic quantization \cite{Damgaard:1987rr} deals with writing field theory expectation values as equilibrium limits of a stochastic process on field space. Here, however, the bare theory is kept as a traditional field theory, and the stochasticity applies to the RG transformation only. See the end of Appendix B for the main ideas of SQ.} The ``time'' $t$ in this equation does not denote physical time, but rather an ``RG time'' which we will call \textit{flow time}, or simply \textit{time}. The noise $\eta_t(p)$ is chosen to be gaussian-distributed according to the measure
\BE \label{noise_distribution}
\dd \mu_{0} (\eta) := \mrm{c}(\mLam_0, \Omega) \exp\Big[ - \frac{1}{2\Omega} \int_I \dd t \; (\eta_t, K_{0}^{-1}\eta_t)\Big] \mathscr{D} \eta,
\EE
where the notation $(\phi, M \chi)$ denotes a quadratic form, written variously as\footnote{We abbreviate $\int_x = \int_{\mathbb{R}^d} \dd^d x$ and $\int_p = \int_{\mathbb{R}^d} \dd^d p / (2\pi)^d$ when no confusion arises.}
\BE
(\phi, M \chi) =  \int_{xy} \phi (x) M(x,y) \chi(y) = \int_{pk} \phi(p) M(p,k) \chi_t(k).
\EE
The cutoff function $K_{0}(p)$ suppresses noise momentum modes greater than $\mLam_0$, e.g. $K_{0}(p) = \me^{-p^2/\mLam_0^2}$ under Schwinger regularization.\footnote{The LE and measure $\dd \mu_0$ can easily be written for a lattice theory, in which case the cutoff function $K_0$ is not necessary, as the lattice naturally regulates the noise at the bare scale.} Expectation values with respect to the noise distribution of functions $O(\eta)$ are defined by
\BE
\mathbb{E}_{\mu_0}[O(\eta)] := \int \! O(\eta) \dd \mu_0 (\eta).
\EE
The first two moments of $\mu_{0}$ are then
\BE
\mathbb{E}_{\mu_0}[ \eta_t(p) ] = 0, \quad  \mathbb{E}_{\mu_0}[ \eta_t(p) \eta_s(k) ]  = 2\pi \Omega \; \delta(t-s) \; (2\pi)^d \delta(p+k) K_{0}(k).
\EE
Later we will take the initial condition $\phi_0 = \varphi$ to be distributed according to a measure $\dd \rho_0 (\varphi)$ corresponding to the bare theory of interest, the cutoff of which is chosen to be $\mLam_0$. Hence, the cutoff for the noise is chosen to match the cutoff of the bare theory.

Turning back to eq. (\ref{noise_distribution}), the constant $\mrm{c}(\mLam_0, \Omega)$ is chosen to normalize $\dd \mu_{0}$ to unity, $\Omega$ is the (dimensionless) variance of the noise, and $I \subset \mathbb{R}$ is an arbitrary time interval large enough to include all desired times. In position space, the Langevin equation takes the form of a stochastic heat equation
\BE
\del_t \phi_t(x) =  - (\omega \phi_t)(x) + \eta_t(x), \quad \phi_0(x) = \varphi(x).
\EE
For the case $\omega(p) = p^2$, one has $(\omega \phi)(x) = -\Delta \phi (x) = -\del_\mu \del_\mu \phi(x)$. In position space, therefore, we see that the equation becomes a stochastic partial differential equation.

The form of the momentum space equation above is a simple field-theoretic generalization of the well-known Ornstein-Uhlenbeck (OU) process (i.e. damped Brownian motion) $q_t$ with Langevin equation and solution \cite{ZinnJustin:2002ru}, respectively,
\BE
\dot q_t = - \omega q_t + \eta_t, \quad q_t = \me^{-\omega t} q_0 + \int_0^t \dd s \; \me^{-\omega(t-s)} \eta_s,
\EE
where $\eta_t$ is gaussian white noise, so it is quite simple to find the solution. One treats the noise term like a non-homogeneous part of the equation, finding
\BE
\phi_t(p) = f_t(p) \varphi(p) + \int_0^t \dd s \; f_{t-s}(p) \eta_s(p),
\EE
where $f_t(p)$ is a generalized momentum space heat kernel of the form
\BE
f_t(p) = \me^{-\omega(p)t}, \quad f_t(z) = \int_p \me^{i p \cdot z} f_t(p).
\EE
In position space, one finds
\BE
\phi_t(x) = (f_t \varphi)(x) + \int_0^t \dd s \; (f_{t-s}\eta_s)(x).
\EE
We will sometimes denote the solution's dependence on initial condition and noise by $\phi_t[\varphi; \eta]$. The first term on the r.h.s. implies that the mean of $\phi_t(x)$ satisfies the free gradient flow equation, i.e. ``heat'' equation, corresponding to the differential operator $\omega$.

\subsection{The Fokker-Planck distribution}

With the explicit solution in-hand, one can compute the probability distribution of fields $\phi$ at time $t$ given $\varphi$ at $t=0$. We say that the Langevin equation \textit{generates} a Fokker-Planck (FP) distribution $P(\phi,t;\varphi,0)$ defined by
\BE \label{Pdef}
P(\phi,t;\varphi,0) := \mathbb{E}_{\mu_0} \big[ \delta(\phi - \phi_t[\varphi;\eta])\big] = \int \mathscr{D} \lambda \; \mathbb{E}_{\mu_0} \big[ \me^{i(\lambda,\phi-\phi_t [\varphi; \eta])}\big].
\EE
From the definition of noise expectations, we then find
\BE
P(\phi, t; \varphi, 0) =  \mrm{c}(\mLam_0, \Omega) \int \! \mathscr{D} \lambda \int \! \mathscr{D} \eta \; \exp\Big[i(\lambda, \phi - \phi_t[\varphi; \eta]) -\frac{1}{2\Omega} \int_I \dd s (\eta_s, K^{-1}_{0} \eta_s)\Big].
\EE
Substituting in the explicit solution for $\phi_t$, the integrand becomes
\begin{align}
\exp&\Big[i(\lambda, \phi - f_t \varphi)  - i \int_0^t \dd s\; (\lambda, f_{t-s} \eta_s) - \frac{1}{2\Omega} \int_I \dd s \; (\eta_s, K_{0}^{-1} \eta_s)\Big] \nonumber \\
& = C \exp\Big[i(\lambda, \phi - f_t \varphi)  - \int_0^t \dd s\; \Big( i (\lambda, f_{t-s} \eta_s) + \frac{1}{2\Omega} (\eta_s, K_{0}^{-1} \eta_s) \Big) \Big],
\end{align}
the constant $C$ involving times $s > t$, which divide out of any noise averages and will now be dropped. The noise integral over relevant $\eta_t$'s is a standard gaussian integral, which yields
\BE
P(\phi, t; \varphi, 0) = \int \! \mathscr{D}\lambda \; \exp\Big[i(\lambda, \phi - f_t \varphi) - \frac{\Omega}{2} \int_0^t \dd s \; (f_{t-s}^\top \lambda, K_{0} f_{t-s}^\top
\lambda)\Big].
\EE
Next, note that the $s$-integral (which does not care about $\lambda$ or $K_{0}$) produces a kernel
\BE \label{Adef}
A_t := \Omega \int_0^t \dd s \; f_{t-s} K_{0} f_{t-s}^\top,
\EE
which in momentum space is given by a diagonal matrix, 
\BE \label{A_momspace}
A_t(p,k) = \Omega (2\pi)^d \delta(p+k)  K_{0}(p) \; \frac{1-\me^{-2 \omega(p) t}}{2\omega(p)}.
\EE
We will sometimes denote $A^{-1}_t$ by $B_t$; the inverse exists by virtue of the restrictions on $\omega(p)$. The remaining $\lambda$-integral is also gaussian, and evaluates to
\BE \label{FP_dist}
P(\phi, t; \varphi, 0) = \big[\det 2 \pi B_t \big]^{\frac{1}{2}} \exp\Big[- \frac{1}{2} \big(\phi - f_t\varphi, B_t (\phi - f_t \varphi)\big)\Big].
\EE
Here we recognize the similarity of this functional to the constraint functional of Wilson and Kogut \cite{Wilson:1973jj} that we worked with in chapter 3, as well as those found in \cite{Wetterich:1989xg, Igarashi:2009tj}. We will call such a functional a \textit{gaussian constraint functional} or a \textit{transition function} (when emphasizing its probabilistic interpretation). In momentum space, the exponent is explicitly
\BE \label{constraint_functional}
- \frac{1}{2} \int_p \; \frac{2\omega(p) K_{0}^{-1}(p)}{1-\me^{-2\omega(p)t}} \big(\phi(p) - \me^{-\omega(p)t} \varphi(p)\big)\big(\phi(-p) - \me^{-\omega(p)t} \varphi(-p)\big).
\EE
One observes that the mean of the field $\phi$ is set to the flowed field $f_t(p) \varphi(p)$ within a functional variance determined by $A_t(p)$. Thus the effect of the stochastic RG transformation is to produce a low-mode fluctuating field, in the sense that the mean value of modes of $\phi$ with $\omega(p) \gg 1/t$ are exponentially suppressed. For $\omega(p) = p^2$, this suggests that the effective cutoff of the resulting theory is roughly $\mLam_t \sim 1/\sqrt{t}$ ; a more precise identification will be made later. 

For reasons explained in the next subsection, we may write the transition function as $P_t(\phi,\varphi)$ rather than $P(\phi,t;\varphi,0)$, and we will sometimes suppress the initial condition by writing $P_t(\phi)$. The transition function is a Green function for the Fokker-Planck equation
\begin{align} \label{FP_transfn}
\frac{\del P_t(\phi)}{\del t} & = \Bnab \circ \Big( \frac{1}{2} \mSig(\phi,t) \Bnab P_t(\phi) + \mathscr{B}(\phi,t) P_t(\phi) \Big), \nonumber \\ 
& \lim_{t \to 0} P(\phi,t; \varphi,0) = \delta(\phi - \varphi),
\end{align}
where the drift vector $\mathscr{B}$ and diffusion matrix $\mSig$ are defined by \cite{Pavliotis:2014}
\begin{align}
\mathscr{B}(\phi,t) & = - \lim_{t' \to t} \frac{1}{t'-t} \int \mathscr{D} \phi' (\phi' - \phi) P(\phi',t';\phi,t), \\
\mSig(\phi,t) &= \lim_{t' \to t} \frac{1}{t'-t} \int \mathscr{D} \phi' (\phi' - \phi) \otimes (\phi' - \phi) P(\phi',t';\phi,t).
\end{align}
A derivation of the FP equation above is provided in Appendix B, where it is demonstrated that such an equation follows from the LE
\BE
\del_t \phi_t = - \mathscr{B}(\phi_t,t) + \eta_t.
\EE
With the explicit solution eq. (\ref{FP_dist}), we compute
\begin{align}
\mathscr{B}(\phi,t) &= \omega \phi, \\
\mSig(\phi,t) &= \Omega K_{0},
\end{align}
as expected. If the initial condition $\varphi$ is distributed according to a measure $\dd \rho_0(\varphi) = \me^{-S_0(\varphi)} \mathscr{D} \varphi$ corresponding to a bare theory, then the effective distribution
\BE
\rho_t(\phi) := \int \! P(\phi,t;\varphi,0) \; \dd \rho_0(\varphi)
\EE
also satisfies the FP equation, with initial condition $\rho_0(\varphi)$. For the specific choice eq. (\ref{LE_momspace}) of LE above, we find
\BE \label{FP_EFT}
\del_t \rho_t(\phi) = \frac{1}{2} K_0 \; \Bnab^2 \rho_t(\phi) + \Bnab \circ \big( \omega \phi \; \rho_t(\phi) \big),
\EE
where we have set $\Omega = 1$.

The drift term $(\omega \phi_t)(x)$ may be regarded as the functional derivative of what we might call a ``flow action''
\BE
\hat S(\phi) = \frac{1}{2} (\phi, \omega \phi) \quad \Rightarrow \quad \del_t \phi_t = - \Bnab\hat S(\phi_t) + \eta_t,
\EE
in which case one would have $\mathscr{B} = \Bnab \hat S$.
For arbitrary choices of $\hat S$, the Langevin equation may become nonlinear and the (still linear) FP equation generalizes to 
\BE \label{FP_seed}
\del_t \rho_t(\phi)= \frac{1}{2} K_{0} \Bnab^2 \rho_t(\phi) +  \Bnab \circ \big( \Bnab \hat S(\phi) \rho_t(\phi) \big).
\EE
Thus we observe that the stochastic process generates ERGE's of the form described in chapter 3. Of course, by writing $\rho_t = \me^{-S_t}$ and letting $\dd \rho_0(\varphi) = \me^{-S_0(\varphi)} \mathscr{D}\varphi$, one recovers functional PDE's for the effective action $S_t(\phi)$ given some bare action $S_0(\varphi)$, similar to the Polchinski equation.

There are many possibilities for how to generalize the scheme presented above. First, one could choose a different distribution for the noise, perhaps even a non-gaussian one. Second, one could generalize the flow action to be arbitrarily complicated in $\phi$, thereby making the Langevin equation non-linear, but these will generate FP distributions which are more difficult to calculate; we will discuss nonlinear RG's at the end of this chapter. For theories whose field variables are in compact spaces, or theories with local symmetries, however, one \textit{must} use non-linear LEs to ensure that the flow preserves the symmetry; such equations will likely resemble those found in the context of stochastic quantization \cite{Damgaard:1987rr, Batrouni:1985jn}.

I remark that the stochastic characterization of RG is a natural one to take. In ordinary stochastic processes, such as Einstein's theory of Brownian motion in 1905, the random noise represents the influence of small-scale degrees of freedom on the large-scale ones: the molecular bath in which the dust particle is submerged imparts random kicks  to the particle. In the case of field theory, we see that the noise plays the role of short-distance degrees of freedom randomly kicking the momentum modes of the field. The drift term enforces the overall damping of high modes, while the noise guarantees that the high modes are made to interact (indirectly) with the low modes, as is apparent from the form of the gaussian constraint functional and the influence of high-mode loops in the effective action that we will describe below.

\subsection{MCRG} Although the transition functional above has the same form as the constraint functionals found in the FRG literature, a notable difference here is that the kernel $B_t = A_t^{-1}$ is \textit{determined} by the associated Langevin equation, having a fixed relation to $\omega$, the choice of drift. Thus, if one wants to change the details of the constraint functional, one must find the appropriate LE. 

The initial condition for the transition function, $P_0(\phi,\varphi) = \delta(\phi - \varphi)$, is guaranteed by the fact that it is generated by a LE with initial condition $\varphi$. As a distribution, it is furthermore normalized such that for all $t \geq 0$,
\BE
\int \! \mathscr{D} \phi \; P_t(\phi,\varphi) = 1,
\EE
and in particular, the integral is independent of the field $\varphi$. These conditions allow one to define the effective theory in a more conventional way by inserting unity into the partition function $Z$ of the bare theory as
\BE
Z = \int \! \dd \rho_0(\varphi)  = \int \! \mathscr{D} \varphi \! \int \! \mathscr{D} \phi \; P_t(\phi,\varphi) \; \me^{-S_0(\varphi)},
\EE
thereby defining a Boltzmann weight of effective (low-mode) fields
\BE \label{eff_action}
\dd \rho_t(\phi) = \frac{1}{Z} \; \me^{-S_t(\phi)} \mathscr{D} \phi , \qquad \me^{-S_t(\phi)} := \int \! \mathscr{D} \varphi \; P_t(\phi,\varphi) \; \me^{-S_0(\varphi)},
\EE
and the partition function remains invariant.

The stochastic process generated by a Langevin equation is a Markov process, so that future states depend only on the present state, so long as the noise at different times are uncorrelated. This kind of feature was desirable at least in Wilson's philosophy of RG, where any particular blocking step could be carried out by knowing only the previous step. In terms of the abstract distribution $P$ this implies, $\forall \; t > s \geq 0$,
\BE
P(t,0) = P(t, s)P(s, 0), \quad \mrm{or} \quad P(\phi, t; \varphi, 0) = \int \! \mathscr{D} \chi \; P(\phi, t; \chi, s) P(\chi, s; \varphi, 0).
\EE
By considering time-homogeneous Langevin equations (i.e. no explicit $t$-dependence in the LE or the noise variance), the transition function depends only on the difference $t-s$, and we can write $P(\phi, t; \chi, s) = P_{t-s}(\phi, \chi)$. \footnote{The noise variance can be chosen to depend on time, but this spoils the convenience of time-homogeneity.} This property may also be directly computed from the definition, eq. (\ref{Pdef}), suitably modified to have the initial condition $\phi_{t\to s} = \chi$. It follows that the set $\{P_t : t\geq 0\}$ form an abelian semigroup of operators and may be written in terms of a generator $\mcal{L}$ as $P_t = \me^{t\mcal{L}}$ \cite{Pavliotis:2014}. We will discuss $\mcal{L}$ in the last section. For now we simply note that $\mcal{L}$ is the adjoint of the functional differential operator appearing in the FP equation.

Next, consider the usual definition of the expectation value of an operator $\mcal{O}$ in the effective theory,
\BE
\langle \mcal{O}(\phi) \rangle_{S_t} := \frac{1}{Z} \int \! \mathscr{D} \phi \; \mcal{O}(\phi) \; \me^{- S_t(\phi)}.
\EE
By inserting the definition eq. (\ref{eff_action}), and noting that
\BE
\int \! \mathscr{D} \phi \; \mcal{O}(\phi) P(\phi, t; \varphi, 0) = \mathbb{E}_{\mu_0} \big[ \mcal{O}(\phi_t[\varphi;\eta]) \big],
\EE
where $\phi_t[\varphi; \eta]$ denotes the solution of the LE, one readily obtains the equality
\BE \label{Equivalence}
\big\langle \mcal{O}(\phi) \big\rangle_{S_t} = \big\langle \mathbb{E}_{\mu_0} \big[\mcal{O}(\phi_t[\varphi; \eta]) \big] \big\rangle_{S_0}
\EE
This formula states the equivalence of a low-mode FRG effective theory and a double expectation value over the bare fields and the random noise. Since the right-hand side may be calculated without knowledge of the effective action, it further constitutes a generalization of MCRG to FRG for all observables. Notice that there are just as many degrees of freedom $\phi$ as there are $\varphi$ (this is especially clear on the lattice). A possible application of this formula to Swendsen-style MCRG will be discussed at the end of the chapter.

In the next section, we will explore various properties of the effective action $S_t(\phi)$ defined above. First, however, one might wonder why the noise average is necessary in eq. (\ref{Equivalence}), when compared with the corresponding statement for a spin-blocked theory \cite{Swendsen:1979gn},
\BE
\big\langle \mcal{O}(\phi) \big\rangle_{S_b} = \langle \mcal{O}(B_b\varphi) \rangle_{S_0},
\EE
where $B_b$ denotes the blocking operator. This is perhaps clarified by the fact that when spin-blocking, there are fewer blocked spins than bare spins, so the blocked expectation values really involve an integration over ``extra'' degrees of freedom, from the perspective of the effective theory; here the role is played by noise. If one were to choose a blocked lattice of the same size as the original, so that the bare Boltzmann factor were integrated against a delta functional over the whole lattice, the resulting blocked action would be trivial, namely, $S_0(B^{-1}_b \phi)$. Likewise in the continuum, it has long been assumed \cite{Wetterich:1989xg} that a pure $\delta$-function constraint functional is not sufficient to define a non-trivial effective action for continuum FRG. Let us elaborate on this. One might have wanted to define the effective action through
\BE \label{delta_constraint}
\me^{-S_t(\phi)} = \int \! \mathscr{D} \varphi \; \delta(\phi - f_t \varphi) \; \me^{-S_0(\varphi)},
\EE
where $f_t\varphi$ is the solution of a gradient flow equation such as
\BE
\del_t \phi_t(x) = \Delta \phi_t(x),
\EE
or some generalization thereof. The problem with this definition is that it generates a trivial effective action, in the sense to be described. In momentum space, the solution is simply $(f_t \varphi)(p) = \me^{-p^2 t} \varphi(p)$, so one can do a linear change of variables in eq. (\ref{delta_constraint}) and compute
\BE
S_t(\phi) = -\mrm{tr} \ln f_t + S_0(f_t^{-1} \phi).
\EE
Hence, the couplings of the new action are exactly computable, and because their dependence on $t$ is trivially determined by how many powers of $\phi$ and $p^2$ appear in each term, without involving any loop corrections, one verifies that the resulting ``effective action'' is not acceptable.

We remark that the inadequacy of eq. (\ref{delta_constraint}) to define an effective action \textit{does not} mean that the observables computed from gradient-flowed fields are not useful for studying certain RG properties of the system. At the end of the next section, in particular, we will describe how gradient-flowed observables are sufficient for studying the \textit{long-distance} properties of an effective theory that \textit{does} have a well-defined effective action.

\section{The effective theory and fixed points} In what follows, the effective action determined by the stochastic RG transformation will be discussed for the example cases of the gaussian model and $\phi^4_3$ theory. We will show that by a rescaling of variables, the existence of an IRFP of the transformation becomes possible. From the point of view of stochastic processes, the result implies that, for $\phi^4_3$ theory, the stationary solutions of the Fokker-Planck equation may be non-gaussian even though the Langevin equation is linear. Lastly, the correlation functions of the effective theory will be related to gradient-flowed correlations.

\subsection{The effective action} That the EFT defined by a gaussian constraint functional for $\Omega > 0$ is nontrivial can be understood as follows. One may insert the expression eq. (\ref{FP_dist}) for the transition function into the definition of the effective action eq. (\ref{eff_action}), and then expand out the exponent of $P_t(\phi,\varphi)$; the part proportional to $\varphi^2$ modifies the bare theory propagator, and the part linear in $\varphi$ acts as a source term with $J = f_t B_t \phi$. The remaining $\phi^2$ term contributes to the $\phi$ propagator. The result is a relation between effective and bare actions:\footnote{In a sense, this constitutes an exact solution to the FP equation, giving the finite-$t$ distribution $\rho_t$ in terms of the cumulants of $\rho_0$. It is the stochastic RG analog of eq. (\ref{lowmode_action}).}
\BE
S_t(\phi) = F_t + \frac{1}{2} (\phi, B_t \phi) - W_{0}^{(t)}(B_t f_t \phi),
\EE
where $F_t$ is due to the normalization of $P_t(\phi,\varphi)$, and $W^{(t)}_0(J) = \ln \langle \me^{(J,\phi)} \rangle_{S_0^{(t)}}$ is the generator of connected Green functions for the bare theory $S_0$ with a modified $t$-dependent inverse propagator
\BE
[\mDelt^{(t)}_{0}]^{-1} := \mDelt_{0}^{-1} + h_t, \quad h_t := f_t B_t f^\top_t.
\EE
Expanding the generator term in $\phi$ yields a formula which allows for the systematic computation of effective vertices,
\BE
W_0^{(t)}(B_t f_t \phi) = \sum_{n=0}^\infty \frac{1}{n!} [W_0^{(t)}]^{(n)} (B_t f_t \phi, \cdots , B_t f_t \phi).
\EE
It is then apparent that the effective action for any finite $t$ is indeed non-trivial, since the vertices of $S_t$ contain the dynamics of the bare theory via the $[W^{(t)}_0]^{(n)}$. 

The scale $\mLam_t$ of the effective theory may be determined by looking at the effective 2-point function at tree level, after isolating the quadratic part of $S_t(\phi)$:
\BE
\langle \phi(p) \phi(-p) \rangle_{S_t}^\mrm{tree} = A_t(p) + f_t^2(p) \mDelt_0(p) = A_t(p) + \frac{\me^{- p^2(a_0^2 + 2 t)}}{p^2 + m_0^2},
\EE
where the inverse cutoff $a_0 = \mLam_0^{-1}$ has been used, and we recall that $A_t$ is given by eq. (\ref{A_momspace}). In position space, the first term decays rapidly at large distances with respect to the first. The second term is a Schwinger-regularized propagator; we therefore observe that the effective cutoff induced by the stochastic RG transformation is
\BE \label{effective_cutoff}
\mLam_t^{-2} = \mLam_0^{-2} + 2t, \quad \mrm{or} \quad \mLam_t = \frac{\mLam_0}{\sqrt{1+ 2\hat t}},
\EE
where the dimensionless flow time $\hat t = \mLam_0^2 t$ has been introduced. The continuous scale factor is therefore $b_t = \sqrt{1 + 2 \hat t}$. We will take another look at the effective correlation functions and the function $A_t$ in the next section.

We can make sense of the odd-looking factors of $f_t$ and $B_t$ that appear in the effective action as follows. First, the additive $h_t$ in the propagator $\mDelt_0^{(t)}$ acts as a sliding IR cutoff for the \textit{bare} theory, since
\BE
\lim_{p\to 0} h_t(p) = \frac{1}{t},
\EE
which means that as $t$ increases, more of the bare field modes get integrated out. For example, in the case of $\phi^4_d$ theory (discussed in more detail in the next subsection), the momentum-independent part of the 1-loop contribution to the amputated effective 4-point vertex in $W^{(t)}_0$ is proportional to\footnote{We choose to consider the mass term in $S_0$ as part of the interaction $V_0(\phi)$ from now on.}
\BE
\int_{\mathbb{R}^d} \frac{\dd^d k}{(2\pi)^d} [\mDelt^{(t)}_{0}(k)]^2 = \int_{\mathbb{R}^d} \frac{\dd^d k}{(2\pi)^d} \; \frac{\me^{-2 k^2 a_0^2}}{\big(k^2 + h_t(k) \big)^2} \; .
\EE
We observe that the presence of $h_t$ in the denominator, combined with the multiplicative bare cutoff function, effectively restricts the domain of integration to $\| p \| \in [\mLam_t, \mLam_0]$, similarly to what one would have found in a standard (sharp) high-mode elimination RG step $\mLam_0 \to \mLam$, where the domain of the integral would be $\| p \| \in [\mLam, \mLam_0]$ (see \cite{Kopietz:2010zz} for details). Next, note that the argument $B_t f_t \phi$ of $W^{(t)}_0$ in $S_t$ implies that the $[W^{(t)}_0]^{(n)}$ vertices are multiplied by a factor of
\BE
B_t(p) f_t(p) = K_{0}^{-1}(p) \frac{2 \omega(p) f_t(p)}{1-f_t^2(p)}
\EE
for each factor of $\phi(p)$. Since the vertices $[W^{(t)}_0]^{(n)}$ are connected $n$-point functions, which have $n$ factors of external propagators $\mDelt_0^{(t)}(p_i) \propto K_0(p_i)$ attached to them, we see that the effective vertices decay like $f_t(p_i) = \me^{-p^2_i t}$ and therefore strongly suppress the $\| p_i \| \gg \mLam_t$ contribution of the $n$-point functions. Moreover, the leading momentum behavior of the products of $B_t f_t$ with $\mDelt^{(t)}_0$ demonstrates that they are, in a sense, \textit{amputated},
\BE
B_t(p) f_t(p) \mDelt^{(t)}_0(p) = 1 - \frac{1}{2} (p^2 t)^2 + O(p^8 t^4),
\EE
in a manner similar to what was found under smooth high-mode elimination. Thus, in sum, the effective vertices are amputated connected $n$-point functions to leading order in external momenta,  which are heavily damped in the UV $(\| p \| \gg \mLam_t)$, and whose loop corrections effectively involve domains of integration $\| p \| \in [\mLam_t, \mLam_0]$. It is also noteworthy that the external momentum dependence implied by the amputation formula above goes like powers of $p^2 / \mLam_t^2$, for $\mLam_t^{-2} \gg \mLam_0^{-2}$, as one expects from the general philosophy of effective field theory.

\subsection{Gaussian fixed point} I'll begin the discussion of possible fixed points of the stochastic RG transformation with the gaussian model, which is explicitly solvable in a manner similar to the Wilson-Kogut ERGE.
Here we consider the existence of a gaussian fixed point in the case where
\BE
S_{\mLam_0}(\varphi) = \frac{1}{2} ( \varphi, M_{\mLam_0} \varphi), \quad M_{\mLam_0}(p) = c p^2 K_0^{-1}(p),
\EE
with drift $\omega(p) = p^2$ in the LE. A straight-forward gaussian integration yields the exact effective action at time $t$:
\BE
S_t(\phi) = \frac{1}{2} \int_p \frac{2}{\Omega} \frac{p^2 K_0^{-1}(p)}{1 + (2/\Omega c - 1) \me^{-2p^2t}} \phi(p) \phi(-p).
\EE
This action has the expected OU process stationary limit as $t\to\infty$,
\BE
S_\infty(\phi) = \frac{1}{\Omega} \int_p p^2 K_0^{-1}(p) \phi(p) \phi(-p),
\EE
which is therefore independent of the choice of $c$ in the bare action. The rescaled effective action, however, has a different limit. Let $p = \mLam_0 \bar p / b_t$, yielding
\BE
S_t(\phi) = \frac{1}{2} \mLam_0^{d+2} b_t^{-d-2} \int_{\bar p} \frac{2}{\Omega} \frac{\bar p^2 K_0^{-1}(\mLam_0 \bar p / b_t)}{1 + (2/\Omega c - 1) \me^{-2\bar p^2 \mLam_0^2 t / b_t^2}} \phi(p) \phi(-p).
\EE
Note that 
\BE
K_0^{-1}(\mLam_0 \bar p / b_t) = \me^{\bar p^2 / b_t^2}, \quad  \me^{-2\bar p^2 \mLam_0^2 t / b_t^2} = \me^{-\bar p^2 (1 - b^{-2}_t)}.
\EE
The asymptotic limit $b_t \to \infty$ then has leading behavior
\BE
S_t(\phi) = \frac{1}{2} \mLam_0^{d+2} b_t^{-d-2} \int_{\bar p} \frac{2}{\Omega} \frac{\bar p^2}{1 + (2/\Omega c - 1) \me^{-\bar p^2}} \phi(p) \phi(-p).
\EE
We see that in order to get a non-uniform distribution ($\rho_t \neq 1$) in the limit, we must look at rescaled fields $\phi(p) = \mLam_0 ^{d_\phi} b_t^{-d_\phi} \mPhi(\bar p)$, leading to a stationary action
\BE\label{SRG_gfp}
S_*(\mPhi) = \lim_{t\to\infty} S_t(\phi)\Big|_{\phi(p) = \mLam_0 ^{d_\phi} b_t^{-d_\phi} \mPhi(\bar p)} = \frac{1}{\Omega} \int_{\bar p} \frac{\bar p^2}{1 + (2/\Omega c - 1) \me^{-\bar p^2}} \mPhi(\bar p) \mPhi(-\bar p).
\EE
Observe that a fixed point exists for every choice of $c$: the GFP exists as a \textit{line} of fixed points, parameterized by the bare coupling. The canonical choice $c=1$ has fixed point (assuming $\Omega=1$)
\BE \label{standard_gfp}
S_*(\mPhi)|_{c=1} = \int_{\bar p} \frac{\bar p^2}{1 + \me^{-\bar p^2}} \mPhi(\bar p) \mPhi(-\bar p).
\EE
Notice that the rescaled effective action has a regularization-independent (indeed, unregularized) fixed point, since the $K_0$ factor disappears in the limit. However, the $\me^{-\bar p^2}$ term, which came from the heat kernel $f_t(p)$, exhibits the scheme-dependence (i.e. choice of flow) of the fixed points thereby obtained.

For the sake of general applicability, we now discuss the parallel derivation in the lattice gaussian model. On the lattice, there is  sharp cutoff $\mLam_0 = \pi / a_0$ and the drift is $\omega(p) = \hat p^2$, with $\hat p_\mu = (2/a) \sin p_\mu a_0 / 2$. The bare action is taken to be
\BE
S_{a_0}(\varphi) = \frac{1}{2} (\varphi, M_{a_0} \varphi), \quad M_{a_0}(p) = c \hat p^2.
\EE
Taking $\Omega = c = 1$ for simplicity, the effective action is then
\BE
S_t(\phi) = \frac{1}{2} \int_p^{\pi/a_0} \frac{2 \hat p^2}{1 + \me^{-2\hat p^2t}} \phi(p) \phi(-p).
\EE
Again, we see that the infinite time limit of the unrescaled effective theory is a simple gaussian model, the lattice OU stationary process.

We expect the effective spacing $a_t$ to behave qualitatively like $a_t^2 = a_0^2 + \mrm{c}_0 t$, for some constant $\mrm{c}_0$, since by inspection of the effective action, we see that the effective theory propagator, although it still has sharp cutoff $a_0$, \textit{further} suppresses the high modes according to $\me^{-2t \hat p^2/a_0^2}$. Now define the rescaled momenta by $p_\mu = \bar p_\mu/a_t = \bar p_\mu / a_0 b_t$ and $b_t = a_t / a_0$. One obtains
\BE
S_t(\phi) = a_0^{-d} b_t^{- d} \int_{\bar p}^{\pi b_t} \frac{(4/a_0^2) \sum_\mu \sin^2 \bar p_\mu / 2b_t }{1 + \exp\big[- 8 t/a_0^2 \sum_\mu \sin^2 \bar p_\mu / 2b_t\big]} \phi(p) \phi(-p).
\EE
Next, define the rescaled fields $\mPhi(\bar p) = b_t^{d_\phi} \phi(p)$ as usual and note that $t \propto b_t^2(1-b_t^{-2})/c_0$. Expanding the lattice momenta $\hat p$ in $\bar p / a_0 b_t$, we see that the $b_t \to \infty$ limit picks out only the leading, continuum-like term $\bar p^2$, and the rest are suppressed by $b_t^{-1}$. No other rescaling will lead to a propagating fixed point theory. It follows that the fixed point action is described by the same action we found in the direct continuum approach, eq. (\ref{standard_gfp}), which is an expression of universality.

We can compute the scaling operators at the GFP as follows. Let $\tau = \ln \mLam_0 / \mLam_t$, and perturb the fixed point as $S_\tau = S_* + \mcal E_\tau$, keeping only first orders in $\mcal E_\tau$. One finds
\BE
\del_\tau \mcal E_\tau = - \frac{1}{2} K_* \Big[ 2 \Bnab S_* \circ \Bnab \mcal E_\tau - \Bnab^2 \mcal E_\tau \Big] - (D-\omega) \mPhi \circ \Bnab \mcal E_\tau,
\EE
which is separable for $\mcal E_\tau = T(\tau) \mcal R(\mPhi)$, leading to $T(\tau) = \me^{\lambda \tau}$ and the eigenvalue equation
\BE
\lambda \mcal R = -\frac{1}{2} K_* \Big[ 2 \Bnab S_* \circ \Bnab \mcal R - \Bnab^2 \mcal R \Big] - (D-\omega) \mPhi \circ \Bnab \mcal R.
\EE
We give as an example the solution for quadratic scaling operators. Letting $\mcal{R}(\mPhi) = \mPhi \circ g \mPhi$ and $S_*(\mPhi) = \frac{1}{2} \mPhi \circ f \mPhi$, for $g(p)$ to be determined, leads to (recall $d_\phi + d/2 = -1$)
\BE
- \lambda g = 4 f g - 2g + p \cdot \nabla_p g - 2 p^2 g,
\EE
or in spherical coordinates, and using $f$ from eq. (\ref{standard_gfp}),
\BE
p \frac{\dd g}{\dd p} = \Big( 2 - \lambda +2 p^2 - \frac{4 p^2}{1 + \alpha \me^{-p^2}} \Big) g.
\EE
The solution is
\BE
g(p) = C_0 \; \frac{p^{2-\lambda}}{(1 + \alpha \me^{-p^2})^2},
\EE
but how do we determine the permissible values of $\lambda$? By demanding analyticity as $p \to 0$. In \cite{Wilson:1973jj}, Wilson and Kogut argue that non-analyticity would lead to unacceptable nonlocality in the perturbed action. Hence, $2 - \lambda = 2m$ for $m \in \mathbb{Z}_+$. The eigenperturbations are then
\BE
g_m(p) = \frac{(p^2)^m}{(1+\alpha \me^{-p^2})^2}.
\EE
The general solution to the full perturbation from $S_*$ is then
\BE
\mcal E_\tau(\mPhi) = \sum_{m = 0}^\infty \veps_m \me^{\lambda_m \tau} \mPhi \circ g_m \mPhi, \quad \lambda_m = 2 - 2m.
\EE
Hence $m=0$ gives the relevant mass deformation, $m=1$ gives the marginal (redundantly so) kinetic term deformation, and $m \geq 2$ gives irrelevant operators.

The analysis above can be extended to higher operators as well. The $\Bnab^2$ term implies that only polynomials in $\mPhi$ are nonzero solutions, however. For the quartic scaling operators, one tries
\BE
\mcal R(\mPhi) = \frac{1}{2} g_2(\mPhi, \mPhi) + \frac{1}{4!} g_4(\mPhi, \dots , \mPhi),
\EE
which leads to a coupled system of PDE's for $g_2$ and $g_4$. We leave its solution as an exercise for the reader.

\subsection{Fixed point in $\phi^4_3$} For the case of interacting $\phi^4_3$ theory, we cannot solve the problem exactly, so we resort to perturbation theory for the sake of comparison to the high-mode elimination RG in chapter 3, and we will find that the two approaches are quite similar.

One might initially think that the effective action, written as an integration against the bare density, eq. (\ref{eff_action}), has a gaussian infinite flow time limit, as
\BE\label{trivlim}
\lim_{t \to \infty} S_t(\phi) = \frac{1}{2} (\phi, B_\infty \phi),
\EE
where $B_\infty(p) = 2 K_{0}^{-1}(p) \omega(p)$, due to the exponential decay of $f_t$. Indeed, it is well-known that the Ornstein-Uhlenbeck process has a gaussian stationary distribution. As we saw in the gaussian model, however, rescaling can make a big difference. In this case we should expect that the fixed point theory is generally interacting.

To understand qualitatively why the gaussian limit is not obtained, note that the properties of the drift $\omega$ imply that the zero mode of the bare field is not suppressed (see eq. (\ref{constraint_functional})); only its variance changes. Since the zero-mode theory is not gaussian, in general, the flowed distribution will also have a non-gaussian zero mode effective action, implying that the long-distance physics is still non-trivial. This would suggest, however, that the infinite-time degrees of freedom do not propagate. To further clarify the situation, we will look at the flow of the most relevant effective couplings as the RG time $t$ increases, and then we will address the role of rescaling of degrees of freedom, finding that the limit of the \textit{rescaled} effective action differs from the gaussian limit, eq. (\ref{trivlim}), obtained above.

We will treat the mass term also as a perturbation. Denoting the coefficient of $p^2$ in the quadratic part of $S_t(\phi)$ by $c_t$, and the momentum-independent parts of the quadratic and quartic terms, respectively, by $m^2_t, \; \lambda_t$, we find
\begin{align}
c_t &= 1 + O(\lambda_0^2), \\
m^2_t & = m^2_0 + \frac{\lambda_0}{2} I^d_0(t) + O(\lambda_0^2, \lambda_0 m^2_0), \\
\lambda_t & = \lambda_0 - \frac{3\lambda_0^2}{2} C^d_0(t) - 2 \lambda_0^2 t I^d_0(t) + O(\lambda_0^3, \lambda_0 m^2_0),
\end{align}
at 1-loop order, where the loop integrals are given by
\begin{align}
I^d_{0}(t) & = \int_{\mathbb{R}^d} \! \frac{\dd^d p}{(2 \pi)^d} \frac{\me^{-p^2 a_0^2}}{p^2 + h_t(p)} = \Omega_d \int_{\mathbb{R}_+} \! \dd p \; p^{d-3} \me^{-p^2 a_0^2} \tanh p^2 t, \\
C^d_{0}(t) & = \int_{\mathbb{R}^d} \! \frac{\dd^d p}{(2\pi)^d} \; \frac{\me^{-2 p^2 a_0^2}}{\big(p^2 + h_t(p) \big)^2} = \Omega_d \int_{\mathbb{R}_+} \! \dd p \; p^{d-5} \me^{-2 p^2 a_0^2} \tanh^2 p^2 t,
\end{align}
and $\Omega_d = S_{d-1} / (2\pi)^d$. The first integral is superficially divergent, but for $a_0 > 0$, it has a finite $t\to\infty$ limit, and one may compute
\BE
t \frac{\dd}{\dd t} I^d_0(t) = \Omega_d \alpha_1 \; t^{1-d/2} + O(t^{-d/2} a_0^2),
\EE
where $\alpha_1 \approx 0.379064$ for $d=3$. The second integral $C^d_0(t)$ exists even for $a_0=0$, and its time derivative is
\BE
t \frac{\dd}{\dd t} C^d_0(t) = \Omega_d \alpha_2 \; t^{2 - d/2} + O(t^{2 - d/2 - \delta} a_0^{2\delta}),
\EE
where $\delta > 0$ and $\alpha_2 \approx 0.594978$ for $d = 3$.\footnote{Recall that $\phi^4_3$ theory is superrenormalizable, having only two superficially divergent diagrams: the snail and the sunset diagrams.} Hence, to 1-loop order, we find for the derivatives of effective couplings
\begin{align}
t \frac{\dd}{\dd t} m^2_t & = \frac{\lambda_0}{2} \Omega_d \alpha_1 \; t^{1-d/2} + O(t^{-d/2} a_0^2), \\
t \frac{\dd}{\dd t} \lambda_t & = - \lambda_0^2 \Omega_d \big( \smallfrac{3}{2} \alpha_2 + 2 \alpha_1 \big) \; t^{2-d/2} + O(t^{2 - d/2 - \delta} a_0^{2\delta}).
\end{align}
These expressions do not clearly indicate any nontrivial fixed-point behavior at this order in perturbation theory. To proceed further, one must cast the flow equations in terms of rescaled dimensionless quantities, as one usually does to study RG flows. We will find below that such quantities naturally arise after a passive momentum and field redefinition.

Now we introduce dimensionless rescaled variables using the effective scale $\mLam_t$ to give dimension \cite{Morris:1993qb}.  Dimensionless momenta $\bar p$ are defined as in chapter 3, by setting
\BE
\bar p = p / \mLam_t.
\EE
The kinetic term in the effective action therefore becomes
\BE
\frac{1}{2} \int_{\bar p} \mLam_t^{d + 2} \bar p^2 \phi(\mLam_t \bar p) \phi(-\mLam_t \bar p).
\EE
This motivates a change of field variables $\phi \to \mPhi$, where $\mPhi$ is dimensionless:
\BE \label{field_cov}
\phi(\bar p \mLam_t) =: \mLam_t^{d_\phi} \mPhi(\bar  p),
\EE
with $d_\phi = -d/2-1$ being the canonical mass dimension of $\phi$ in momentum space. After doing so, the kinetic term is of the canonical form
\BE
\frac{1}{2} \int_{\bar p} \bar p^2 \mPhi(\bar p) \mPhi(- \bar p)
\EE
at 1-loop order, while the mass and quartic terms pick up factors of $\mLam_t$ which define dimensionless couplings $r_t, \; u_t$ by
\BE
r_t := \mLam_t^{-2} m^2_t, \qquad u_t := \mLam_t^{d-4} \lambda_t.
\EE
We note that these rescalings are all quite familiar when written in terms of the scale factor
\BE
b_t := \frac{\mLam_0}{\mLam_t} \quad \Rightarrow \quad r_t = b_t^2 \hat m_t^2, \quad u_t = b_t^{4-d} \hat \lambda_t,
\EE
reflecting that the mass and the 4-point coupling are relevant at the gaussian fixed point (hats denote quantities rendered dimensionless with $\mLam_0$).

Next, we compute the RG flow equations which describe how the dimensionless variables change with the flow time $t$. In the expression for the derivatives above, one replaces $m^2_0$ and $\lambda_0$ by $m^2_t$ and $\lambda_t$, valid at this order in perturbation theory. The derivatives of the dimensionless couplings with respect to $b$ (dropping $t$-subscripts) are then
\begin{align}
b \frac{\dd r}{\dd b} &= 2 r + \beta_1 u, \\
b \frac{\dd u}{\dd b} &= (4 - d) u - \beta_2 u^2,
\end{align}
up to terms of order $b^{-2}$, since $t = \frac{1}{2} \mLam_t^{-2}(1-b^{-2})$, and where $\beta_1 = 2^{\frac{1}{2}} \Omega_3 \alpha_1, \; \beta_2 = 2^{\frac{1}{2}} \Omega_3 (\frac{3}{2} \alpha_2 + 2 \alpha_1)$ in $d=3$. As $b \to \infty$, the second equation has a nontrivial stationary solution $u_*$, and implies a corresponding critical value $r_*$, which for $d = 3$ are given, at 1-loop order, by $u_* \approx 8.46, \; r_* \approx -0.12$. Linearizing about the fixed point and computing the left-eigenvalues $y_a$ of the stability matrix, one finds that $y_2 = 2, \; y_4 = - 1$, which are crude approximations to the precisely-known values $y_2 = 1.58831(76), \; y_4 = -0.845(10)$ at the Wilson-Fisher fixed point \cite{Hasenbusch:1999mw}. This is our third and final derivation of the WFFP.

The values at 1-loop order from sharp high-mode elimination combined with epsilon expansion in \cite{Kopietz:2010zz} are $y_2 = 1.67, \; y_4 = -1$, which, however, treats the mass non-perturbatively. As a step in that direction, we can extend the analysis above to include terms of order $ru$. This brings in several more non-1PI diagrams, leading to a system of ODE's given by
\begin{align}
b \frac{\dd r}{\dd b} & = 2 r + \mcal{K}_{01} u + \mcal{K}_{20} r^2 + \mcal{K}_{11} r u, \nonumber \\
b \frac{\dd u}{\dd b} &  = u + \mcal{L}_{20} u^2 + \mcal{L}_{11} ur,
\end{align}
where the coefficients in 3d are
\BE
\mcal{K}_{01} = 2^{\frac{1}{2}} \Omega_3 \alpha_1, \quad \mcal{K}_{20} = -1, \quad \mcal{K}_{11} = 2^{+\frac{1}{2}} \Omega_3 (\alpha_1 - \smallfrac{1}{2} \alpha_2),
\EE
\BE
\mcal{L}_{20} = - 2^{\frac{1}{2}} \Omega_3 (\smallfrac{3}{2} \alpha_2 + 2 \alpha_1), \quad \mcal{L}_{11} = -4.
\EE
Setting the $b$-derivatives to zero yields, of course, two fixed points. One is gaussian, and the other is the WFFP, whose couplings are determined from
\BE
u_* = - \frac{1+\mcal{L}_{11} r_*}{\mcal{L}_{20}}, \quad 0 = -\frac{\mcal{K}_{01}}{\mcal{L}_{20}} + \Big[2 - \frac{\mcal{K}_{01} \mcal{L}_{11}}{\mcal{L}_{20}}  - \frac{\mcal{K}_{11}}{\mcal{L}_{20}} \Big] r_* + \Big[ \mcal{K}_{20} - \frac{\mcal{K}_{11} \mcal{L}_{11}}{\mcal{L}_{20}} \Big] r_*^2.
\EE
Expanding the couplings near the WFFP as
\BE
r = r_* + \delta r, \quad u = u_* + \delta u,
\EE
one can linearize the flow equations about the WFFP, finding
\BE
b \frac{\dd}{\dd b}
\begin{bmatrix}
\delta r \\
\delta u
\end{bmatrix}
=
\begin{bmatrix}
2 + 2 \mcal{K}_{20} r_* + \mcal{K}_{11} u_* & \mcal{K}_{01} + \mcal{K}_{11} r_* \\
\mcal{L}_{11} u_* & 1 + 2\mcal{L}_{20} u_* + \mcal{L}_{11} r_*
\end{bmatrix}
\begin{bmatrix}
\delta r \\
\delta u
\end{bmatrix}
.
\EE
By computing the left-eigenvalues, one finds modified exponents $y_2 \approx 1.63, \; y_4 \approx -1.33$; we stress that our formalism is not expected to do any better than the epsilon expansion.

Thus we observe the existence of an IR fixed point in perturbation theory, as we expect in $\phi^4_3$ theory. If we worked to $O(\lambda_0^2)$, we would find, as usual, the necessity of including a wave function renormalization factor $\zeta_t = b_t^{d_\phi} c_t^{1/2}$ to normalize the kinetic term coefficient, so that eq. (\ref{field_cov}) is replaced by
\BE\label{finalrescale}
\phi(\bar p \mLam_t) = \mLam_t^{d_\phi} c_t^{-1/2} \mPhi(\bar  p) = \mLam_0^{d_\phi} \zeta_t^{-1} \mPhi(\bar p),
\EE
which modifies the scaling dimension $\Delta_\phi$ of $\phi$ to include an \textit{anomalous} dimension $\gamma_\phi = O(u_t^2)$, which has a non-zero $t\to\infty$ limit.

The existence of an IR fixed point for the dimensionless, rescaled effective action implies that the expectation values of rescaled effective observables
\BE
\langle \mPhi(\bar p_1) \cdots \mPhi(\bar p_n) \rangle_{S_t} = b_t^{n \Delta_\phi} \mLam_0^{-nd_\phi} \langle \phi(p_1) \cdots \phi(p_n) \rangle_{S_t}
\EE
can have nontrivial infinite flow time limits. In terms of the stochastic RG transformation of section 2, this is written as
\BE
\langle \mPhi(\bar p_1) \cdots \mPhi(\bar p_n) \rangle_{S_t} = b_t^{n \Delta_\phi} \mLam_0^{-nd_\phi}  \big\langle \mathbb{E}_{\mu_0} \big[\phi_t(p_1) \cdots \phi_t(p_n) \big] \big\rangle_{S_0}.
\EE
Since the stochastic RG transformation was generated by a linear Langevin equation, it may be surprising to find that by simply rescaling the correlation functions, one can arrive at a non-gaussian stationary distribution of the Fokker-Planck equation. We also note that the quantities $\mLam_0^{-d_\phi} \phi_t$ correspond directly to the dimensionless field variables one would obtain by numerical integration of the LE on lattice.

Lastly, the Fokker-Planck equation for the stochastic RG transformation may be written in dimensionless form following the procedure outlined in chapter 3. Using $\del_t = b^{-1}\del_b$, one finds the rescaled equation\footnote{We saw in chapter 3 that including the full wave function renormalization $\zeta_t$ modifies $D$ and the diffusion and drift terms. In this case, $\zeta_t$ cancels in the drift term, but survives in the diffusion as $\zeta_t^2$. In conventional FRG, this is typically accounted for by a redefinition $C_t = \zeta_t^{-2} C'_t$. We see two options for accounting for it in SRG, if we wish to have a simple rescaled FP equation. First, we can let $K_0 \to  \zeta_t^{-2} K_0$, rendering the noise variance time-dependent, which must be input by hand in a simulation. A bolder solution may be to consider the field-dependent diffusion matrix $\mSig(\phi) = K_0 \phi \otimes K_0 \phi$, which then implies a total cancellation of $\zeta_t$ factors upon rescaling. Amusingly, such a stochastic process is the field-theoretical generalization of \textit{geometric Brownian motion} \cite{Pavliotis:2014}, which is used in stock market modeling under the name of \textit{Black-Scholes equation}.}
\BE
\del_\tau \rho + D\mPhi \circ \Bnab \rho = \frac{1}{2} K_b \Bnab^2 \rho + \bar \omega \mPhi \circ \Bnab \rho,
\EE
where $\tau = \ln b$ has been defined, and $K_b(\bar p) = \me^{-\bar p^2 / b^2}$. In terms of the flowing action,
\BE
\del_\tau S_\tau = -\frac{1}{2} K_b \Big[ \Bnab S_\tau \circ \Bnab S_\tau - \Bnab^2 S_\tau \Big] - (D-\omega) \mPhi \circ \Bnab S_\tau.
\EE
It can be checked that the GFP eq. (\ref{SRG_gfp}) is a solution to this equation.\footnote{By choosing $\omega$ to be a higher polynomial in $p^2$, we expect that the exotic fixed points discussed in \cite{Wilson:1974mb,Kuti:1994ii} may become accessible.} We see that an explicit time-dependence enters via $K_b(\bar p)$, but in the limit $b \to \infty$, $K_b \to 1$.

\subsection{Correlation functions}
Wilson and Kogut demonstrated a relation between effective $n$-point functions and the bare $n$-point functions in their FRG scheme \cite{Wilson:1973jj}. Recently, the authors of \cite{Sonoda:2019ibh} have noted that this relation is an equivalence between effective correlations and gradient-flowed correlations. In the context of the stochastic approach here, the corresponding relation is given in terms of generators $W(J)$ of connected Green functions by
\BE
W_t(J) = \frac{1}{2}(J,A_t J) + W_0(f_t J),
\EE
where $A_t$ is given by (eq. \ref{Adef}). This relation is simply derived by shifting $\phi' = \phi - f_t \varphi$ in eq. (\ref{eff_action}) and using the definition of the generator,
\BE
\me^{W_t(J)} := \frac{1}{Z_0} \int \! \mathscr{D} \phi \; \me^{-S_t(\phi) + (J,\phi)},
\EE
with $Z_0$ being the free theory partition function \cite{Kopietz:2010zz, ZinnJustin:2002ru}. It follows that the 2-point functions of $S_t$ and $S_0$ are related by
\BE
W^{(2)}_t = A_t + f_t W^{(2)}_0 f_t,
\EE
and higher $n$-points are related by
\BE
W^{(n)}_t(\chi, \dots, \chi) = W^{(n)}_0(f_t \chi, \dots, f_t \chi)
\EE
in multilinear notation. The function $A_t(x,y)$ is determined by the choice of Langevin equation. In the case $\omega(p) = p^2$, for example, one finds an expression in terms of upper incomplete gamma functions
\BE
A_t(z,0) = \frac{1}{8 \pi^{d/2} z^{d-2}} \Big[\Gamma\Big(\frac{d}{2}-1, \frac{z^2}{4 a_t^2}\Big) - \Gamma\Big(\frac{d}{2}-1, \frac{z^2}{4 a_0^2}\Big) \Big],
\EE
where the inverse effective cutoff $a_t = \mLam_t^{-1}$ was used. For large separations $\| z \| \gg a_t$, this quantity decays as a gaussian. The effective propagator is therefore equal to the gradient-flowed propagator asymptotically in $x-y$ (so long as the correlation length $\xi \gg a_t$):
\BE
\langle \phi(x) \phi(y) \rangle_{S_t} \longrightarrow \langle (f_t\varphi)(x)(f_t\varphi)(y) \rangle_{S_0}.
\EE
Note also that if no cutoff function were imposed on the gaussian noise $\eta_t$, there would be a short-distance singularity in $A_t(z,0)$, regardless of whether the bare theory was regulated.

The connected correlators of composite operators also are simply related to their gradient flow counterparts, except we must be careful to define the generators of their $m$-point functions properly. For example, in the case of $\mcal{O} = \phi^2$, the generator of correlators $W_t^{(0,m)}$ is defined by \cite{ZinnJustin:2002ru, Amit:1984ms}
\BE
\me^{W_t(L)} := \frac{1}{Z_0} \int \! \mathscr{D} \phi \; \me^{-S_t(\phi) + \frac{1}{2}(L,\phi^2)}.
\EE
By inserting the definition of $S_t(\phi)$, one may compute the relation between effective and bare generators exactly, as the integrals involved are gaussian. We note, however, that given the simplicity of the Langevin equation, we can just as easily use the explicit solution $\phi_t[\varphi; \eta]$ to compute expectations. For example, the 2-point correlator of the $\phi^2$ composite operator is
\BE
\langle \phi^2(x) \phi^2(y) \rangle_{S_t}^\con = \langle (f_t\varphi)^2(x)(f_t\varphi)^2(y) \rangle_{S_0}^\con +  A_t(x-y)\langle (f_t\varphi)(x)(f_t\varphi)(y) \rangle_{S_0}^\con + 2 A_t(x-y)^2,
\EE
where the connected part of a correlator of local operators $A, \; B$ is defined by
\BE
\langle A(x) B(y) \rangle^\con := \langle A(x) B(y) \rangle - \langle A(x) \rangle \langle B(y) \rangle,
\EE
which again shows the asymptotic equivalence of effective and gradient-flowed quantities.

In sum, what we have found is that the correlation functions of composite operators in the effective theory are equal to the gradient-flowed correlators, up to terms proportional to powers of $A_t(x-y)$, which itself is determined by the drift term $\omega$. Thus, so long as the drift is chosen to imply an exponentially decaying $A_t$, the flowed observables are sufficient to determine the long-distance properties of the effective theory.

\section{Ratio formulas}

The fact that the transition functional $P_t$ satisfies the Fokker-Planck equation implies that observables at finite $t$ satisfy 
\BE
\frac{\del}{\del t} \langle \mcal{O}(\phi) \rangle_{S_t} = \langle \mcal{L} \mcal{O}(\phi) \rangle_{S_t},
\EE
where the \textit{generator} $\mcal{L}$ of the Markov process is a linear differential operator given by
\BE
\mcal{L} = \frac{1}{2} \mSig(\phi,t) \Bnab \circ \Bnab  - \mathscr{B}(\phi,t) \circ \Bnab.
\EE
For the flow we have been considering, the generator takes the form
\BE
\mcal{L} = \frac{1}{2} K_0 \Bnab  \circ \Bnab - \omega \phi \circ \Bnab,
\EE
where $\omega$ is (minus) the laplacian operator. We remark that corresponding equations for the rescaled observables may easily be written.

After a small timestep $\epsilon$, then, successive observables are related by
\BE
\langle \mcal{O}(\phi) \rangle_{S_{t+\epsilon}} = \langle \mcal{O}(\phi) \rangle_{S_{t}} + \epsilon \langle \mcal{L} \mcal{O}(\phi) \rangle_{S_t} + O(\epsilon^2).
\EE
Applied to $n$-point functions, the formula reads\footnote{This formula corresponds to the spin-blocking equation
\BE
\langle B_b \varphi (m_1) \cdots B_b \varphi (m_n) \rangle_{S} = \langle \varphi (m_1) \cdots \varphi(m_n) \rangle_{S} + O(\varepsilon / \Delta m),
\EE
where $B_b \varphi(m) = b^{-d}\sum_\varepsilon \varphi(m+\varepsilon)$ is the blocking operator, $\varepsilon \leq b$, and $\Delta m$ stands for the differences $|m_i - m_j| \gg b \; \forall i \neq j$. This follows from the usual correlator scaling relations of \textit{rescaled} spins $\varphi_b(n/b) := b^{\Delta_\phi} (B_b \varphi)(n)$,
\BE
\langle \varphi_b (m_1/b) \cdots \varphi_b (m_n / b) \rangle_{S_b} = b^{n \Delta_\phi} \langle \varphi (m_1) \cdots \varphi(m_n) \rangle_{S} + O(\varepsilon / \Delta m),
\EE
that one finds in textbooks, e.g. \cite{Cardy:1996xt, Amit:1984ms}. See chapter 1 for more details.
}
\BE \label{npoint_step}
\langle \phi(x_1) \cdots \phi(x_n) \rangle_{S_{t+\epsilon}} = \langle \phi(x_1) \cdots \phi(x_n) \rangle_{S_{t}} + O(\epsilon).
\EE
Writing both sides in terms of the rescaled theory variables, $\phi(x) \propto \mLam_t^{\Delta_\phi} \mPhi(\bar x)$, where the dimless position $\bar x$ is defined by $x = \mLam_t^{-1} \bar x$, one finds
\BE
\mLam_{t+\epsilon}^{n\Delta_\phi} \langle \mPhi(\bar x_1) \cdots \mPhi(\bar x_n) \rangle_{S_{t+\epsilon}} = \mLam_t^{n\Delta_\phi} \big[ \langle \mPhi( \bar y_1) \cdots \mPhi(\bar y_n) \rangle_{S_{t}} + O(\epsilon)\big].
\EE
Motivated by the definition of scale changes $b_t = \mLam_0 / \mLam_t$ with respect to the bare scale, we introduce the \textit{relative} scale change $b_\epsilon(t) := b_{t+\epsilon} / b_t = \mLam_t / \mLam_{t+\epsilon}$. Since the rescaled positions at different scales, $\bar x$ and $\bar y$, refer to the \textit{same} dimensionful position $x$ defined at the bare scale (i.e. in units of $a_0 = \mLam_0^{-1}$), it follows that $\bar y = b_\epsilon \bar x$, and we may write the previous formula as
\BE
 \langle \mPhi(\bar x_1) \cdots \mPhi(\bar x_n) \rangle_{S_{t+\epsilon}} = b_\epsilon(t)^{n\Delta_\phi}  \big[\langle \mPhi(b_\epsilon \bar x_1) \cdots \mPhi(b_\epsilon \bar x_n) \rangle_{S_{t}} + O(\epsilon)\big],
\EE
To the extent that we may neglect the $O(\epsilon)$ terms (which we justify in appendix C), we therefore find a familiar RG scaling relation,
\BE \label{phi_ratios}
 \langle \mPhi(\bar x_1) \cdots \mPhi(\bar x_n) \rangle_{S_{t+\epsilon}} \approx b_\epsilon(t)^{n\Delta_\phi} \langle \mPhi(b_\epsilon \bar x_1) \cdots \mPhi(b_\epsilon \bar x_n) \rangle_{S_{t}}.
\EE
This formula is the stochastic RG analogue of a spin-blocked correlator scaling relation.

The RG scaling property of correlations of scaling operators $\mcal{R}_a$ now follows from an argument identical to that of chapter 3, but now we may relate it to gradient flow. Writing the rescaled variables in eq. (\ref{scalingops}) in terms of $\phi$, and by using the MCRG equivalence between expectations of $\phi$ and $f_t \varphi$, one finds that the factors of $b_\epsilon^{n\Delta_\phi}$ cancel, and the remaining gradient-flowed quantities satisfy a ratio formula:
\BE
\frac{\langle \mcal{R}_a[b_{t+\eps}^{\Delta_\phi} f_{t+\epsilon} \varphi(x)] f_{t+\epsilon} \varphi(x_1) \cdots f_{t+\epsilon} \varphi(x_n) \rangle_{S_0}}{\langle \mcal{R}_a[b_t^{\Delta_\phi} f_{t} \varphi(x)] f_{t} \varphi(x_1) \cdots f_{t} \varphi(x_n) \rangle_{S_0}} \approx b_\epsilon(t)^{\Delta_a}.
\EE
To reiterate, the position arguments in the numerator and denominator are the \textit{same} physical positions in units of $a_0$. We have therefore produced an alternative derivation of correlator ratio formulas of the sort used in chapter 2 in GFRG, making no use of spin-blocking analogies.

\section{Concluding remark}

In Wilson and Kogut's 1973 review, they express the hope that ``a longer range possibility is that one will be able to develop approximate forms of the transformation which can be integrated numerically; if so, then one may be able to solve problems which cannot be solved in any other way.'' One may consider the discrete spin-blocking MCRG of Swendsen and the numerical integration of truncated ERGE's as an actualization of their wish. I believe that the framework of stochastic RG presented in this chapter may provide another actualization, perhaps one even closer to their wish, as it constitutes a direct discretization of the ``blocking'' that leads to their constraint functional.

\section{Future directions}

To wrap up our discussion of stochastic RG, we will now speculate on a few applications which will be pursued in future work.

\subsection{Nonlinear RG's} In 1974, Wilson and Bell (WB) defined and studied the difference between linear and nonlinear RG transformations \cite{Bell:1974vv}. A \textit{linear} RG transformation is one that relates blocked and bare spins linearly, an example of which is the usual transformation
\BE
\varphi_b(x_b) = \frac{b^\Delta}{b^d} \sum_{\veps} \varphi(x + \veps),
\EE
as well as the SRG transformation with drift $\omega(p)$ presented in this chapter. By contrast, a \textit{nonlinear} RG transformation relates blocked and bare spins in a nonlinear way. An example is the majority-rule transformation in the Ising model, whereby one sets the block spins to be $\pm 1$ determined by whichever type is the majority of the block. Another example would be any transformation that involves a projection of the transformed variables back to the original target space of the theory, such as the blocking transformations defined on link variables in gauge theories. An equivalent characterization is that the $n$-point functions of the blocked and bare theory are not linearly related.

The motivation for systematically analyzing these two types of transformation is the following. Linear RG's require the fine-tuning of some parameter, $b^\Delta$ in the block-spin case, in order for the transformation to have a fixed point, whereas the Ising majority-rule and gauge theory blockings seem to automatically produce fixed points without any such tuning. A blocking step $\sigma \to \sigma'$ of the nonlinear transformation WB chose to analyze was implemented via constraint functional:
\BE \label{WB_NL}
\me^{-S'(\sigma')} = \int \mathscr{D} \sigma \;  \exp \Big( - \frac{a}{2} \| \sigma' - b \sigma - c \sigma^3 \|^2 - S_0(\sigma) \Big),
\EE
with sharp cutoffs on the momentum integrals and using the notation $\| \psi \|^2 = (\psi, \psi)$. As opposed to the linear constraint functional, we see that the mean of $\sigma'$ is set to $b\sigma + c \sigma^3$ rather than just $b\sigma$. WB carried out their analysis in perturbation theory, demonstrating that the nonlinear transformation had a fixed point at first order in $c$, which was slightly displaced from the gaussian fixed point with $c=0$. They also reproduced the behavior of the majority-rule transformation: for a certain region of $(b,c)$ parameter space, the transformation had a fixed point with no requirement of tuning $b$. By computing the stability of the new fixed point to perturbations, they determined that the (``nonphysical'') RG eigenvalue associated with field rescalings became negative to first order in $c$, whereas for the linear transformation it was exactly marginal, at the gaussian fixed point, thus explaining the non-necessity of tuning $b$.

The framework of stochastic RG seems particularly well-suited to studying nonlinear RG's both analytically and numerically. This is because, for a discrete time-step $\eps$, one can show that the transition functional produced by a LE
\BE
\del_t \phi_t = - \mathscr{B}(\phi_t) + \eta_t
\EE
is given by \cite{ZinnJustin:2002ru}
\BE
P(\phi,t+\eps;\varphi,t) = C \exp\Big( - \frac{1}{2\eps \Omega} \| \phi - \varphi + \eps \mathscr{B}(\varphi) \|_{K_0}^2 \Big),
\EE
where we write $\| \psi \|^2_K = (\psi, K \psi)$. (This result also allows one to derive a path integral representation of the stochastic process, a feature used extensively in stochastic quantization \cite{Damgaard:1987rr}.) The choice $\mathscr{B}(\phi) = -\Delta \phi + c \phi^3$ reproduces an SRG analogue of WB's transformation. Moreover, we see that such a transformation would be the stochastic  generalization of the nonlinear gradient flow equations that were studied by Fujikawa and Suzuki \cite{Fujikawa:2016qis}. Now, we remarked in chapter 2 that such flows were typically not renormalized by a renormalization of the bare parameters of the theory. It is not clear how to interpret their result in the context of FRG;  nonrenormalizability has been argued to not be a problem in Wilsonian RG. What \textit{is} clear is that the flow must be regularized in any case. The noise is regularized by $K_0$, but the product $\phi^3(x)$ can lead to $\delta(0)$ singularities. A natural choice which guarantees a Markov property is to replace $\phi^3 \to (K_0 \phi)^3$, so that the interacting $\phi^4$ flow equation is
\BE
\del_t \phi_t(x) = \Delta \phi_t(x) - c (K_0 \phi_t)^3(x).
\EE 
This choice would replace
\BE
\delta(x-y) \longrightarrow K_0(x-y) = \frac{\me^{-(x-y)^2/4 a_0^2}}{(4\pi a_0^2)^{d/2}}
\EE
upon functional differentiations ($\Bnab \circ \mathscr{B}$) which arise in the Fokker-Planck equation.

The existence of new fixed points can be analyzed by expanding the flowing action to first order in $c$ in the ERGE for a nonlinear SRG, which reads:
\BE
\del_\tau S_\tau = -\frac{1}{2} K_b \Big[ \Bnab S_\tau \circ \Bnab S_\tau - \Bnab^2 S_\tau \Big] + \mathscr{B}_b \circ \Bnab S_\tau - \Bnab \circ \mathscr{B}_b - D \mPhi \circ \Bnab S_\tau.
\EE
After solving for the new fixed point, one can compute its stability as in the analysis of scaling operator perturbations to the GFP under SRG. A problem which arises for $\phi^4_d$ interacting flow is that, after rescaling, the drift takes the form
\BE
\mathscr{B}_b(\mPhi) = -\bar \del^2 \mPhi(p) + c b^{4-d} (K_b \mPhi)^3.
\EE
For $d=3$, we see that the presence of $b$ prevents writing down an equation for the fixed point action. Two options suggest themselves as possible solutions: (1) Consider only canonically marginal flow actions $\hat S$ (where $\mathscr{B}_b = \Bnab \hat S$), e.g., $\phi^6$ rather than $\phi^4$ in 3d, and (2) Give time-dependence to $c$ by redefining $c \to c b^{d-4}$, but this would spoil the Markov property. However, it is not obvious that the Markov property is a necessity in RG transformations, rather than merely a convenience. The extension to an interacting bare $\phi^4_3$ theory presents a further challenge to the analysis, but the possibility of numerical implementation presents itself as a nonperturbative means of performing the study. And of course, one could apply the derivative expansion of FRG to study the problem analytically, as well. The viability of these options is being pursued by the author.

\subsection{Stochastic MCRG} The continuity of SRG naturally suggests a method for implementing a smooth counterpart to the Swendsen equations described in chapter 1. One begins with the observation that, for any observable $\mcal{O}$, the path integral representation of its expectation value implies
\BE
\frac{\dd}{\dd t} \langle \mcal{O}(\phi) \rangle_{S_t} = - \langle \mcal{O}(\phi) \dot S_t(\phi) \rangle_{S_t}^\con,
\EE
where $S_t$ is the effective action. Writing it as a linear combination of (volume-averaged) operators, and assuming all time-dependence is confined to the couplings, we have
\BE
S_t(\phi) = \sum_i g_i(t) S_i(\phi).
\EE
This leads to
\BE
\frac{\dd}{\dd t} \langle \mcal{O}(\phi) \rangle_{S_t} = - \sum_j \dot g_j(t) \langle \mcal{O}(\phi) S_j(\phi) \rangle_{S_t}^\con.
\EE
Letting $\mcal{O} = S_i$, we find a \textit{continuous} cousin of the Swendsen equations, eq. (\ref{Swendsen_Eqns}),\footnote{Actually, the derivation leading to the Swendsen equation can be carried over line by line, by differentiating with respect to couplings rather than $t$. Both approaches would be interesting to pursue}
\BE\label{cont_swendsen}
\frac{\dd}{\dd t} \langle S_i(\phi) \rangle_{S_t} = - \sum_j \dot g_j(t) \langle S_i(\phi) S_j(\phi) \rangle_{S_t}^\con.
\EE
The expectation values on either side can be measured in a lattice simulation using the stochastic MCRG equivalence,
\BE
\langle \mcal{O}(\phi) \rangle_{S_t} = \langle \mathbb{E}_\mu \big[ \mcal{O}(\phi_t[\varphi;\eta]) \big] \rangle_{S_0}.
\EE
The derivative $\del_t \langle \mcal{O} \rangle$ can be measured either by discretization of the $t$-derivative, or using the Markov property and computing $\langle \mcal{L} \mcal{O} \rangle$. Thus, eq. (\ref{cont_swendsen}) enables one to measure the beta functions $\dot u_i$ of couplings in the action by inverting the matrix of $\langle S_i S_j \rangle^\con$ correlations on the vector of derivatives $\del_t \langle S_i \rangle$. SRG would offer a serious advantage over conventional MCRG, since the latter is necessarily confined to a few blocking steps on any given lattice size, whereas only the integrator step size $\eps$ limits the continuity of SRG.

One is generally more interested in the flow of the scaling variables $u_a$ corresponding to scaling operators $\mcal{R}_a$, however, since one expects that
\BE \label{scaling_deriv}
b_t \frac{\dd u_a(t)}{\dd b_t} = y_a u_a(t).
\EE
The derivatives of the $u_a$ may be constructed from knowledge of $\dot g_i(t)$ and $g_i(t)$ as follows. The scaling variables are linear combinations of couplings $g_i$:
\BE \label{scaling_LC}
u_a = \sum_i c_{ai} g_i.
\EE
The matrix $\bo C = [c_{ai}]$ can be computed numerically by diagonalization of the mixed action operator correlations, since
\BE
\langle S_i S_j \rangle^\con = \sum_{ab} c_{ai} \langle \mcal R_a \mcal R_b \rangle^\con c_{bj},
\EE
and $\langle \mcal{R}_a \mcal{R}_b \rangle^\con$ is diagonal near the IRFP. A caveat is that these expectations are true only of the rescaled effective theory, so that $S_i(\mPhi) = b_t^{m_i\Delta_\phi + \ell_i - d} S_i(\phi_t)$, where $m_i$ is the number of factors of $\phi$ in $S_i$, and $\ell_i$ the number of derivatives. Thus, one must input a value of $\Delta_\phi$, as was necessary in the diagonalization method in chapter 2. To eliminate such a systematic, it is therefore desirable to have determined if the nonlinear SRG can be achieved without needing to tune the rescaling factor, as suggested by Wilson and Bell. Alternatively, one could attempt to measure $\Delta_\phi$ using the correlator method of chapter 2.

One still needs values of the flowing couplings $g_i(t)$ in order to make use of eqs. (\ref{scaling_deriv},\; \ref{scaling_LC}). This can be done by numerically integrating the coupling beta functions obtained from eq. (\ref{cont_swendsen}) up to the desired flow time:
\BE
g_i(t) = g_i(0) + \int_0^t \dd s \; \dot g_i(s),
\EE
knowing that the $t=0$ couplings are the bare couplings one simulates with. The implementation of this procedure is therefore somewhat involved. One needs a fine discretization of the flow time to reduce numerical integration errors in the $g_i(t)$, one needs to know the exact functional form of $b_t$ on the lattice, and one needs to input a value for $\Delta_\phi$ (at least in the linear RG case). Further, a truncation error is introduced by necessarily considering only a finite number of action operators $S_i$, although this would be systematically improvable. Last, and certainly not least, one must perform a double-ensemble average to compute the stochastic expectation values, coming from the integration of a Langevin equation and the bare ensemble average, thereby producing two separate sources of statistical error. On the bright side, the elements of the ensemble of Langevin integrations would be automatically uncorrelated, making \textit{those} errors simple to calculate.

\chapter{Summary}

In this work we have developed new, continuous RG transformations, in the continuum and on the lattice, based on gradient flow and the Langevin equation, and explained their relationship with Wilsonian RG in the form of block-spin RG and functional RG. In fact, RG methods based on GF or LE's essentially constitute the implementation of functional RG on the lattice.

In chapter 2, we saw how to define the GFRG transformation, and described how to measure scaling dimensions of local operators in lattice simulations by virtue of the correlator scaling laws associated with the RG transformation. We applied the method in two scalar field theories, $\phi^4_3, \; \phi^4_2,$ and a 12-flavor SU(3) gauge theory in four dimensions, thereby displaying the general viability of GFRG not only across various physical systems with different field content, but also across various spacetime dimensions. In the 3d scalar model, we produced a numerical determination of the leading four scaling dimensions of the theory, including the $\phi^3$ scaling dimension $\Delta_3$. In the gauge theory, we produced an estimate of the fermion mass anomalous dimension, as well as a first lattice determination of the baryon anomalous dimension. We noted that the method will be applied to 3-dimensional noncompact QED in future work, and that it is already being applied in some interesting 4d systems by other groups, as well.

In chapter 4, we demonstrated an equivalence between functional RG transformations and stochastic processes, based on the observation that functional RG equations for effective actions have the same form as Fokker-Planck equations which govern stochastic processes. The viability of the stochastic RG (SRG) transformation was checked in $\phi^4_3$ theory, where the Wilson-Fisher fixed point was observed in perturbation theory. This result furthermore implied that, from the perspective of stochastic processes, the stationary distribution of the field theoretical Ornstein-Uhlenbeck process (with $\phi^4$ theory initial condition) is non-gaussian, up to a rescaling.  An equivalence of long-distance correlation functions of the SRG effective theory and gradient-flowed correlators was found, which permitted a reinterpretation of the GFRG transformation from chapter 2 as an implication of the SRG transformation. 

Lastly, we speculated on a few possibilities for future work. SRG seems to provide a natural framework in which to study nonlinear RG transformations, which may be practically more useful than the linear RG's we simulated in chapter 2 if it indeed eliminates the requirement of a finely-tuned rescaling factor $b_t^\Delta$ for the field variables being transformed. The continuity of SRG furthermore suggests that a continuous version of Swendsen's MCRG can be carried out, which would allow for the measurement of the RG eigenvalues $y_a$ in a manner distinct from the correlator ratio method of chapter 2.

\bibliographystyle{JHEP}
\bibliography{RG_GF}

\appendix

\chapter{Finite volume heat kernel}

Solving the heat equation in a periodic box of size $L^d$ with initial condition $\phi(0,x) = \varphi(x)$, and with continuous $x\in[0,L]^d$, one finds
\BE
\phi(t,x) = \int_{[0,L]^d} \dd^d y \; K_L(x,t;y,0) \varphi(y),
\EE
where
\BE \label{momspacekernel}
K_L(x,t;y,0) = \frac{1}{L^d} \sum_{n\in\mathbb{Z}^d} \exp\Big[i  p_n \cdot(x-y) - p_n^2 t\Big], \quad p_n = \frac{2\pi n}{L},
\EE
and $K_L$ denotes the finite size heat kernel. This series is that of an elliptic theta function of the third kind, $\Theta_3(z,q)$:
\BE
\Theta_3(z,q) = \sum_{n\in\mathbb{Z}} \me^{2inz} q^{n^2} = 1 + 2 q \cos 2z + 2q^4 \cos 4z + O(q^9).
\EE
Letting $q = \me^{-\tau}$, we write $\Theta_3(z|\tau) := \Theta_3(z, q=\me^{-\tau})$, so
\BE
\Theta_3(z|\tau) = \sum_{n\in\mathbb{Z}} \me^{2inz} \me^{-n^2 \tau} = 1 + 2 \me^{-\tau} \cos 2z + 2\me^{-4\tau} \cos 4z + O(\me^{-9\tau}).
\EE
The heat kernel is then observed to be
\BE
K_L(x,t;y,0) = \frac{1}{L^d} \prod_{\mu=1}^d \Theta_3 \Big(z = \frac{\pi}{L} (x-y)_\mu | \tau = \Big(\frac{2\pi}{L}\Big)^2 t \Big).
\EE
We can derive a more useful formula by applying the Appell transformation identity for theta functions,
\BE
\Theta_3(z|\tau) \equiv \sqrt{\frac{\pi}{\tau}} \; \me^{-\frac{z^2}{\tau}} \; \Theta_3\big(\smallfrac{\pi z}{i\tau} | \smallfrac{\pi^2}{\tau} \big),
\EE
which implies ($x\in[0,L]$)
\BE
\sum_{n\in\mathbb{Z}} \exp\Big[i  p_n x - p_n^2 t\Big] \equiv \frac{L}{\sqrt{4\pi t}} \; \me^{-\frac{x^2}{4t}} \Big[ 1 + 2 \sum_{n=1}^{\infty} \me^{-\frac{L^2 n^2}{4t}} \cosh \frac{nLx}{2t} \Big].
\EE
After this rewriting, we find that the finite size heat kernel is the standard gaussian kernel plus a series that depends on $L^2/t$:
\BE
K_L(z,t;0,0) = \frac{\me^{-\frac{z^2}{4t}}}{(4\pi t)^{\frac{d}{2}}} \prod_{\mu=1}^d \Big[1 + 2 \sum_{n_\mu=1}^{\infty} \me^{-\frac{L^2 n_\mu^2}{4t}} \cosh \frac{L n_\mu z_\mu }{2t} \Big],
\EE
where now $z = x - y$ and $x,y\in[0,L]^d$. Letting the infinite volume kernel be $K := K_{L\rightarrow\infty}$, we have the leading behavior
\BE
K_L(z,t;0,0) = K(z,t;0,0)\Big[ 1 + 2\sum_{\mu=1}^d \me^{-\frac{L^2}{4t}} \cosh \frac{L z_\mu }{2t} + O(\me^{-\frac{L^2}{t}}) \Big]. 
\EE
Solutions $\phi$ in the finite box therefore have a leading infinite volume term, plus $z/L$- and $L^2/t$-dependent finite size terms. A straight-forward estimation for $z < L/2$ implies that the growing term in the hyperbolic cosine is always smaller than the inverse multiplicative factor of $\me^{n^2L^2/4t}$, so that those terms are suppressed for all $z$ as $n$ increases.

\chapter{The Fokker-Planck equation}

A \textit{stochastic process} is, in non-rigorous terms, a sequence of random variables $X_k \in \mathbb{R}$, indexed by $k$, whose probability density $\rho_k(x)$ may evolve along the sequence. Such processes may be discrete, so $k \in \mathbb{Z}$, or continuous, with $k \in I \subset \mathbb{R}$. In the continuous case, it is usually helpful to imagine that the sequence corresponds to time evolution, and the $k$ label is denoted by $t$ instead. We refer the reader to \cite{Pavliotis:2014} for a mathematical account of stochastic processes and to \cite{Glimm:1987ng} for their application in rigorous field theory.

A common way to define continuous stochastic processes $X_t$ is through a Langevin equation (LE), which formally defines the time-evolution of the variables $X_t$ with initial condition $X_{t \to 0} = X_0$. The example of the Ornstein-Uhlenbeck (OU) process was described in chapter 4. For field theoretical processes $\phi_t$, there is a LE for every position (or momentum) label, $\phi_t(x)$. The general form of such LE's is\footnote{It turns out that this equation is not mathematically well-defined: the derivative $\del_t \phi_t$ generally does not exist, in the sense that $\mathbb{E}[(\phi_{t+\veps} - \phi_t)^2] = O(\veps)$. Instead, the rigorous form is given by the ``integral'' form $\dd \phi_t = B_t \dd t + \sigma_t \dd W_t,$ where $W_t$ is the Weiner process. We shall not dwell on this subtlety, however. See \cite{Pavliotis:2014} for details.}
\BE
\del_t \phi_t(x) = -B_t(x) + (\sigma_t \eta_t)(x),
\EE
where $B_t$ is the \textit{drift} vector and $\sigma_t$ is the ``square-root'' \textit{diffusion} matrix, both of which will depend, in general, on $\phi_t$, and can in principle depend explicitly on $t$. The noise $\eta_t(x)$ is (usually) taken to be a gaussian-distributed random variable with mean and covariance given by
\BE
\mathbb{E}_\mu [ \eta_t(x) ] = 0, \quad \mathbb{E}_\mu [ \eta_t(x) \eta_s(y) ] = 2 \pi \delta(t-s) K(x,y),
\EE
where $\mu$ is the measure of the noise distribution and $K(x,y)$ is some kernel ($\Omega \delta(x-y)$ in the simplest case, or a regulator for the noise). Such noise is uncorrelated at different times $t,\;s$ by virtue of the delta function.

Observables of the process $\phi_t$ with initial condition $\varphi$ are the expectation values of functions of the field,
\BE
\mathbb{E}_\mu [ \mcal{O}(\phi_t) ] = \int \dd \mu \; \mcal{O}(\phi_t).
\EE
For a field theory, this is a functional integral. If the initial condition itself is a random variable distributed according to a measure $\dd \rho_0$, then full expectation values may be written as
\BE
\langle \mathbb{E}_\mu [ \mcal{O}(\phi_t) ] \rangle_{\rho_0} = \int \dd \rho_0 \int \dd \mu \; \mcal{O}(\phi_t) = \int \mathscr{D} \phi \; \mcal{O}(\phi) \rho_t(\phi).
\EE
In the last equality we have introduced the time-dependent density $\rho_t(\phi)$. It is important to distinguish the solution $\phi_t = \phi_t(\varphi_0; \eta)$ from the dummy variables $\phi$ that are integrated over in the expectation value. It is clear from the equalities above that $\rho_t(\phi)$ is expressible as
\BE
\rho_t(\phi) = \int \mathscr{D} \varphi \; \mathbb{E}_\mu[\delta(\phi - \phi_t)] \rho_0(\varphi),
\EE
where $\rho_0(\varphi)$ is the initial density of the field variables. The expectation value in the integrand defines the \textit{transition function} $P(\phi,t;\varphi,0)$, which plays a vital role.

The LE determines the time evolution of the stochastic process $\phi_t$, and it is therefore expected to generate a time evolution of the distribution $\rho_t$. This evolution is described by the Fokker-Planck (FP) equation. It can be derived either abstractly by defining certain properties of diffusion processes and computing its time evolution directly, or it can be computed as a limit of a discrete process. We will describe the latter method, based on the derivation in \cite{Batrouni:1985jn}, as it is more useful in the context of lattice simulations.

The discrete form of the equation involves a discretized noise $\tilde \eta_t$, and can be defined by the Euler step
\BE
\phi_{t+\veps}(x) = \phi_t(x) - \veps B_t(x) + \sqrt{\veps} (\sigma_t \tilde \eta_t)(x) = \phi_t(x) - F_\veps(\phi_t, \tilde \eta_t; x),
\EE
where we have defined $F$ for convenience. The noise is assumed to have a gaussian distribution at all times,
\BE
\dd \mu(\eta) = C \exp\Big[ - \frac{1}{2\Omega} \sum_{i=0}^N (\tilde \eta_{t_i}, \tilde \eta_{t_i}) \Big],
\EE
where $t_i = i \veps$, and the noise might be regularized in some fashion, be it a lattice or a cutoff function. We now see how the discrete equation suggests the continuous one: by letting $\tilde \eta_i = \sqrt{\veps} \eta_{t_i}$, the discrete sum becomes a Riemann sum approximating a continuous time integral, and the LE is formally recovered.

The distribution of fields at time $t+\veps$ given $\varphi$ at $t$ is the incremental transition function, and it is defined by
\BE
P(\phi, t+\veps; \varphi, t) := \mathbb{E}_{\mu} \big[ \delta( \phi - \varphi + F_\veps(\varphi, \eta_t)) \big].
\EE
Suppressing the $\eta$ argument of $F_\veps$, the expectation of an observable at $t+\veps$ given $\varphi$ at $t$ is therefore
\BE
\int \mathscr{D} \phi \; \mcal{O}(\phi) \mathbb{E}_{\mu} \big[ \delta( \phi - \varphi + F_\veps(\varphi)) \big].
\EE
Next, let's introduce some more tensor notation. For an $n$-tensor $T$, we can write its contraction with $n$ vectors $v_k$ using an ``interior product'' $\mcal{I}$ as
\BE
\mcal{I}_v T := T(v, \dots ), \quad \mrm{and} \quad \mcal{I}_{v_n} \cdots \mcal{I}_{v_1} T = T(v_1, \dots, v_n).
\EE
If all the $v_k$ are identical, we can write $\mcal{I}^{(n)}_v T = T(v, \dots , v)$. For example, the expansion of a generating function $W(J)$ in $J$ can be written as
\BE
W(J) = \sum_{n=0}^\infty \frac{1}{n!} \; W^{(n)} (J, \dots, J) = \sum_{n=0}^\infty \frac{1}{n!} \; \mcal{I}^{(n)}_J W^{(n)}.
\EE
The interior product notation is more useful when $v$ is a differential operator which can act on $T$, so that we can write the derivatives on the left side of $T$ and therefore preserve the ``acting to the right'' convention, as we will observe in eq. (\ref{intprod_example}) below.\footnote{This is a deviation from the conventions of differential geometry.}

By changing variables $\phi = \phi' - F_\veps$ and assuming we may expand $\mcal{O}$ about $\phi'$, we obtain
\begin{align}
\int \mathscr{D} \phi' \mathbb{E}_{\mu} & \big[ \delta( \phi' - \varphi) \mcal{O}(\phi' - F_\veps(\varphi)) \big]  \nonumber \\
& = \sum_{n = 0}^\infty \frac{(-1)^n}{n!} \int \mathscr{D} \phi' \delta( \phi' - \varphi) \mathbb{E}_{\mu} \big[ \mcal{I}_{F_\veps(\varphi)}^{(n)} (\Bnab_{\phi'} \otimes \cdots \otimes \Bnab_{\phi'} \mcal{O}(\phi')) \big] \nonumber \\
& = \sum_{n = 0}^\infty \frac{(-1)^n}{n!}\mathbb{E}_{\mu} \big[ \mcal{I}_{F_\veps(\varphi)}^{(n)} (\Bnab_\varphi \otimes \cdots \otimes \Bnab_\varphi \mcal{O}(\varphi)) \big].
\end{align}
Now integrate against the initial distribution $\rho_t(\varphi)$, integrate by parts $n$ times per term in the expansion, and then relabel $\varphi = \phi$:
\begin{align} \label{intprod_example}
 \sum_{n = 0}^\infty \frac{1}{n!} \int \mathscr{D} \phi \; \mcal{O}(\phi) \mcal{I}_{\sBnab}^{(n)} \Big( \mathbb{E}_{\mu} \big[ F_\veps(\phi) \otimes \cdots \otimes F_\veps(\phi) \big] \rho_t(\phi) \Big).
\end{align}
This being true for arbitrary $\mcal{O}(\phi)$, we identify the distribution at $t + \veps$:
\BE \label{discrete_dist}
\rho_{t+\veps}(\phi) = \sum_{n = 0}^\infty \frac{1}{n!} \mcal{I}_{\sBnab}^{(n)} \Big( \mathbb{E}_{\mu} \big[ F_\veps(\phi) \otimes \cdots \otimes F_\veps(\phi) \big] \rho_t(\phi) \Big).
\EE

To obtain the $O(\veps)$ contribution to the expansion of eq. (\ref{discrete_dist}), it is sufficient to calculate the first two moments of $F_\veps$. First note that
\begin{align}
\mathbb{E}_\mu \big[ \sigma_t(\phi) \eta_t \big] & = 0, \nonumber \\
\mathbb{E}_\mu \big[ \sigma_t(\phi) \eta_t \otimes \sigma_t(\phi) \eta_t \big] & = \veps \Omega \; \sigma_t K_0 \sigma^\top_t.
\end{align}
The first two moments are then
\begin{align}
\mathbb{E}_\mu\big[ F_\varepsilon(\varphi) \big] & = \veps B_t(\varphi), \nonumber \\
\mathbb{E}_\mu\big[ F_\veps(\varphi) \otimes F_\veps (\varphi) \big] & = \veps^2 B_t(\varphi) \otimes B_t(\varphi) + \veps \Omega \; \sigma_t(\varphi) K_0 \sigma^\top_t(\varphi),
\end{align}
since expectations of a product of an odd number of $\eta$'s vanish. This implies the absence of half-integer powers of $\veps$ in eq. (\ref{discrete_dist}). The leading terms in the expansion are then
\BE
\rho_{t+\veps}(\phi) = \rho_t(\phi) + \veps \Big[ \Bnab \circ \big( B_t(\phi) \rho_t(\phi) \big) + \frac{1}{2} \Omega \;  \mcal{I}_{\sBnab} \mcal{I}_{\sBnab} \big( \sigma_t(\phi) K_0 \sigma_t^\top(\phi) \rho_t(\phi) \big) \Big] + O(\veps^2).
\EE
By dividing by $\veps$ and taking the limit $\veps \to 0$, we obtain the Fokker-Planck equation
\BE
\del_t \rho_t(\phi) = \Bnab \circ \big( B_t(\phi) \rho_t(\phi) \big) + \frac{1}{2} \Omega \;  \mcal{I}_{\sBnab} \mcal{I}_{\sBnab} \big( \mSig_t(\phi) \rho_t(\phi) \big),
\EE
where the diffusion matrix is given by
\BE
\mSig_t(\phi) = \sigma_t(\phi) K_0 \sigma_t^\top(\phi).
\EE
In the standard case of field-independent diffusion, $\sigma_t(\phi) = \mathbb{I}$, one recovers the simplest type of FP equation:
\BE
\del_t \rho_t(\phi) = \Bnab \circ \big( B_t(\phi) \rho_t(\phi) \big) + \frac{1}{2} \Omega K_0 \Bnab \circ \Bnab \rho_t(\phi).
\EE
By writing it in its ``continuity'' form,
\BE
\del_t \rho_t(\phi) = \frac{1}{2} \Omega K_0 \Bnab \circ \Big[ \Bnab \rho_t(\phi) + 2 \Omega^{-1} K_0^{-1} B_t(\phi) \rho_t(\phi) \Big],
\EE
one observes that an equilibrium distribution $\rho_*(\phi)$ exists if $B_t$ has a limit $B_*$ as $t \to \infty$,
\BE
\rho_*(\phi) = \mcal{N} \exp \Big[ - \frac{2}{\Omega} \int_\gamma K_0^{-1} B_* (\phi') \circ \dd \phi' \Big],
\EE
where $\gamma$ is a path in field space ending at $\phi$, and of course requires some care to define properly. In the case of \textit{dissipative} drift, one has $B_t (\phi) = \Bnab S(\phi)$, so that
\BE
\rho_*(\phi) = \mcal{N} \exp \Big[ - \frac{2}{\Omega} S_0(\phi) \Big],
\EE
an observation that motivates the enterprise of stochastic quantization (SQ) \cite{Damgaard:1987rr}.\footnote{For the curious reader, we refer also to the older work of Nelson \cite{PhysRev.150.1079} for deeper connections between quantum mechanics and stochastic processes.} In SQ, one \textit{defines} a euclidean quantum field theory with regularized action $S_0(\phi)$ as the equilibrium limit of a stochastic process with drift $B = \Bnab S$ and noise regulated by $K_0$. SQ was studied extensively in the 1980's and may even be implemented numerically to simulate lattice field theories. In the lattice context, the discrete form of the distribution, eq. (\ref{discrete_dist}), implies an $\veps$-dependent equilibrium distribution that may be systematically ``improved'' by subtracting appropriate $O(\veps)$ terms in the discrete LE. See \cite{Batrouni:1985jn} for details.

\chapter{Action of $\mcal{L}$ on $n$-point functions}

Here we demonstrate that the $O(\epsilon)$ terms in eq. (\ref{npoint_step}) decay like a gaussian at large distances. For the $n$-point functions of $\phi$, the action of $\mcal{L}$ yields
\begin{align} \label{L_action1}
\langle \mcal{L}[\phi(x_1) \cdots \phi(x_n)] \rangle_{S_t} & = \sum_{i \neq j}^n K_0(x_i - x_j) \langle \phi(x_1) \cdots \hat \phi(x_i) \cdots \hat \phi(x_j) \cdots \phi(x_n) \rangle_{S_t} \\ 
& + \sum_{i=1}^n \langle \phi(x_1) \cdots \Delta_i \phi(x_i) \cdots \phi(x_n) \rangle_{S_t}, \nonumber
\end{align}
where $\hat \phi(x_i)$ means that $\phi(x_i)$ is absent from the product, and $\Delta_i$ is the laplacian with respect to $x_i$. The kernel $K_0(x_i - x_j)$ decays as a gaussian for $\| x_i - x_j \| \gg a_0$. From the MCRG equivalence, both expectation values may be written in terms of gradient-flow $n$-point functions, up to terms involving $A_t(x_i - x_j)$, which decay like $\me^{-x_{ij}^2 \mLam^2_t}$. For the gradient flow terms, note that an insertion of $\Delta_i (f_t \varphi)(x_i)$ satisfies the heat equation
\BE
\Delta_i (f_t \varphi)(x_i) = \del_t (f_t \varphi)(x_i),
\EE
and therefore the second sum in eq. (\ref{L_action1}) may be written as
\BE
\sum_{i=1}^n \langle (f_t \varphi)(x_1) \cdots \Delta_i (f_t \varphi)(x_i) \cdots (f_t \varphi)(x_n) \rangle_{S_0} = \frac{\del}{\del t} \langle (f_t \varphi)(x_1) \cdots (f_t \varphi)(x_n) \rangle_{S_0}.
\EE
The right-hand side typically decays fast at large distances. For example, in the $d=3$ massless gaussian model, the flowed 2-point function is given by
\BE
\langle (f_t \varphi)(x) (f_t \varphi)(y) \rangle_{\mrm{g}} = \frac{\mrm{erf}(z \mLam_t/2)}{4\pi z},
\EE
and its time derivative decays at large distances like $\me^{-z^2 \mLam_t^2/4}$. This is a reflection of the fact that gradient flow does not suppress the zero-mode of $\varphi$. We therefore expect it to hold also in the case of interacting theories at criticality, where there are no other scales for $t$ to couple to at large distances; this was observed empirically in chapter 2, where we saw that the ratios of $\phi\phi$ correlators exhibit minimal movement with flow time.

Applied to composite operators such as $\phi^2$ and $\phi^4$, one finds that the application of $\mcal{L}$ is not exponentially decaying, but rather implies mixing among correlations. This is expected, as such monomial operators are not in fact scaling operators; we expect that only certain linear combinations of the monomial operators will exhibit exponential decay like what we have observed for $\mcal{L}(\phi\phi)$.


\end{document}